\documentclass[useAMS,usenatbib]{mnras}		         

\usepackage[]{natbib}
\usepackage{graphicx}	
\usepackage{deluxetable}
\usepackage{amsmath}
\usepackage[T1]{fontenc}		
\usepackage[usenames]{color}			
\definecolor{jgreen}{rgb}{0.0, 0.42, 0.24}
\definecolor{grey}{rgb}{0.5, 0.5, 0.5}
\definecolor{edit}{rgb}{0.0,0.0,0.0}
\definecolor{edit}{rgb}{1.0,0.0,0.0}

\newcounter{species}
\def\ion#1#2{\setcounter{species}{#2}#1$\;${\sc\roman{species}}\relax}

\newcommand{\kms}{km~s$^{-1}$}

\newcommand{\Msun}{M$_{\odot}$}

\newcommand{\OIIIw}{[O$\,\textsc{iii}]$~$\lambda$5007}
\newcommand{\OIIIdblt}{[O$\,\textsc{iii}]$~$\lambda\lambda$4959,5007}
\newcommand{\OIsky}{[O$\,\textsc{i}]$~$\lambda$5577}

\newcommand{\Hb}{{H}$\beta$}

\def\lsim{\lower0.3em\hbox{$\,\buildrel <\over\sim\,$}}
\def\gsim{\lower0.3em\hbox{$\,\buildrel >\over\sim\,$}}

\newcommand{\pone}{Paper~\textsc{I}}
\newcommand{\ptwo}{Paper~\textsc{II}}

\def\kms{\,km~s$^{-1}$}      

\def\sun{\hbox{$\odot$}}

\def\lesssim{\mathrel{\hbox{\rlap{\hbox{%
 \lower4pt\hbox{$\sim$}}}\hbox{$<$}}}}
\def\gtrsim{\mathrel{\hbox{\rlap{\hbox{%
 \lower4pt\hbox{$\sim$}}}\hbox{$>$}}}}

\title[Supermassive black hole binaries]{A Large Systematic Search for Close Supermassive Binary and Rapidly Recoiling Black Holes - III. Radial Velocity Variations}
\author[J. C. Runnoe et al.]
{\parbox{\textwidth}{Jessie C. Runnoe$^{\,1}$\thanks{Current address: Department of Astronomy, University of Michigan, 1085 S. University, Ann Arbor, MI 48109-1107; e-mail: \texttt{runnoejc@umich.edu}},
Michael Eracleous$^{\,1,2,3}$, 
Alison Pennell$^{\,4}$,
Gavin Mathes$^{\,1,5}$,
Todd Boroson$^{\,6}$,
Steinn Sigur{\dh}sson$^{\,1}$, 	
Tamara Bogdanovi\'c$^{\,2}$,
Jules P. Halpern$^{\,7}$,
Jia Liu$^{\,7,8}$,
and Stephanie Brown$^{\,1}$}
\vspace{0.4cm}\\ \\
\parbox{\textwidth}{$^{1}$Department of Astronomy \& Astrophysics, and Institute for Gravitation and the Cosmos, The Pennsylvania State University, 525 Davey Lab, University Park, PA 16802, USA\\
$^{2}$Center for Relativistic Astrophysics, School of Physics, Georgia Institute of Technology, Atlanta, GA 30332, USA \\
$^{3}$Department of Astronomy, University of Washington, Box 351580, Seattle, WA 98195, USA \\
$^{4}$Department of Electrical Engineering, The Pennsylvania State University, 342 Information Sciences and Technology Building, University Park, PA 16802, USA\\
$^{5}$Department of Astronomy, New Mexico State University, Las Cruces, NM 88003, USA \\
$^{6}$Las Cumbres Observatory Global Telescope Network, Goleta, CA 93117, USA \\
$^{7}$Columbia Astrophysics Laboratory, Columbia University, 550 West 120th Street, New York, NY 10027-6601, USA \\
$^{8}$Department of Astrophysical Sciences, Princeton University, Princeton, NJ 08544, USA 
}}

\begin{document}		

\date{Preprint 2016 July 14}

\pagerange{\pageref{firstpage}--\pageref{lastpage}} \pubyear{2016}

\maketitle

\label{firstpage}

\begin{abstract}
We have been spectroscopically monitoring 88 quasars selected to have broad \Hb\ emission lines offset from their systemic redshift by thousands of km~s$^{-1}$. By analogy with single-lined spectroscopic binary stars, we consider these quasars to be candidates for hosting supermassive black hole binaries (SBHBs). In this work we present new radial velocity measurements, typically 3--4 per object over a time period of up to 12 years in the observer's frame. In 29/88 of the SBHB candidates no variability of the shape of the broad \Hb\ profile is observed, which allows us to make reliable measurements of radial velocity changes. Among these, we identify three objects that have displayed systematic and monotonic velocity changes by several hundred km~s$^{-1}$ and are prime targets for further monitoring. Because the periods of the hypothetical binaries are expected to be long, we cannot hope to observe many orbital cycles during our lifetimes. Instead, we seek to evaluate the credentials of the SBHB candidates by attempting to rule out the SBHB hypothesis. In this spirit, we present a method for placing a lower limit on the period, and thus the mass, of the SBHBs under the assumption that the velocity changes we observe are due to orbital motion. Given the duration of our monitoring campaign and the uncertainties in the radial velocities, we were able to place a lower limit on the total mass in the range $4.7\times10^4-3.8\times10^8$~\Msun, which does not yet allow us to rule out the SBHB hypothesis for any candidates.
\end{abstract}


\section{Introduction}\label{sec:intro}
Supermassive black hole binaries (SBHBs) are thought to be a common, if not inevitable, outcome of hierarchical galaxy evolution scenarios \citep[e.g.,][]{menou01,volonteri03,hopkins06} in which most massive galaxies host a central supermassive black hole \citep[BH; e.g.,][]{kormendy95}. We now know of many examples of dual active galactic nuclei (AGNs) in interacting galaxies, with separations of a few kpc \citep{comerford09b, wang09, smith10, liu10b, liu10a, shen11b, fu11a, fu12, barrows12, comerford12, comerford13, comerford15, muller-sanchez15, mcgurk15, koss16}. These are thought to represent the initial stage of the evolution of SBHBs, as described early on by \citet{begelman80}. However, there are no known unambiguous cases of {\it close, bound} SBHBs, with separations of 1~pc or less, which would represent the late stages of the evolution of such systems. A great deal of effort has recently been devoted to finding close SBHBs because they afford tests of supermassive black hole and galaxy co-evolution scenarios and because they are the progenitors of low-frequency gravitational waves detectable by future space-based missions \citep{amaro-seoane12,arzoumanian14,sesana15}.

\citet{begelman80} describe in detail the evolution of such a SBHB via processes that extract energy and angular moment from the system. Initially, the supermassive black holes sink to the centre of the gravitational potential well in the merger remnant and the binary orbit decays due to dynamical friction. The duration of this stage is relatively short, $\sim 10^6$~years. Once the orbital velocity becomes comparable to the stellar velocity dispersion the SBHB is considered ``hard.''  Scattering events with stars in the nucleus are now the dominant angular momentum loss mechanism for the SBHB. If scattered stars are not replaced, the SBHB evolves slowly to separations of $\sim 1$~pc. At separations on the order of $10^{-2}$ or $10^{-3}$~pc, the emission of gravitational waves takes over and the SBHB evolves to coalescence on a time scale of $\sim$a~few$\times 10^8\;$years. It is not clear whether the process of scattering stars alone can bring the SBHB to the point where gravitational wave emission becomes efficient within a Hubble time. The 3-body scattering process requires a sufficient supply of stars that occupy a certain region of phase space and as these are depleted, the binary separation may stall at $\sim1\;$pc. Solutions to this ``last parsec problem'' include replenishing the population of stars available for scattering \citep[e.g.,][]{yu05}, non-spherical potentials \citep[e.g.,][]{yu02a,merritt04a,berczik06,khan13,vasiliev13,vasiliev14,vasiliev15,holley15}, or interactions between the SBHB and a reservoir of gas \citep[e.g.,][]{armitage02,escala04,dotti07,dotti09b,cuadra09,lodato09}. 

In light of this evolutionary scenario, it is particularly interesting to search for SBHBs with sub-parsec orbital separations. The chances of finding systems in this regime may be  high because, for likely evolutionary pathways, the residence time is relatively long ($\gtrsim 10^8$~years). However, the details of the evolution of systems through these separations are not known, so the detection of such SBHBs would also contribute much needed constraints on relevant physical models. Here, we implicitly consider a model that involves interaction of the SBHB with a gaseous circumbinary disc \citep[e.g.][]{cuadra09}. This disc acts as a source of fuel for the BHs and as a means of removing angular momentum from the system \citep[see, for example, the models of][]{artymowicz96,hayasaki07,cuadra09}. Thus, accretion on to one or both of the BHs creates an observational signature that one can exploit to search for such systems. Depending on the mass ratio of the two BHs, either one or both of them may accrete at a substantial rate. If the mass ratio differs significantly from unity, the secondary (less massive) BH has easier access to the gas in the circumbinary disc, while the primary resides close to the centre of mass of the system.
	
Two methods have been widely used to search for SBHBs, under the assumption that at least one of the BHs is actively accreting. One approach is to search for variability in the light curves of AGN, assuming that the accretion is modulated on time-scales of the order of the orbital period of the SBHB. Thus far, these searches have identified candidates with periods of order several years or less \citep[e.g.,][]{graham15a,graham15b,liu15,charisi16}. If these are SBHBs, they likely correspond to the last phase of binary evolution where separations are small enough that the decay of the orbits is driven by gravitational wave radiation. A significant uncertainty in this approach arises from  the fact that regular quasar light curves exhibit red noise \citep[e.g.,][and references therein]{macleod10}, blurring the distinction between a periodic signal and normal quasar variability when only a few oscillations of the light curve are observed \citep[see discussion by][]{vaughan16}. The second approach is based on an analogy to spectroscopic binary stars and seeks to identify periodic variability in the offsets of quasar broad emission lines \citep[e.g.,][]{gaskell96a} indicative of bulk motion through space of one or both BHs and associated broad-line regions (BLRs). The double-line spectroscopic binary case has been tested and rejected for quasars with double-peaked broad emission lines \citep{eracleous97,liu16}. Quasars whose broad emission lines are displaced by a substantial amount represent the single broad line spectroscopic binary case, which remains a viable way to look for the signature of orbital motion using quasar spectra. To be sure, this method does have drawbacks and is subject to several uncertainties, which we discuss in later sections of this paper. A ``hybrid'' approach has also been used to look for SBHBs in nearby Seyfert galaxies with long and well-sampled photometric and spectroscopic time series \citep[e.g., NGC~4151 and NGC~5548; see][]{bon12,bon16,li16}. In this approach one looks for periodic behavior in the light curves and also tries to find regular variability of the broad Balmer line profiles on the same period. This approach is subject to the same uncertainties resulting from red noise in the light curves. Moreover, interpretation of the variability of the line profiles involves fitting them with linear combinations of several Gaussian components and assigning physical meaning to these components, which is highly subjective at best.

Starting with \citet{boroson10}, a significant effort has been invested in observational searches for close, SBHBs via the single broad line spectroscopic binary test. The effort has been directed towards both the population of quasars with broad lines located at their systemic velocities \citep[in the context of the SBHB hypothesis such cases would correspond to binaries in conjunction; see][]{shen13a,ju13,wang16} and those where the broad emission lines are offset from the rest frame by thousands of kilometers per second \citep{tsalmantza11,eracleous12,decarli13,liu14}. So far, these searches have sought to test the SBHB hypothesis for candidates using multiple (generally two) spectra taken years apart to measure or constrain the projected acceleration. The main caveat to this approach is that normal quasar spectral variability, that can mimic the sought after acceleration, has not been characterized on these time-scales. Thus, in order to gain confidence in these results a long-term spectroscopic monitoring campaign is necessary. Long-term spectroscopic monitoring would also help test the recoiling BH hypothesis for the offset of the broad emission lines of the quasars. Recoils of up to several thousand km~s$^{-1}$ can result from the merger of two BHs and can displace the product of the merger from the centre of the host galaxy \citep[e.g.,][]{campanelli07a,campanelli07b,baker08}. These BHs can accrete either from a reservoir of gas they carry with them or from gas that they encounter after they are ejected \citep[e.g.][]{loeb10,guedes11,blecha11,blecha16}. On time scales of order decades we do not expect substantial {\it and monotonic} velocity variations in the emission lines from such systems. Such is the case with QSO E1821+643, where the lack of velocity variations over a period of 24 years has been used to argue in favor of the recoiling BH over the SBHB scenario \citep{shapovalova16}, although it is possible to explain the observations without invoking either.

As part of a systematic, long-term effort we have been spectroscopically monitoring the 88 SBHB candidates of \citet[][hereafter \pone]{eracleous12} since 2009 in order to test the SBHB and recoiling BH hypotheses. In \citet[][hereafter \ptwo]{runnoe15} we presented the followup spectra, measurements of spectroscopic properties from all the spectra, and an analysis of the variability of the {\it integrated fluxes} of the broad \Hb\ lines. We found that the continuum and broad \Hb\ lines of SBHB candidates have the same variability properties as those of typical quasars of the same luminosity and redshift. We concluded that, if the line variability is driven by continuum variability, then the BLRs of the SBHB candidates have sizes comparable to those of typical quasars. The goal of this paper, the third in the series, is to present the radial velocity curves derived from the spectra we presented in \ptwo. As we noted in section~2 of \pone, one may expect radial velocity changes of up to a few hundred km~s$^{-1}$ over the course of our monitoring period, assuming that the secondary BH is active (see the physical picture described in \pone\ and re-iterated above). In Section~\ref{sec:data} we describe how the sample was identified and its properties. The shift measurements and methodology are presented in Section~\ref{sec:xc} with the most significant caveats to this approach discussed in Section~\ref{sec:caveats}. We analyse the radial velocity curves in the context of the SBHB hypothesis in Section~\ref{sec:interpretation} and discuss them in the context of previous work in Section~\ref{sec:discussion}. Our results are summarized in Section~\ref{sec:summary}. Throughout this work, we adopt a cosmology with $H_0 = 73$ km s$^{-1}$ Mpc$^{-1}$, $\Omega_{\Lambda} = 0.73$, and $\Omega_{m} = 027$.

\section{Sample and Data}
\label{sec:data}

The SBHB candidates were selected from SDSS DR7 quasars with $z<0.7$ to have broad \Hb\ emission lines that were offset by $\gsim 1,000\;{\rm km\;s}^{-1}$ relative to the rest frame set by the \OIIIw\ narrow line. We note that some redshifts have been refined since that data release. Most notably, the redshift of J092712 was listed as 0.697 in DR7 putting this object in the redshift range of our sample, while its new redshift is 0.713. Potential candidates were initially identified via the spectral principal component analysis described by \citet{boroson10} and were added to the sample after a single offset peak in the broad line was verified by visual inspection. The final sample includes 88 candidates, with redshifts of $0.077 < z < 0.713$ and absolute V-band magnitudes of $-21.25 < V < -26.24$ after Galactic extinction corrections but not K corrections. Additional details on the sample selection and properties are available from \pone. In comparison to other similar surveys, the objects in our sample and the sample of \citet{decarli13} are at the high end of the distribution of broad H$\beta$ velocity offsets. The objects in the samples of \citet{shen13a} and \citet{ju13} have velocity offsets close to zero by selection. The objects in the sample of \citet{liu14} have intermediate offsets (typically less than 1,000~\kms)\footnote{In the context of the single broad line spectroscopic binary analogy, described in \S\ref{sec:intro}, and assuming circular orbits and a relatively narrow range of masses, periods, and orbital inclinations the objects with the highest velocity offsets may correspond to orbital phases close to quadrature, where the residence times are large while the accelerations are small. Here, we are referring to the velocity and acceleration projected along the line of sight (for circular orbits the true velocity and acceleration have a constant magnitude and vary in direction with orbital phase). In contrast, the objects with velocity offsets close to zero may be close to conjunction, where the residence times are small and the accelerations are large. Objects with intermediate velocity offsets may represent intermediate orbital phases.}.

We have been monitoring the candidates spectroscopically with a variety of ground-based facilities since late 2009. In addition to the original 88 SDSS spectra, we have obtained 320 follow-up spectra spanning time intervals from a few weeks to 12 years in the observer's frame. Each object has been observed at least twice and many have three or more observations. All of the data were presented in Papers I and II; we refer the reader to those papers for details of our observations and data reduction. 

The wavelength calibration is critical for measuring radial velocity shifts so we summarize it here. For each observing run, we derived a low-order (order$\;<5$) polynomial wavelength solution based on between 20 and 60 arc emission lines. We required that the standard deviation of the residuals from this fit be less than 0.1 pixel, which set the uncertainty in the relative velocity scale for each spectrum at 6400~\AA, the median wavelength of \Hb\ in our sample, to be between 5 and 11~km~s$^{-1}$, depending on the telescope and instrument. The absolute wavelength scale was set with the help of the \OIsky\ night sky line, which was recorded simultaneously with the quasar spectrum. After applying heliocentric velocity corrections and converting the wavelengths scale to vacuum values, we aligned the spectra of the same object by comparing the wavelengths of \OIIIdblt\ doublet lines and using the cross-correlation method described in \pone. The uncertainty in this alignment translates directly into how well we can determine the relative velocity of the broad \Hb\ line between epochs. This uncertainty is smaller than 55~km~s$^{-1}$ for all objects where we detect a significant velocity shift and smaller than 35~km~s$^{-1}$ for 80 per cent of these objects.

The first-epoch velocity offset in the broad \Hb\ emission line, which anchors the radial velocity curve, was measured from the SDSS spectra of our objects and reported in \pone. In summary, the peak was located by fitting the region around it with a Gaussian. To assess the uncertainties resulting from the finite signal-to-noise ratio (S/N) and possible asymmetries of the profile near the peak, we repeated the fitting exercise several times, changing the range of pixels that we fitted each time. In the end we took the average peak wavelength from our trials to be the best estimate of the true value and we adopted the full range of peak wavelengths from those trials as a measure of the uncertainty. As such, the error bars on the peak velocity measured from the SDSS spectrum effectively indicate the 99 per cent confidence limits. Velocity changes are then measured {\it relative to the first-epoch spectrum}, as we describe in the next section.

\section{Broad \Hb\ radial velocity variations}

\subsection{Measurement of velocity variations}
\label{sec:xc}

\begin{figure}
\begin{center}
\includegraphics[width=8.5 truecm]{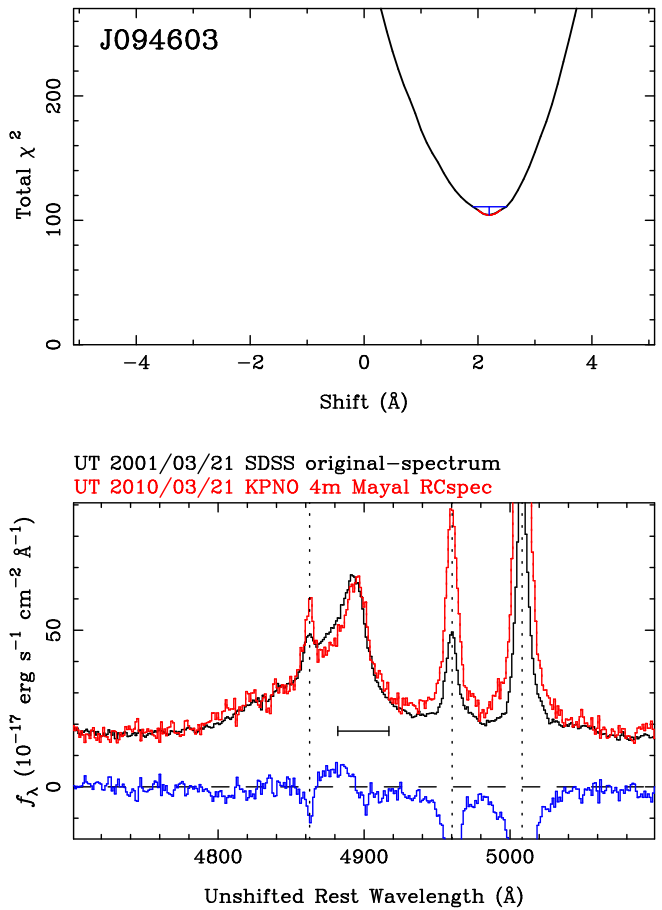}
\end{center}
\caption{An example of a diagnostic plot for J094603 showing a representative shift measurement described in Section~\ref{sec:xc}. Top: The curve of $\chi^2$ as a function of shift. The minimum, in red, identifies the best shift (in this case $2.2\pm0.3\;$\AA), while the blue horizontal bar shows the 99 per cent confidence interval on the shift measurement. Bottom: The shift in the top panel is measured between the SDSS spectrum (black) and the follow-up spectrum (red). The residual spectrum after subtraction of the later spectrum from the earlier one is shown in blue at the bottom of the frame. The dotted lines indicate the rest-frame wavelengths of the \Hb\ and \OIIIdblt\ lines, and the horizontal black bar identifies the wavelength window used for the $\chi^2$ measurement.}
\label{fig:xc}
\end{figure}

\begin{figure*}[t]
\centerline{
  \includegraphics[width=8.9cm]{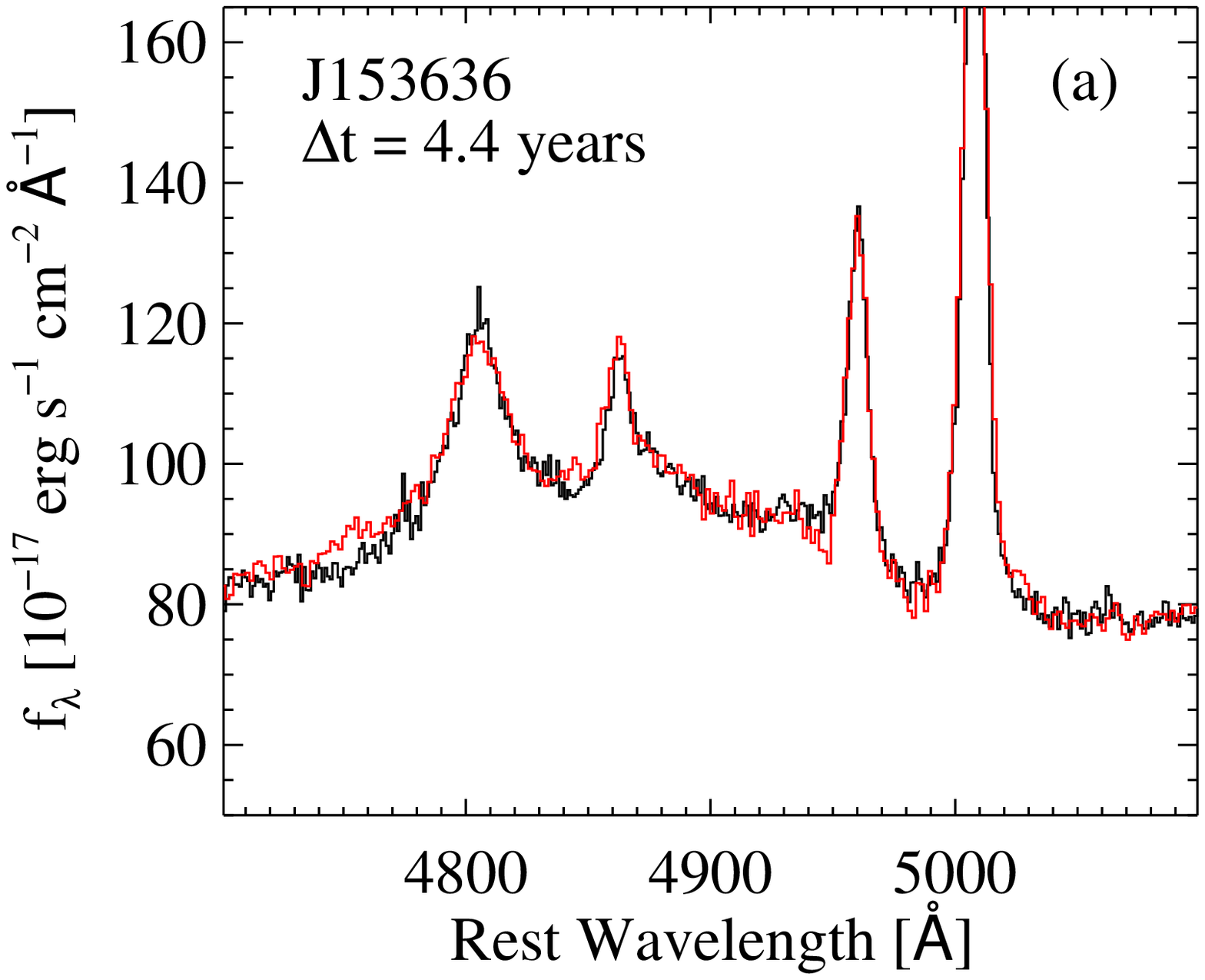}
  \includegraphics[width=8.9cm]{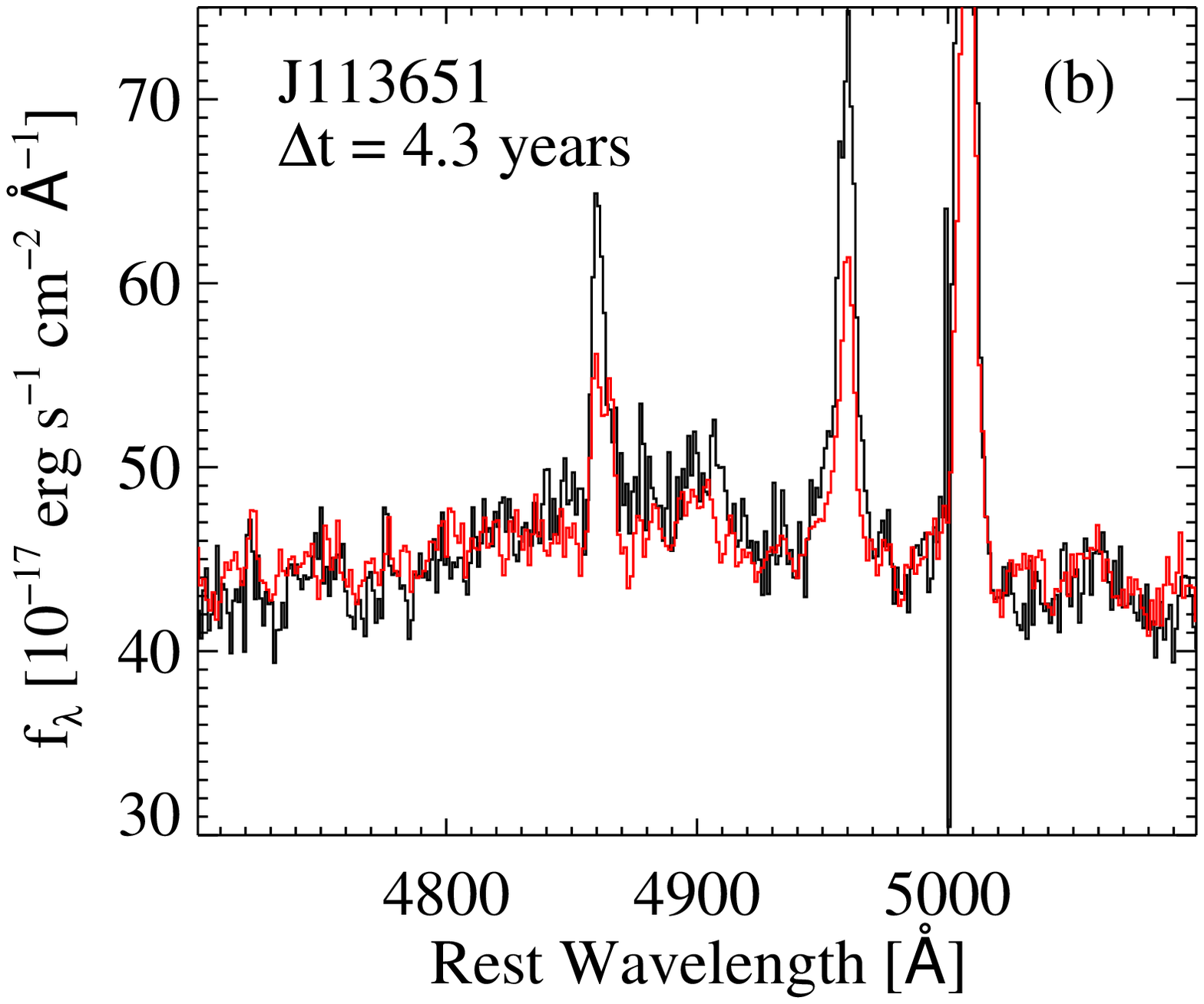}
}
\centerline{
  \includegraphics[width=8.9cm]{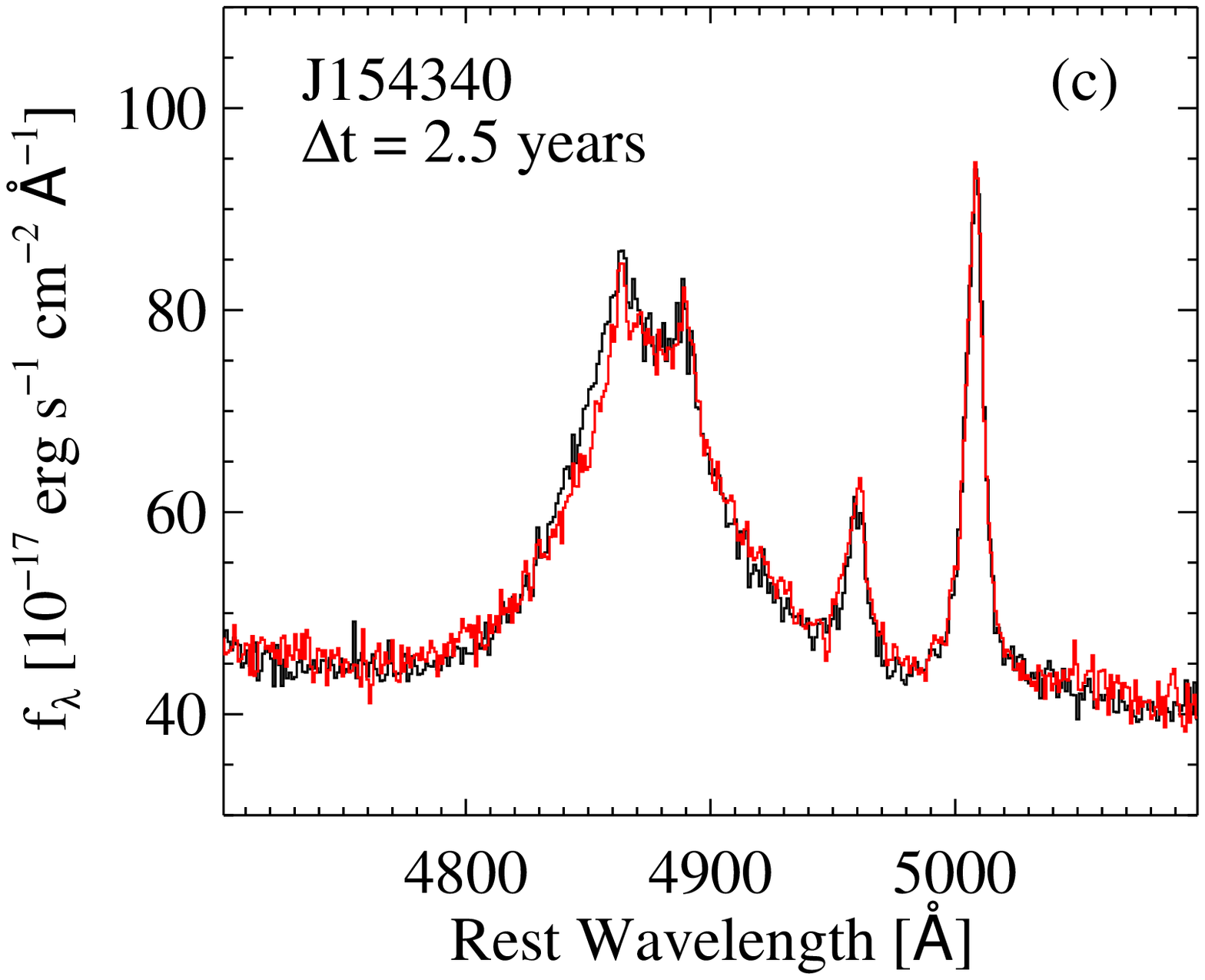}
  \includegraphics[width=8.9cm]{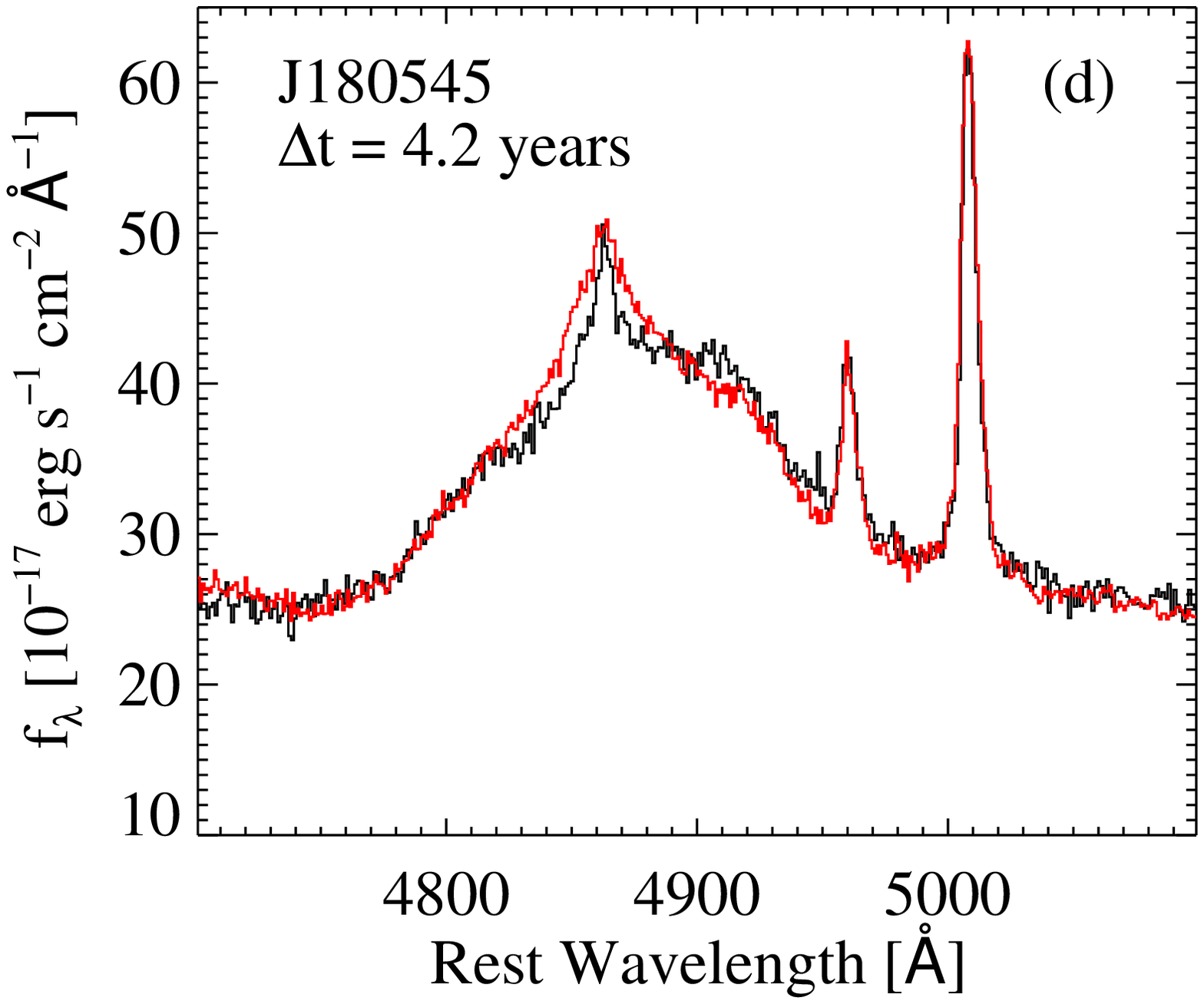}
}
  \caption{Examples of broad H$\beta$ profiles that yield shifts consistent with zero or unreliable shifts (categories 2--5 according to the last paragraph of \S\ref{sec:xc}). Each panel shows the SDSS spectrum in black and a subsequent spectrum in red. The subsequent spectra were shifted and scaled to match the continuum level and broad line flux, thus affording an easy visual comparison of the broad line profiles. The time intervals given in each panel are measured in the rest frame of the source. (a) A case where the profile has remained stable between two epochs thus the shift is consistent with zero within uncertainties. (b) An example of low-S/N profiles that do not yield restrictive shifts. (c) An example of slight profile shape variability that increases the uncertainty in the shift measurement. (d) An example of substantial profile variations that prevent us from measuring radial velocity changes.} 
\label{fig:profvar}
\end{figure*}

The first step in measuring the velocity shifts of the broad \Hb\ emission line from the initial offset observed in the SDSS spectrum is to apply the $\chi^2$ cross-correlation method of \pone. In this approach, we shift the follow-up spectrum past the original SDSS spectrum in small steps. At each step, we rebin the observed follow-up spectrum to the wavelength scale of the SDSS spectrum and renormalize it to match the flux scale of the SDSS spectrum, as described by equation~(6) of \pone. This normalization effectively highlights changes in the shape of the broad-line profile, but it hides changes in the broad-line flux and continuum. These can still be identified from the narrow lines in Figures~\ref{fig:xc} and \ref{fig:profvar}; an apparent mismatch in the strength of the narrow lines indicates that the broad-line flux has changed between observations. Finally, we measure the $\chi^2$ statistic in a window that includes emission from the broad \Hb\ line and avoids the narrow lines, effectively excluding their contribution from the cross-correlation signal. In this way, we build a $\chi^2$ curve as a function of shift and can identify the best shift and uncertainties at the 68 per cent and 99 per cent confidence levels (`$1\sigma$' and `$2.6\sigma$' respectively). Since we make this measurement on the observed spectra, the resulting shifts do not depend on any parametric model for the emission-line profiles. We also repeat the measurement with the SDSS spectrum rebinned to the wavelength scale of the follow-up spectrum, so that the shift is measured twice per follow-up spectrum. An example of a diagnostic plot for one velocity shift measurement is shown in Figure~\ref{fig:xc}. The measured shifts are reported in Table~\ref{tab:rv_tab} where we give the average between the two measurements calculated for each follow-up spectrum and the `$1\sigma$' and `$2.6\sigma$' uncertainties. 

\begin{table*}
\begin{minipage}{14truecm}
\caption{Results of Cross-correlation Analysis of Broad H$\beta$ Lines (Abridged)$^{\dagger}$ \label{tab:rv_tab}}
\renewcommand{\thefootnote}{\alph{footnote}}
\begin{tabular}{lcclccc}
            	&           &                                                            & \Hb\ Offset in       & \Hb\ Shift Relative                             	&                              	&                            \\
 Object 	& Date  & $\Delta t_{\rm rest}$\footnotemark[1] & SDSS Spectrum & to SDSS Spectrum\footnotemark[2] 	& $2.6\sigma$ Error\footnotemark[3] 	& $1\sigma$ Error\footnotemark[3] \\
 (SDSS J) & (UT)  & (years) 				      		& (km s$^{-1}$)      & (km s$^{-1}$)				 	& (km s$^{-1}$)		& (km s$^{-1}$) \\
\hline
001224&2001/08/20& \phantom{0}\nodata &  \phantom{0}$-$1870$\pm$80\phantom{0} & \nodata & \nodata & \nodata \\
 & 2009/12/16&\hphantom{0}7.2& \phantom{000$-$}\nodata & $-120$ & $+50/{-}40$ & $\pm20$ \\
 & 2010/10/12&\hphantom{0}7.9& \phantom{000$-$}\nodata & $-170$ & $+30/{-}40$ & $\pm20$ \\
 & 2014/08/28&11.3& \phantom{000$-$}\nodata & $-120$ & $+40/{-}70$ & $+30/{-}40$ \\
002444&2000/12/22& \phantom{0}\nodata &  \phantom{00}$-$700$\pm$100 & \nodata & \nodata & \nodata \\
 & 2009/12/18&\hphantom{0}7.5& \phantom{000$-$}\nodata & $(100)$ & $\pm200$ & $+100/{-}90$ \\
 & 2014/08/29&11.5& \phantom{000$-$}\nodata & \nodata & \nodata & \nodata\tablenotemark{d} \\
015530&2001/09/16& \phantom{0}\nodata &  \phantom{0$-$}1400$\pm$100 & \nodata & \nodata & \nodata \\
 & 2009/12/17&\hphantom{0}7.7& \phantom{000$-$}\nodata & $(100)$ & $\pm200$ & $\pm100$ \\
 & 2011/08/31&\hphantom{0}9.2& \phantom{000$-$}\nodata & $(-170)$\tablenotemark{e} & $+100/{-}80$ & $\pm40$ \\
 & 2014/08/28&12.0& \phantom{000$-$}\nodata & $-390$ & $+60/{-}100$ & $+20/{-}30$ \\
\hline
\end{tabular}
$^{\dagger}${This is an abbreviated version of this table, shown here for content.  The full table can be found in the electronic version of the article.}
\footnotetext[1]{Time intervals are given in the rest frame.}
\footnotetext[2]{Numbers without brackets denote a statistically significant shift, while brackets indicate a shift measurement that is consistent with zero at 99\% confidence.  A positive shift means that the post-SDSS spectrum is shifted towards longer wavelengths.}
\footnotetext[3]{Uncertainties on the broad \Hb\ shift at 90\% and 68\% confidence ($2.6\sigma$ and $1\sigma$, respectively).  These are used in the test described in Section~\ref{sec:interpretation} to determine minimum periods that are consistent with the radial velocity curves.}
\footnotetext[4]{The broad H$\beta$ profile varied substantially between observations, thus a meaningful shift could not be determined. These variations may include large changes in the profile shape or changes in the width that cause the two wings of the line to move in opposite directions.}
\footnotetext[5]{Variability of the broad H$\beta$ line profile affects our ability to determine a shift and increases the uncertainties.}
\footnotetext[6]{The low S/N in the follow-up spectrum does not allow us to get meaningful constraints on the shift.}
\footnotetext[7]{The velocity offsets for this object are measured with respect to the {\it strong} set of [\ion{O}{3}] lines in its spectrum. See the results of \citet{decarli14} and associated discussion regarding the redshift of the host galaxy of this quasar.}
\end{minipage}
\end{table*}


Following \pone, we carried out simulations to check that the uncertainties resulting from the cross-correlation method were not underestimated. In these simulations we cloned the higher-S/N spectrum in each pair and injected in it synthetic noise to match that of the lower-S/N spectrum. We then cross-correlated the two spectra. After 1,000 trials we examined the distribution of shifts about zero to verify that the range encompassing 99 per cent of the trials agreed with the 99 per cent confidence limits derived from the $\chi^2$ cross-correlation method. Indeed, the agreement was excellent.

To guard against profile shape variability at least two authors visually inspect all the diagnostic plots for each shift measurement, and a third performs this task if the first two provide differing visual classifications. Visual inspection is used in conjunction with the minimum value of $\chi^2$ from the cross-correlation procedure to determine whether profile shape variability has influenced the shift measurement. Moreover, we require that the two shift measurements that result from rebinning the SDSS or follow-up spectrum are consistent with each other in magnitude and have opposite signs. Thus, measurements that appear to be affected by profile shape variability are deemed unreliable, while those that appear not to be affected by profile variations are hereafter referred to as reliable.

\begin{figure*}
\centerline{
  \includegraphics[width=8.9cm]{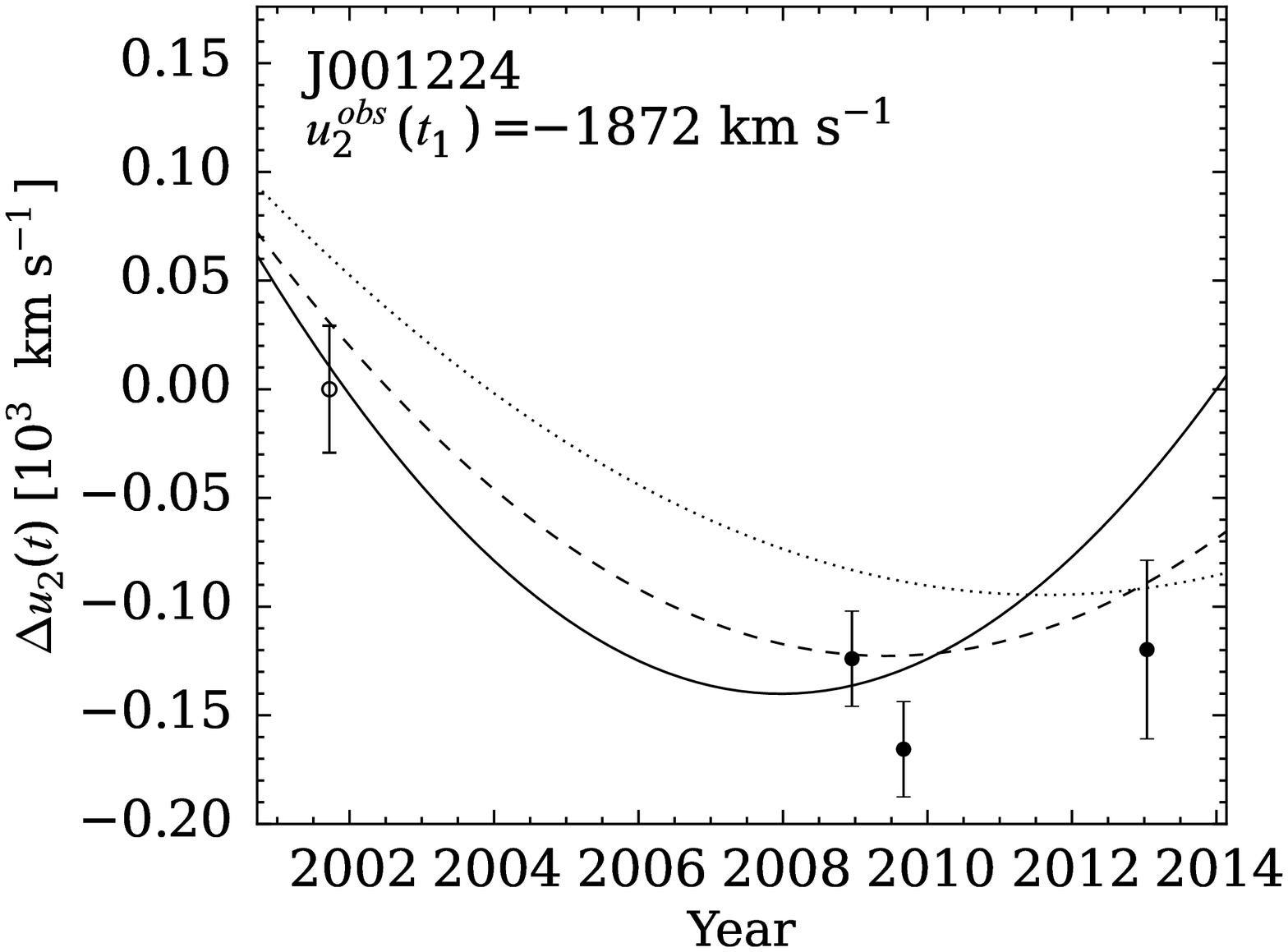} 
  \includegraphics[width=8.9cm]{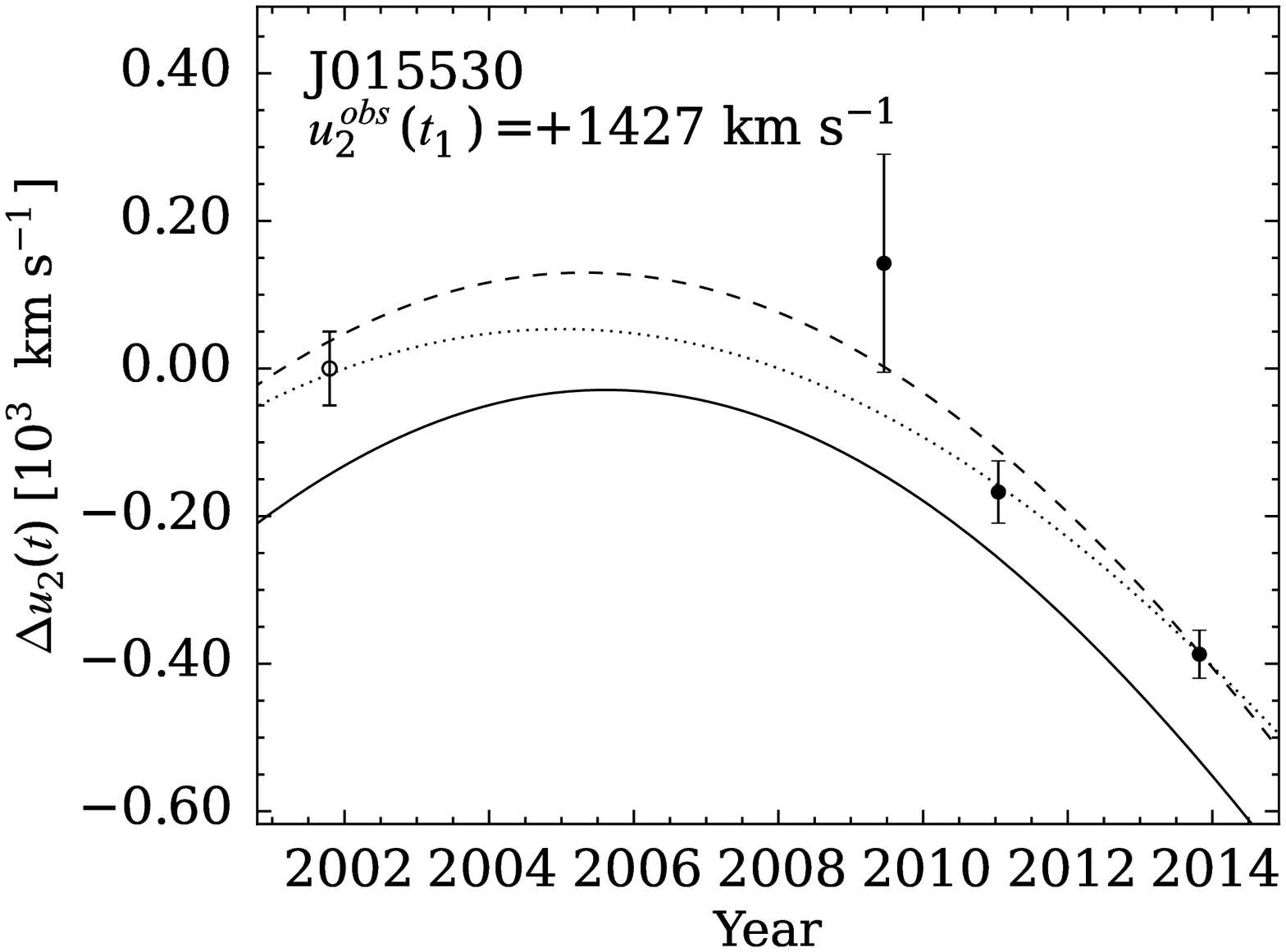} 
}
\centerline{
  \includegraphics[width=8.9cm]{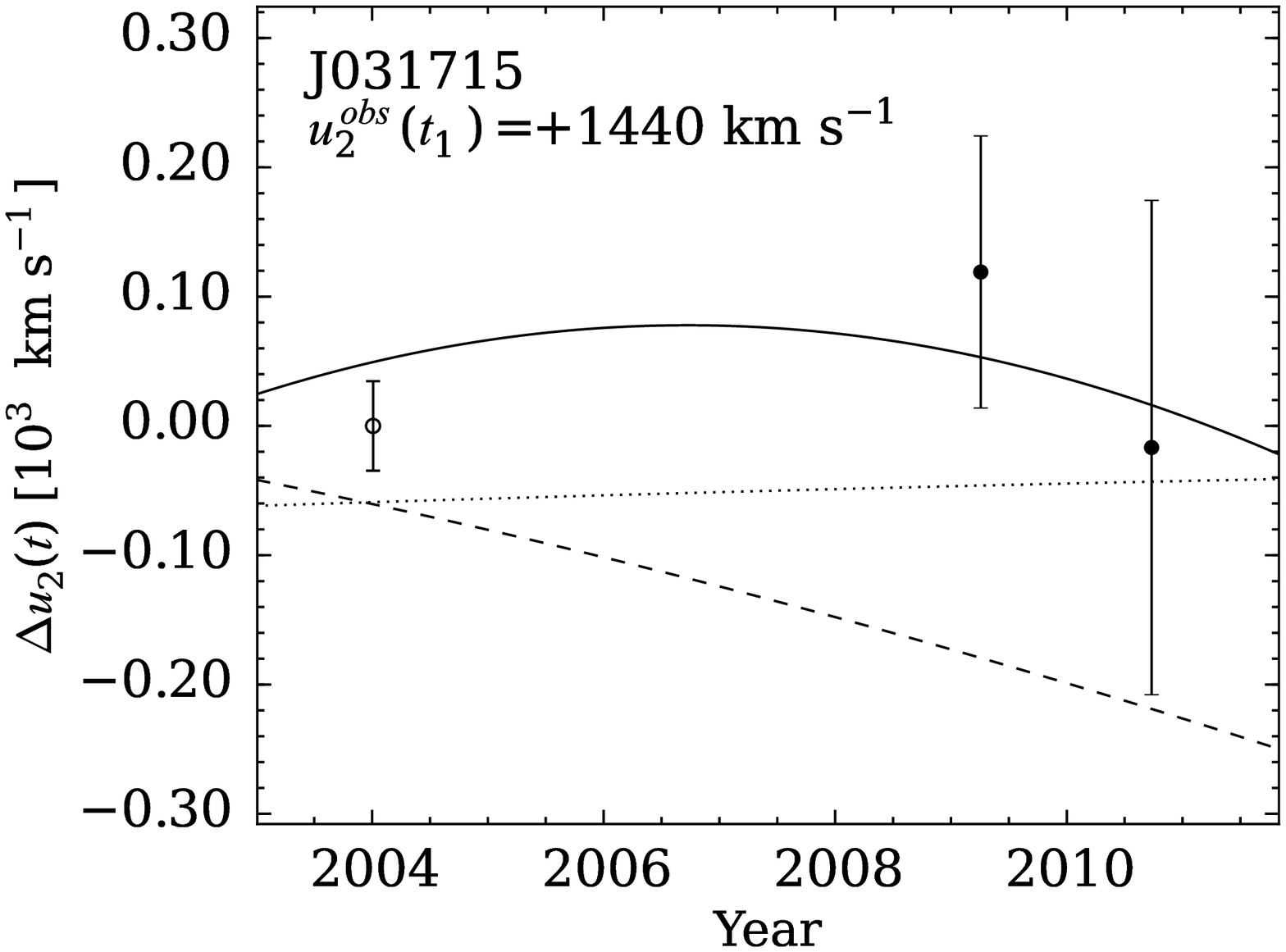} 
  \includegraphics[width=8.9cm]{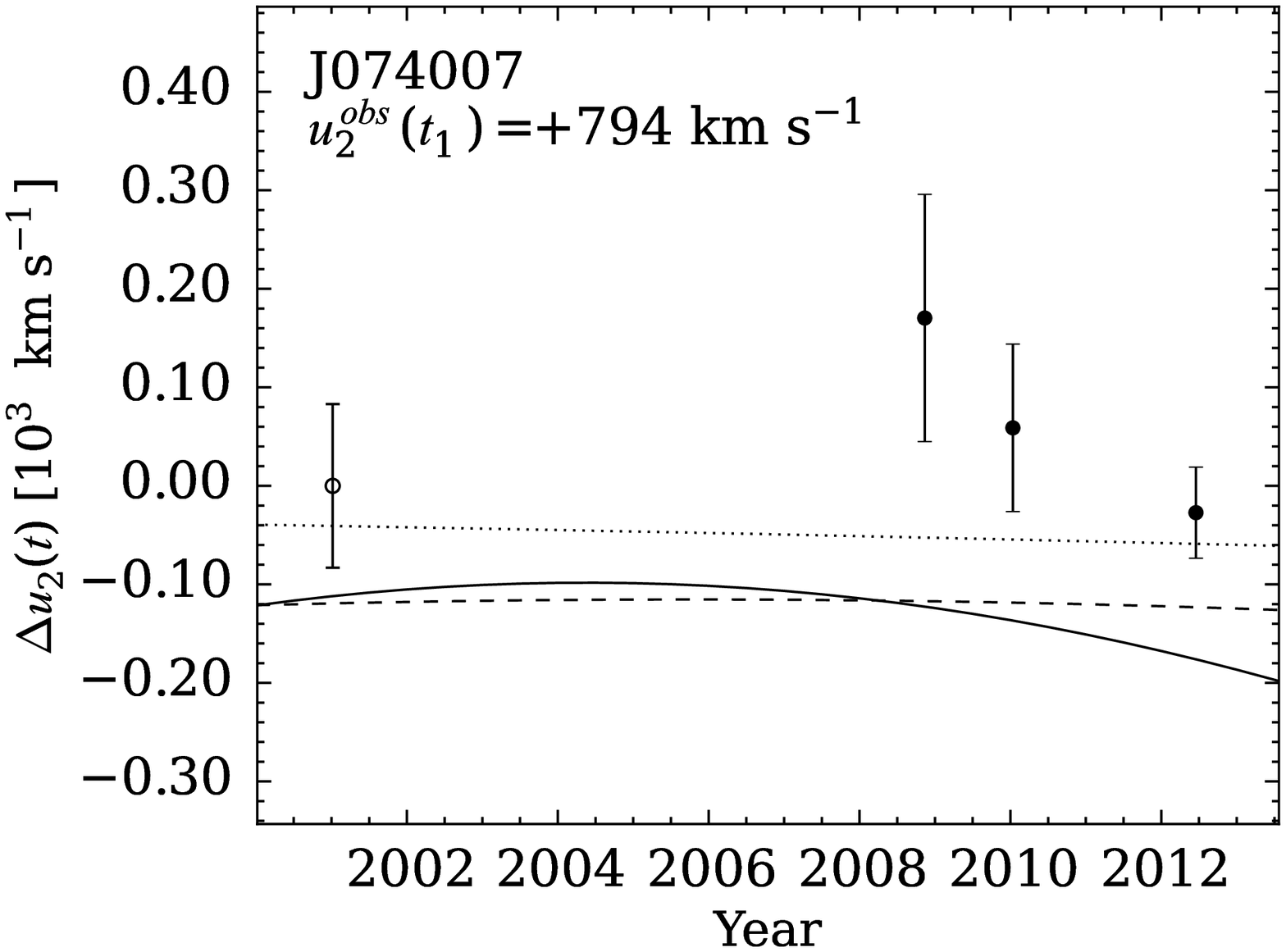} 
}
\centerline{\hskip 0.5truein
  \vbox{\hsize=3.truein \caption{Shifts in the velocity offset of the broad \Hb\ emission line versus time relative to the SDSS spectrum for the 29 SBHB candidates with at least three reliable radial velocity measurements. The parameter $u_{2}^{obs}(t_{1})$ is the velocity offset of the broad \Hb\ lines relative to the narrow lines in the SDSS spectrum (see Section~\ref{sec:interpretation} for an explanation of the notation). The open circle shows the time of the SDSS observation and solid circles represent subsequent shift measurements. In all cases, velocity shift measurements are adopted after verifying by visual inspection that there is no profile shape variability (see discussion in \S\ref{sec:xc}). The error bars correspond to the ``$1\sigma$'' (68 per cent confidence) intervals. The dotted, dashed, and solid lines represent the sinusoids associated with the 68th, 90th, and 99th percentile minimum periods determined from the MCMC simulations described in Section~\ref{sec:interpretation}. \vspace{0.0truein}} \label{fig:zbrv}}\hspace{0.05truein}
  \includegraphics[width=8.9cm]{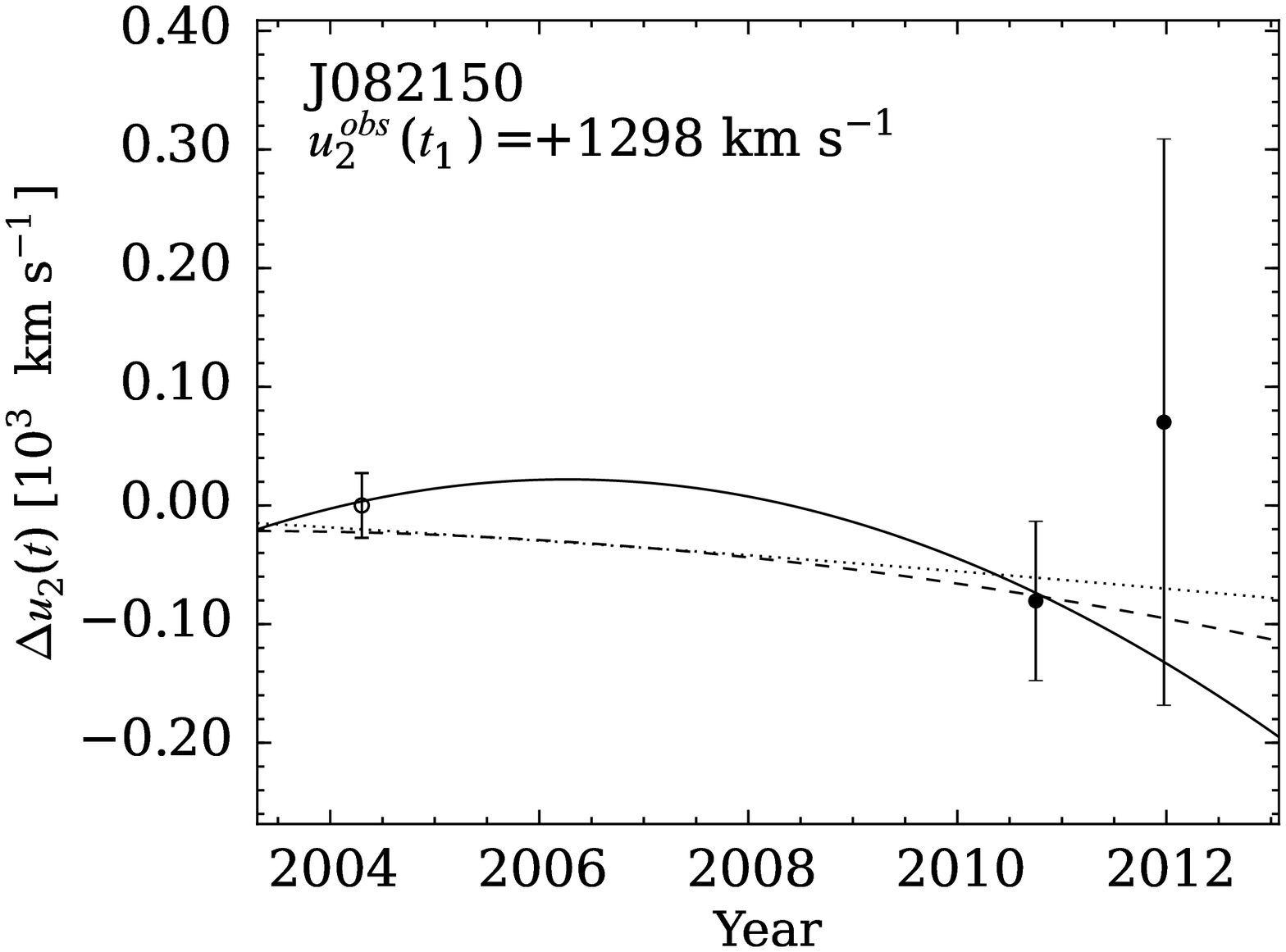} 
}
\clearpage
\end{figure*}

Following \pone, we place each shift measurement into one of five categories, noted in Table~\ref{tab:rv_tab}: (1) we reliably detect a statistically significant shift, (2) we reliably determine that the shift is consistent with zero within measurement uncertainties, (3) the S/N in the spectrum is too poor to obtain a meaningful shift measurement (the cross-correlation algorithm is fooled by noise spikes), (4) we obtain a shift measurement, albeit with larger uncertainties because of slight profile shape variability, (5) the profile varied substantially between observations and no shift measurement can be made. An example of profiles yielding a reliable shift measurement (category 1) is shown in Figure~\ref{fig:xc}. Examples of profiles from categories 2--5 are shown in Figure~\ref{fig:profvar}. It is specifically the measurements from categories 1 and 2 that were deemed ``reliable.'' We find that 29/88 objects have at least three reliable measurements. Ignoring the low S/N measurements, we can build reliable radial velocity curves for these objects that are useful for further analysis. We show these radial velocity curves in Figure~\ref{fig:zbrv}. It is also noteworthy that 39/88 objects exhibit substantial profile shape variability in at least one spectrum (i.e., category 5 above). While we cannot rule out the SBHB hypothesis for objects whose broad H$\beta$ profiles have varied substantially, we are unable to measure the radial velocity changes in these cases.

\begin{figure*}
\centerline{
  \includegraphics[width=8.9cm]{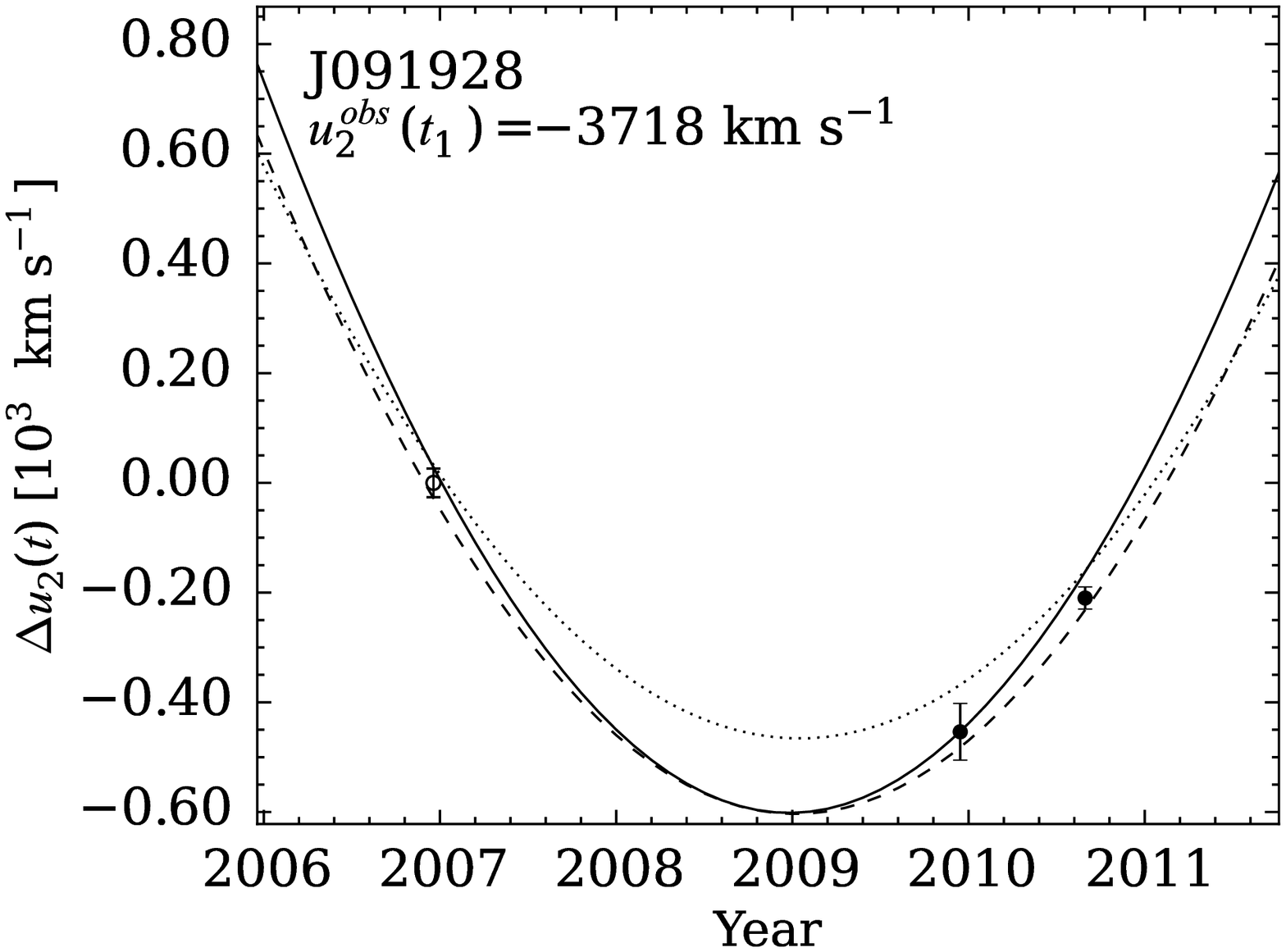} 
  \includegraphics[width=8.9cm]{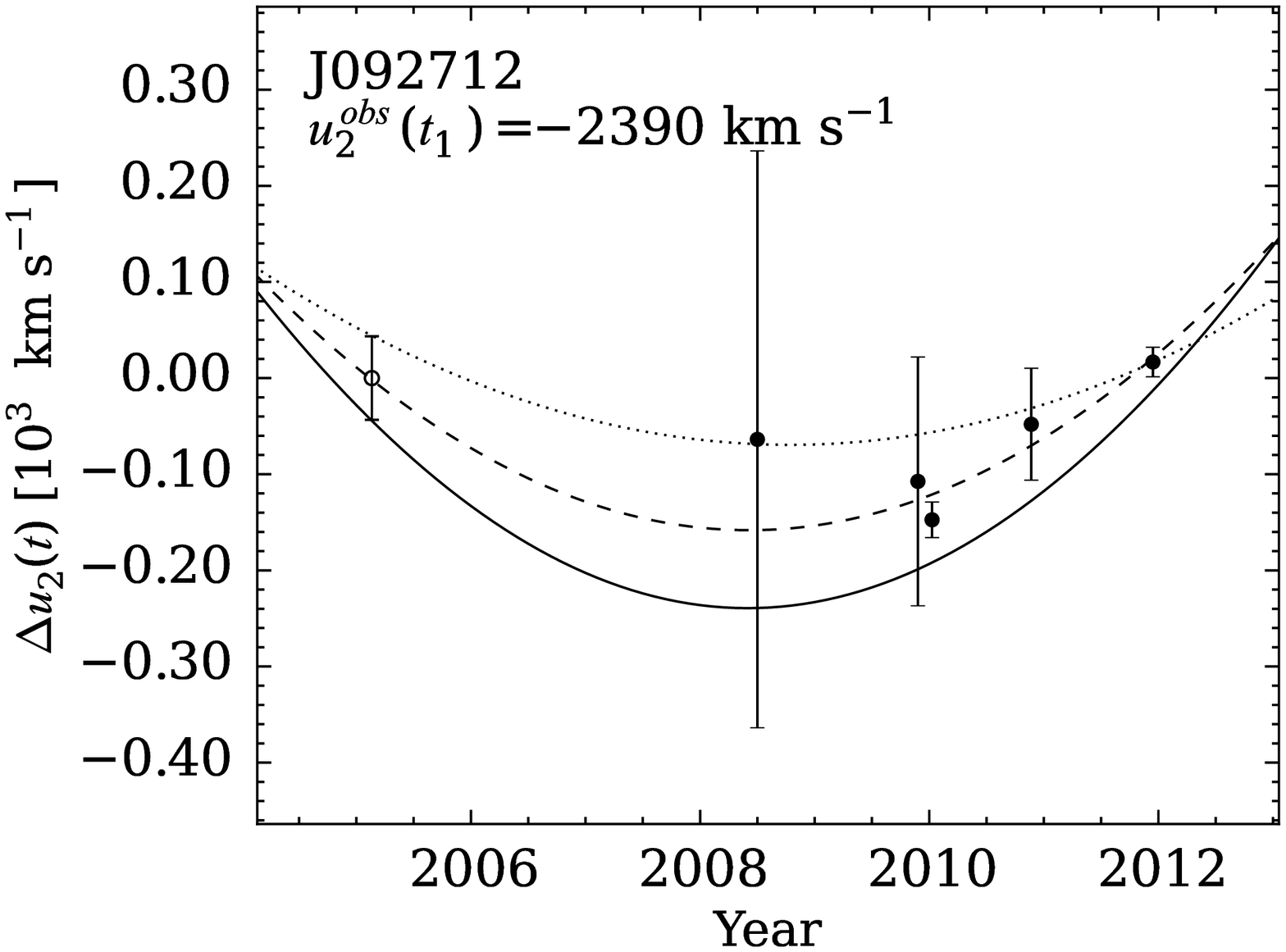} 
}
\centerline{
  \includegraphics[width=8.9cm]{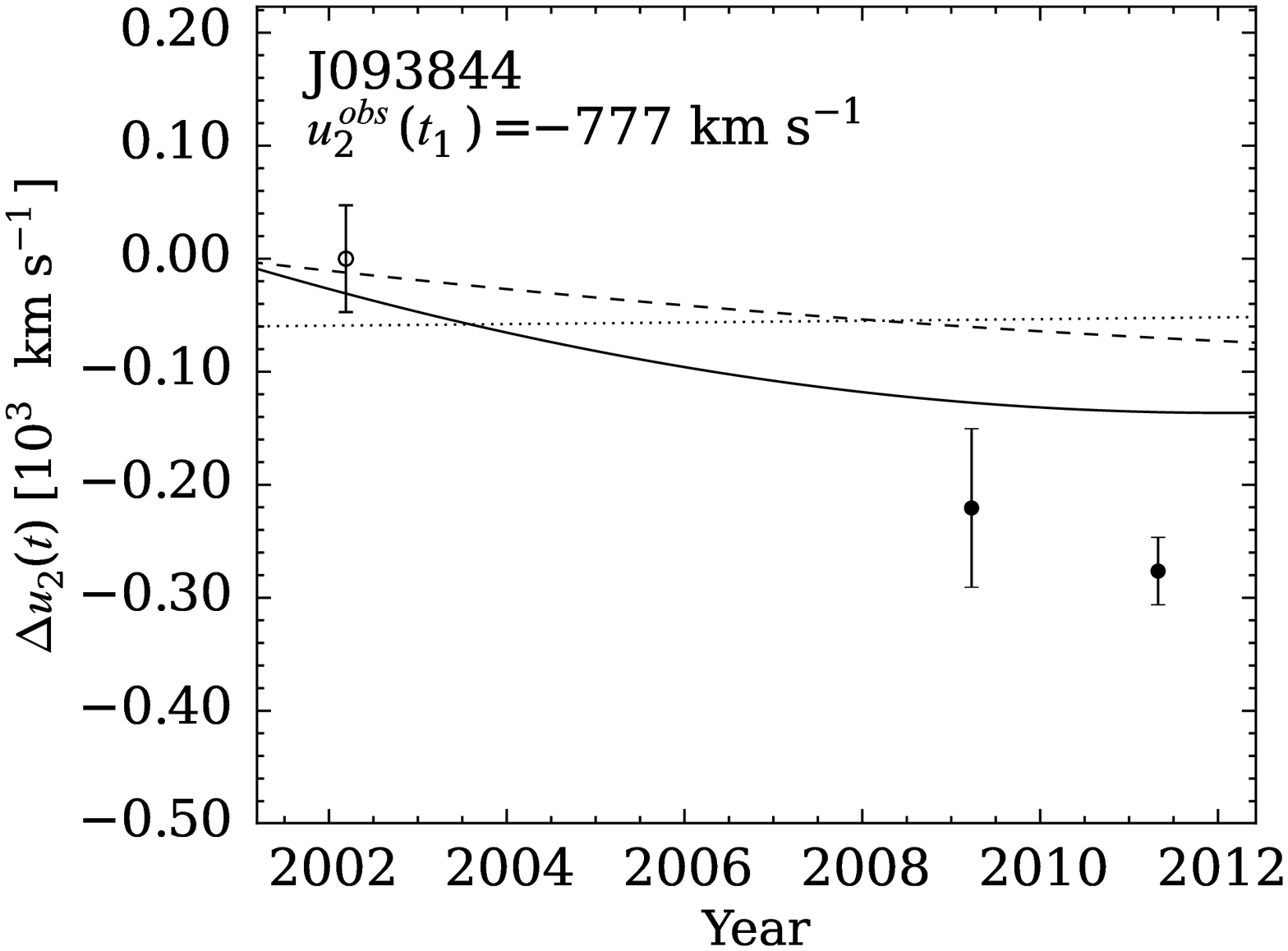} 
  \includegraphics[width=8.9cm]{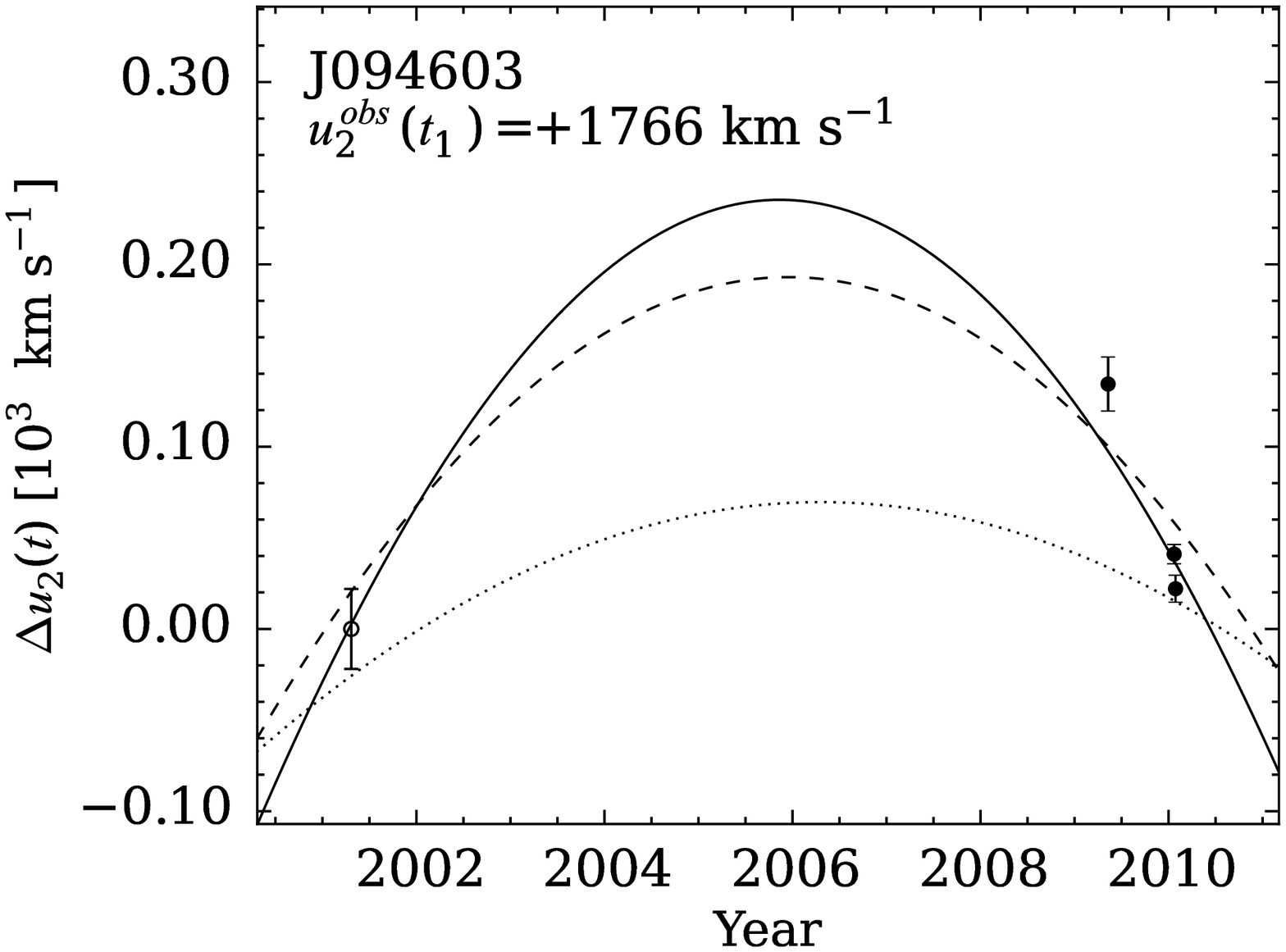} 
}
\centerline{
  \includegraphics[width=8.9cm]{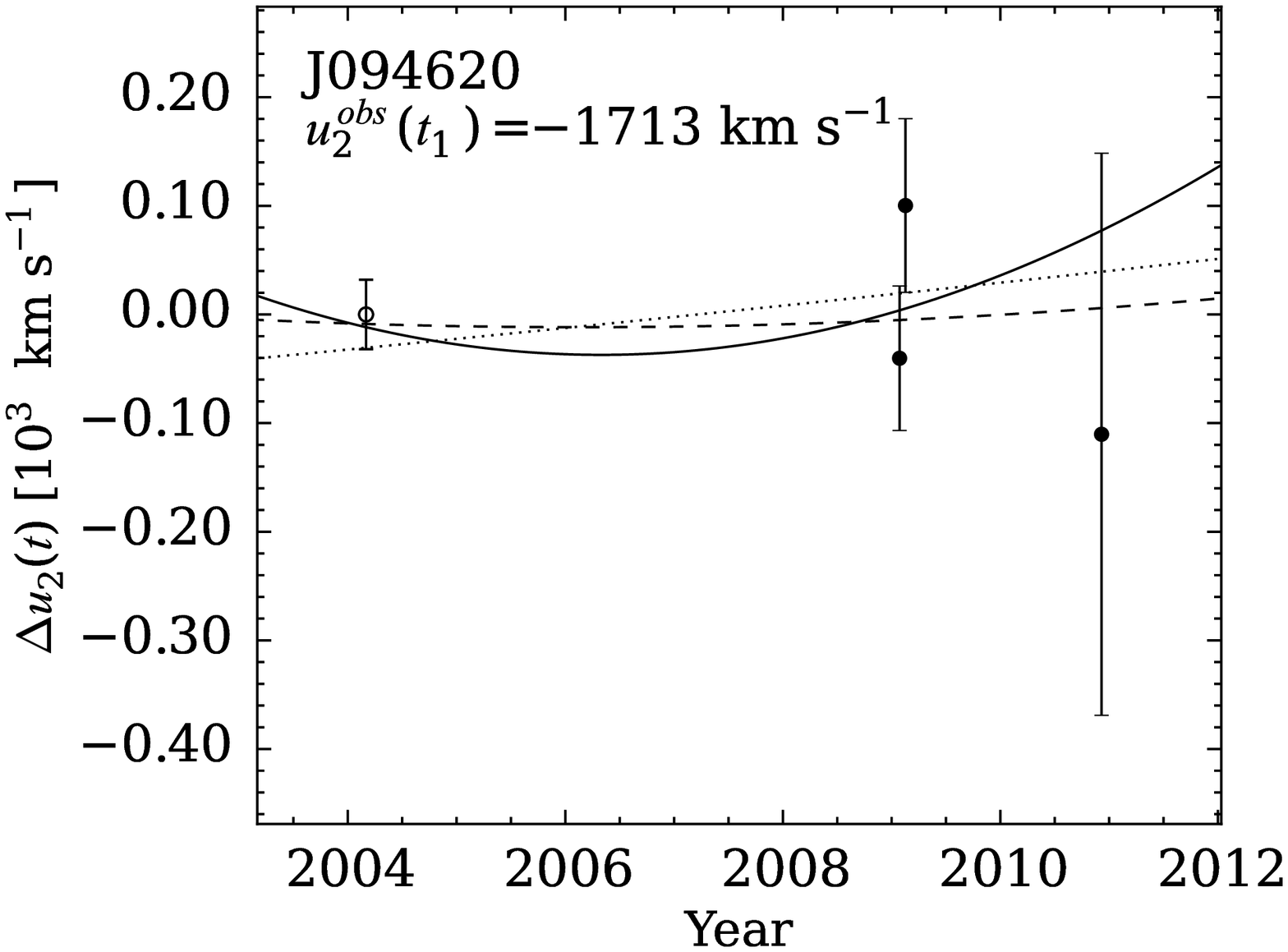} 
  \includegraphics[width=8.9cm]{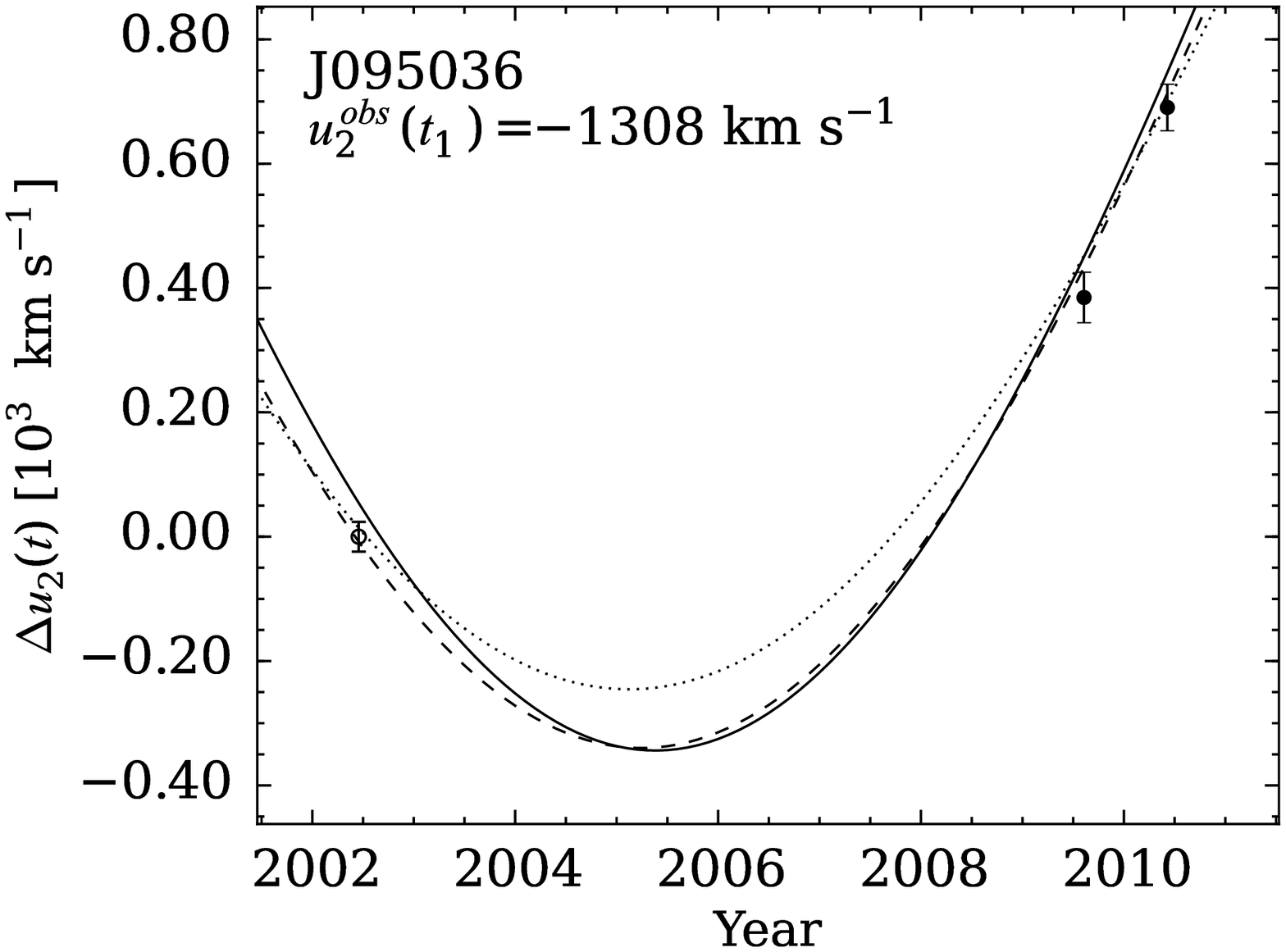} 
}
\centerline{Figure~\ref{fig:zbrv}. -- Continued.}
\end{figure*}

\begin{figure*}
\centerline{
  \includegraphics[width=8.9cm]{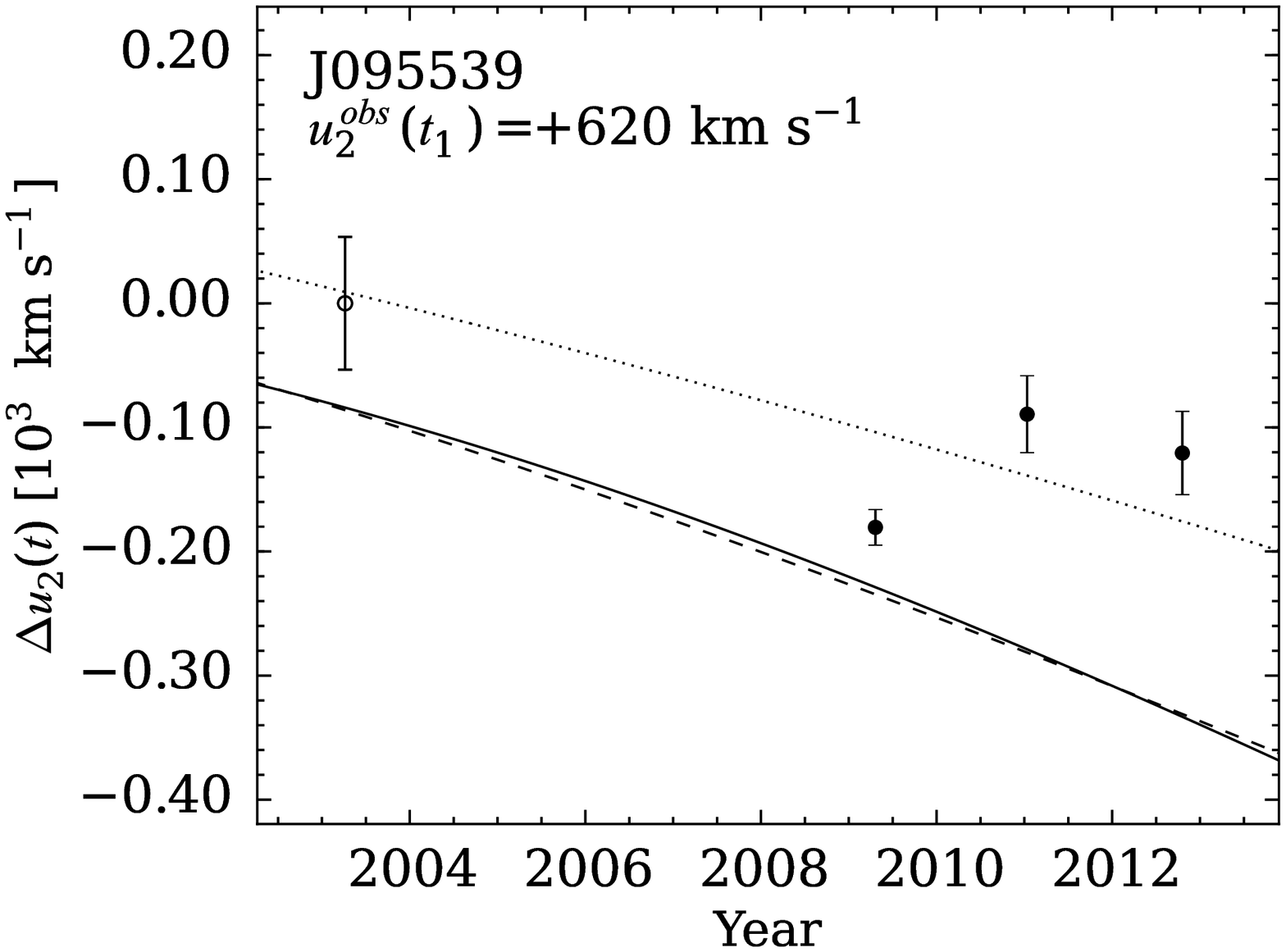} 
  \includegraphics[width=8.9cm]{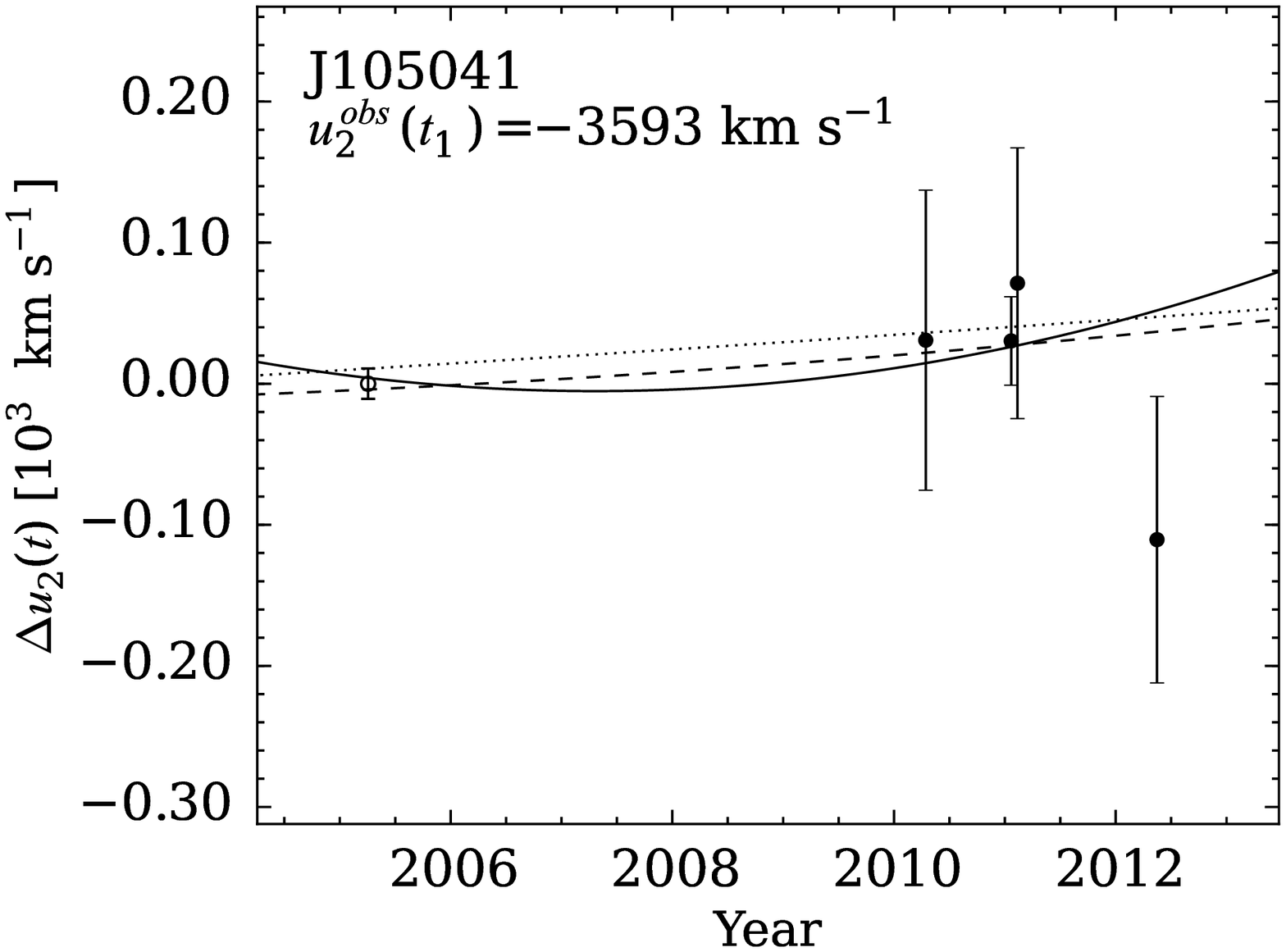} 
}
\centerline{
  \includegraphics[width=8.9cm]{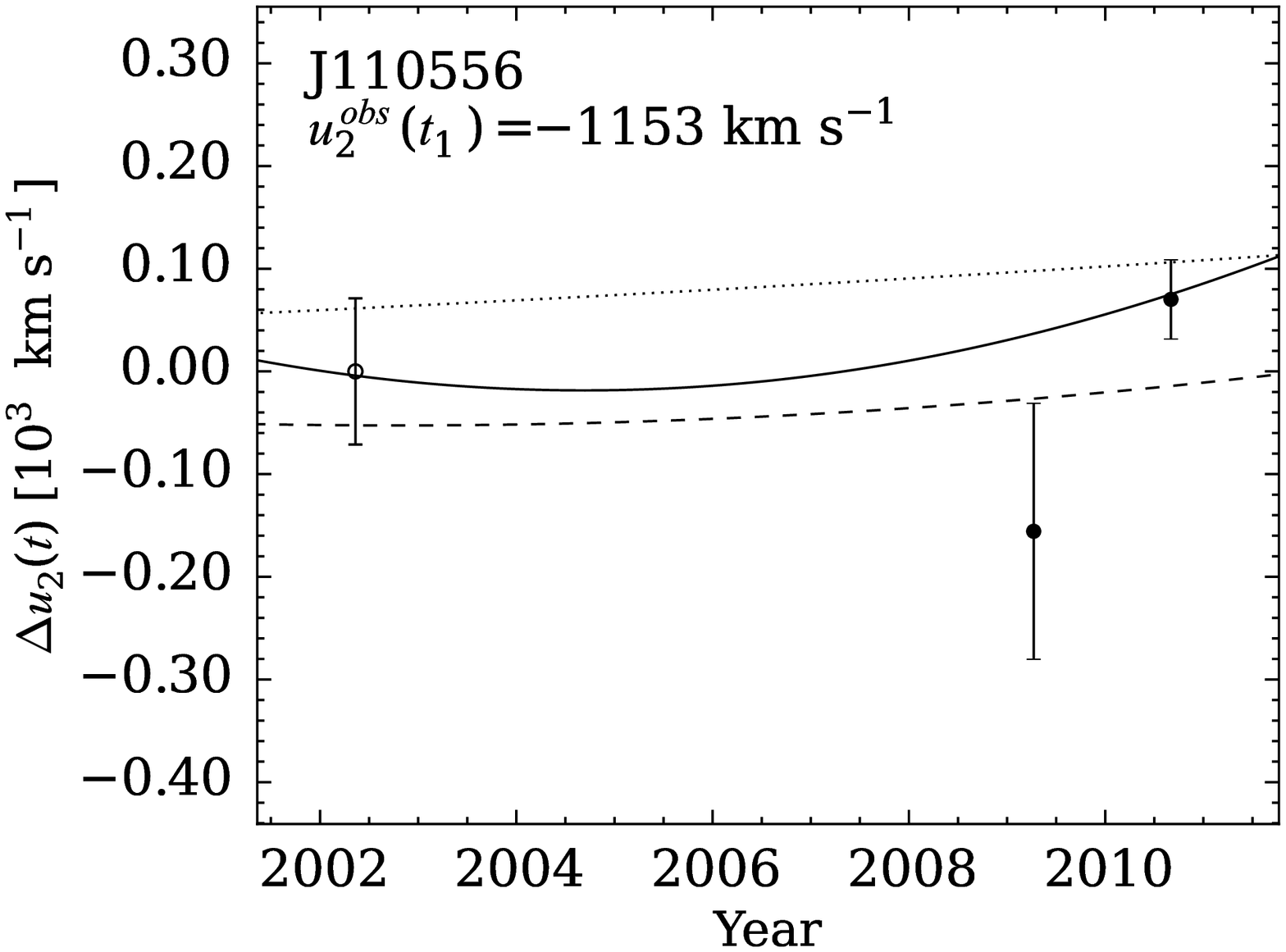} 
  \includegraphics[width=8.9cm]{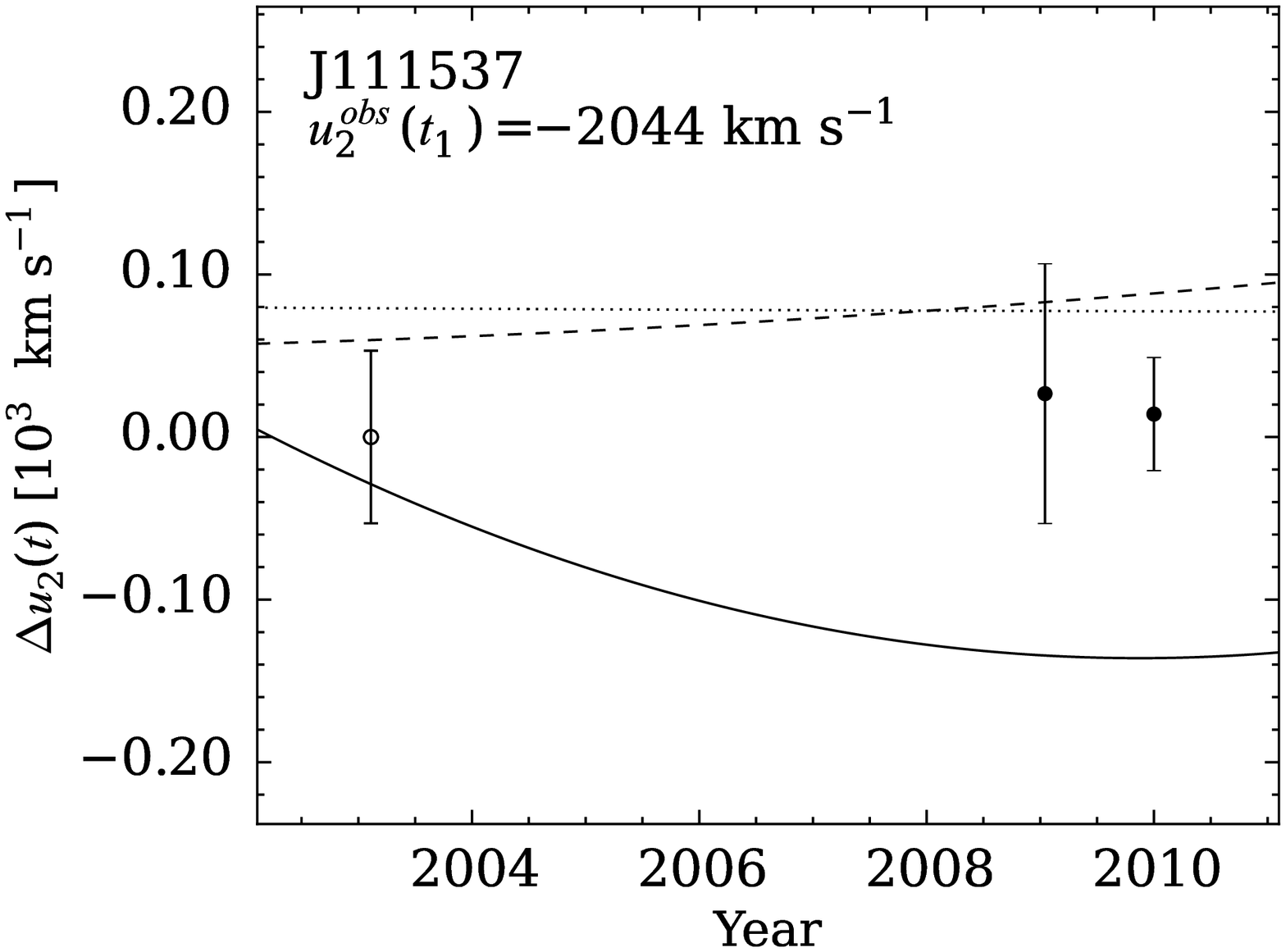} 
}
\centerline{
  \includegraphics[width=8.9cm]{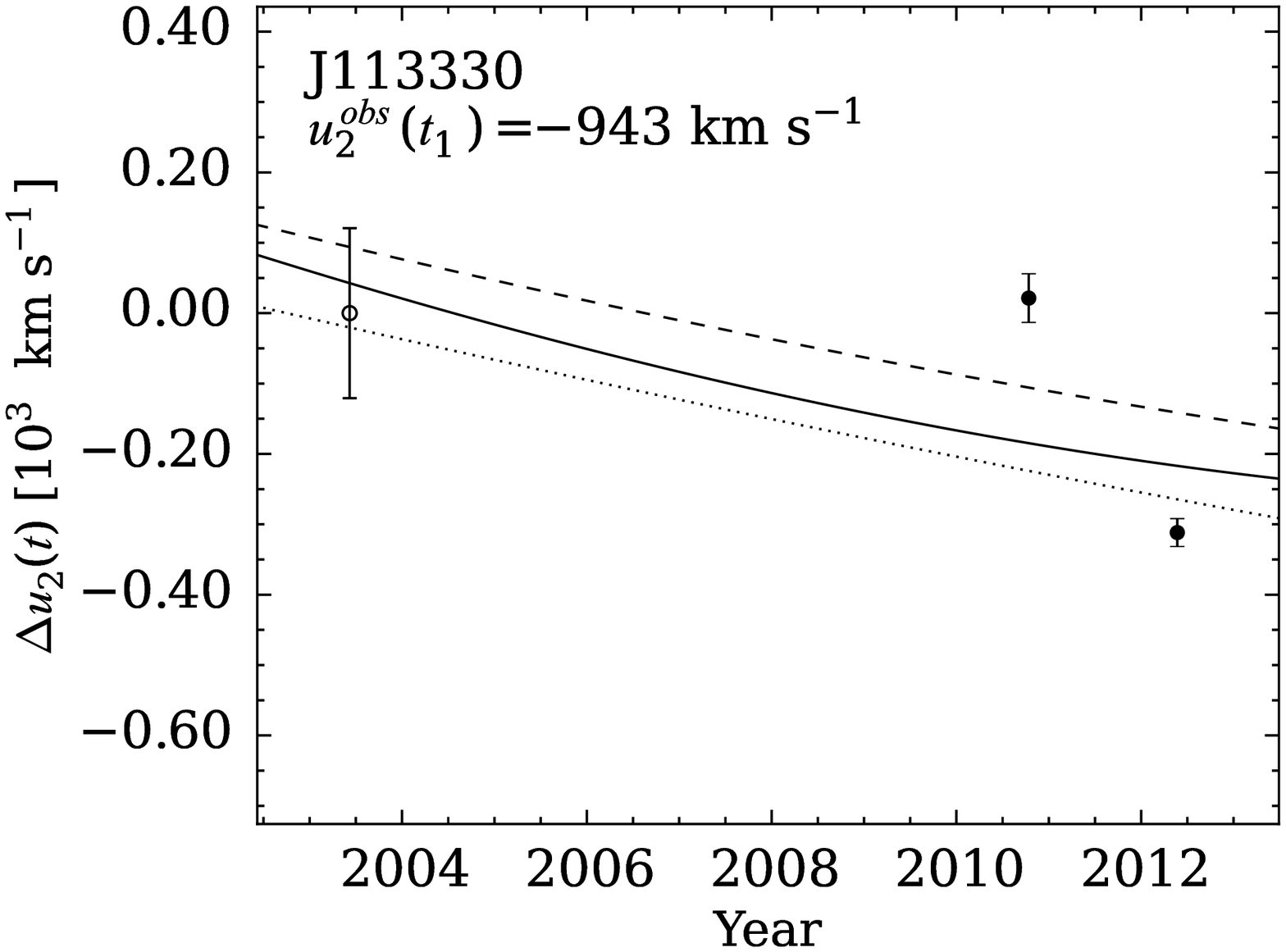} 
  \includegraphics[width=8.9cm]{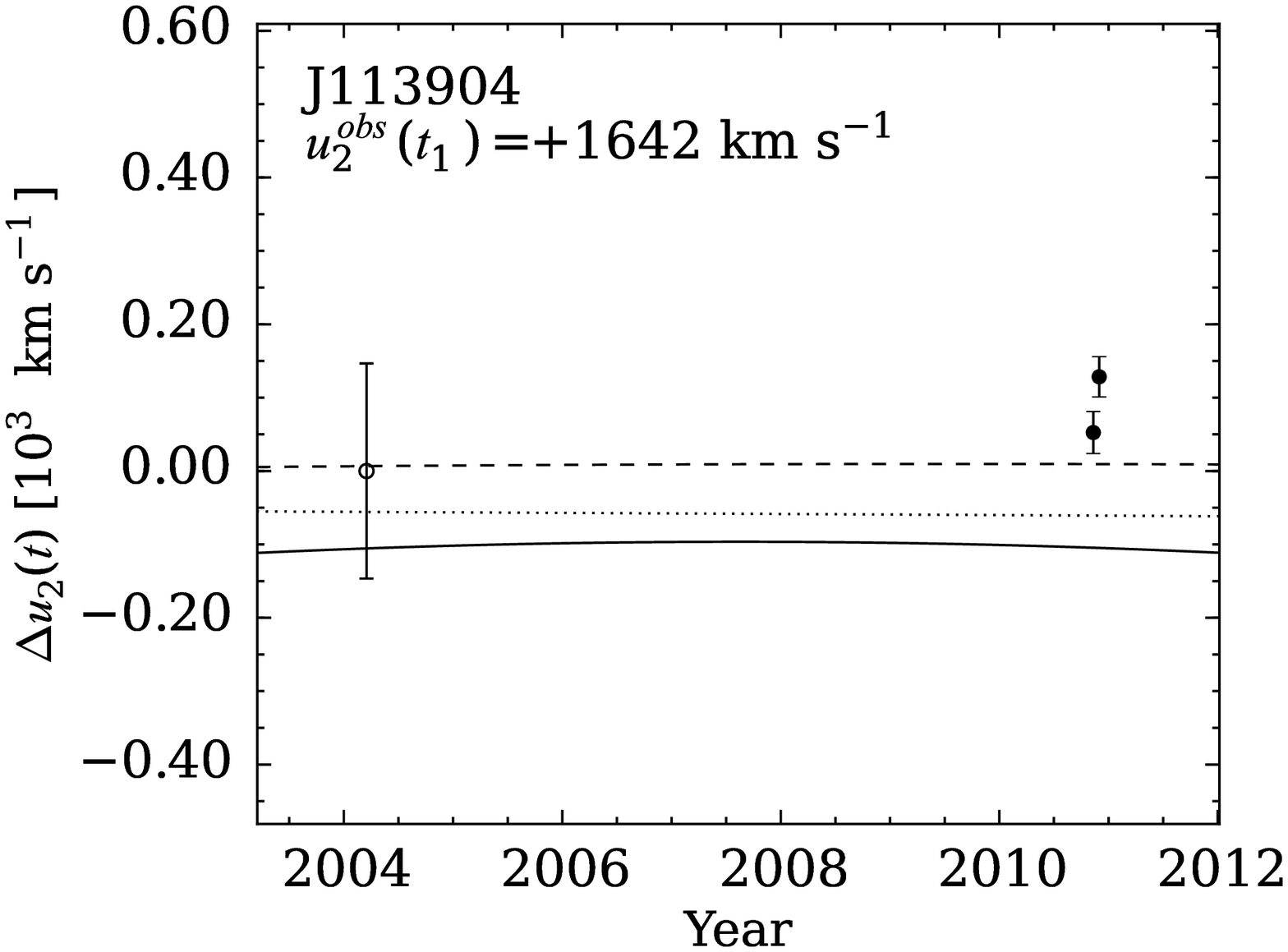} 
}
\centerline{Figure~\ref{fig:zbrv}. -- Continued.}
\end{figure*}

\begin{figure*}
\centerline{
  \includegraphics[width=8.9cm]{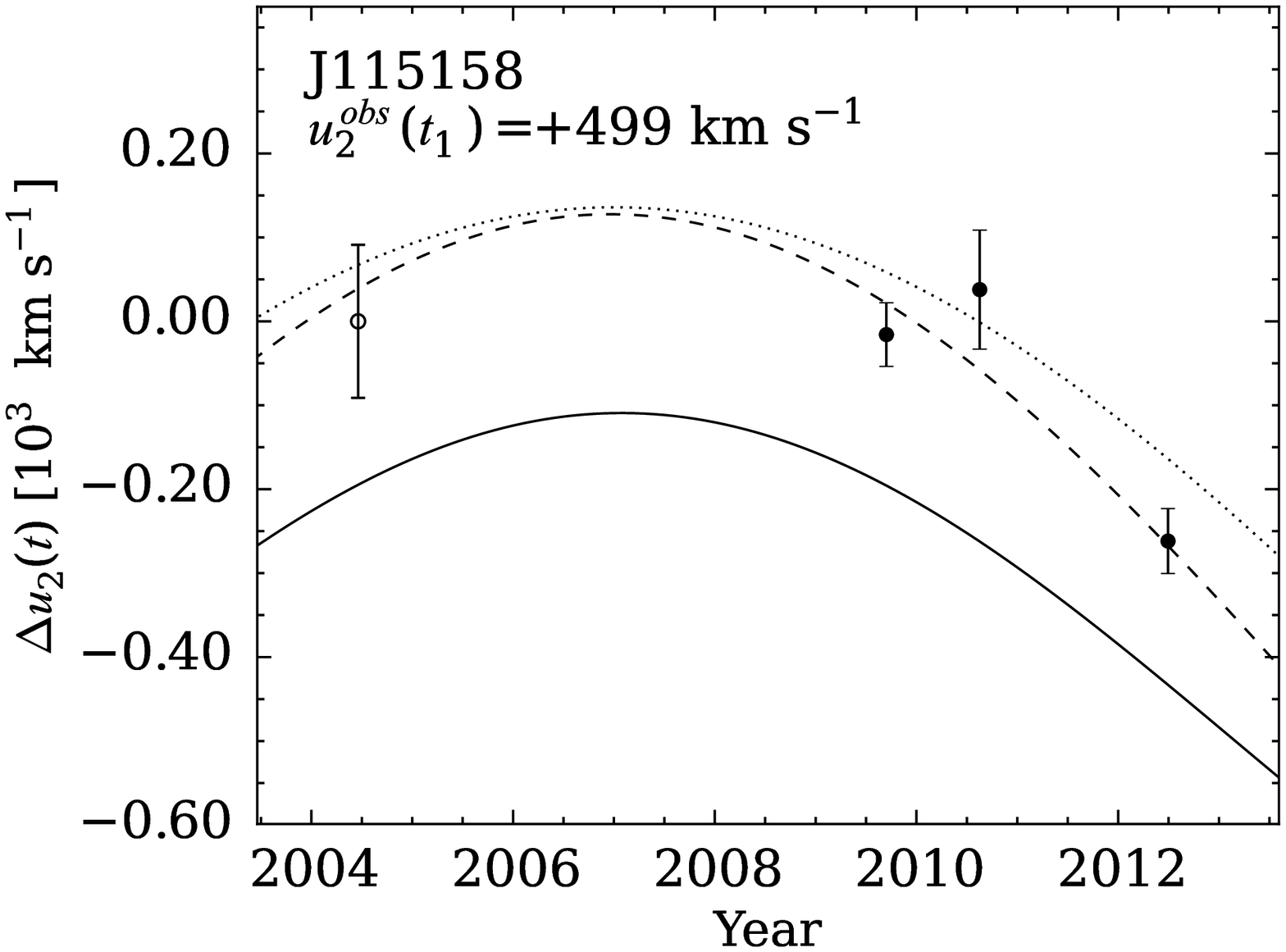} 
  \includegraphics[width=8.9cm]{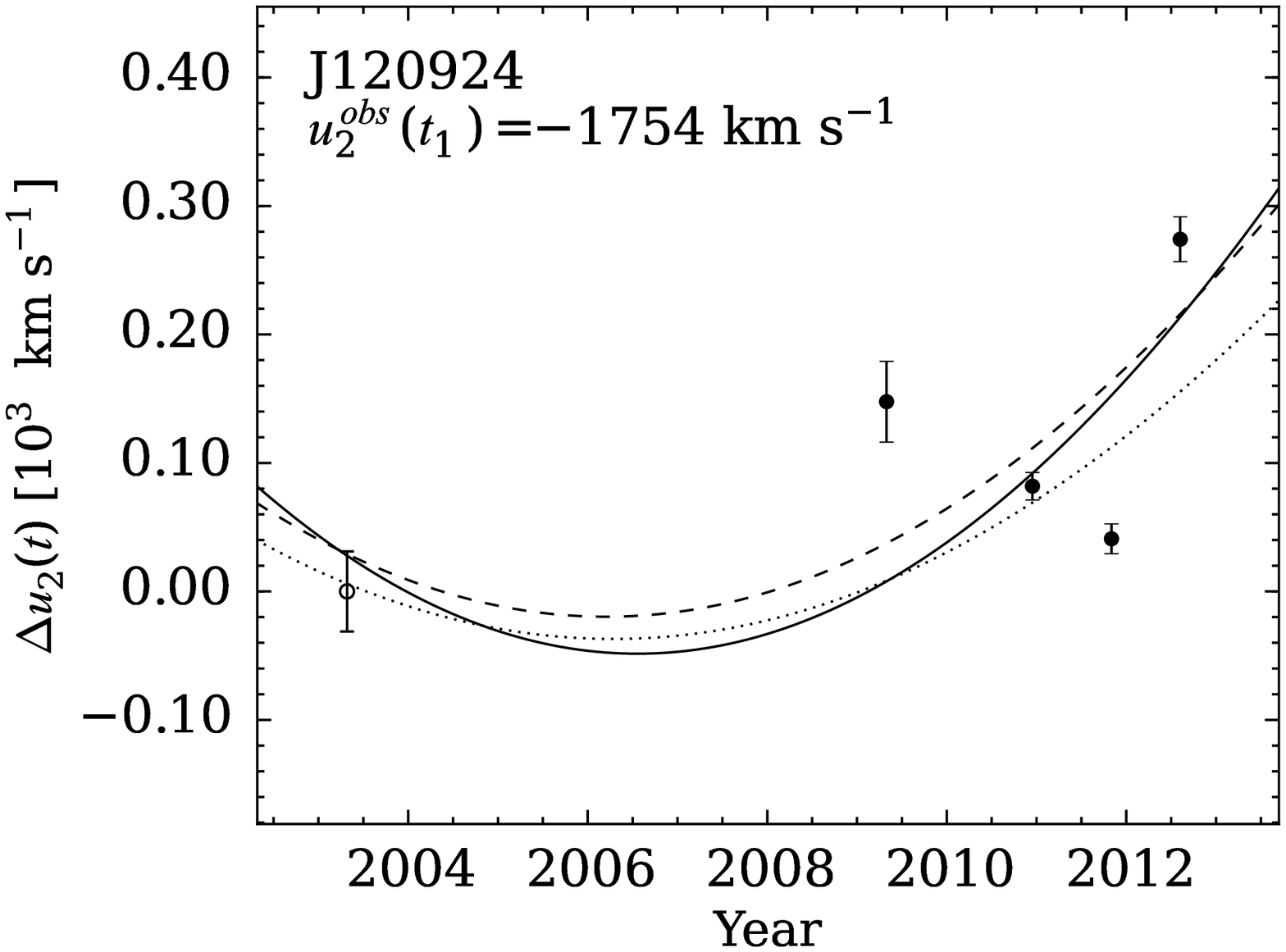} 
}
\centerline{
  \includegraphics[width=8.9cm]{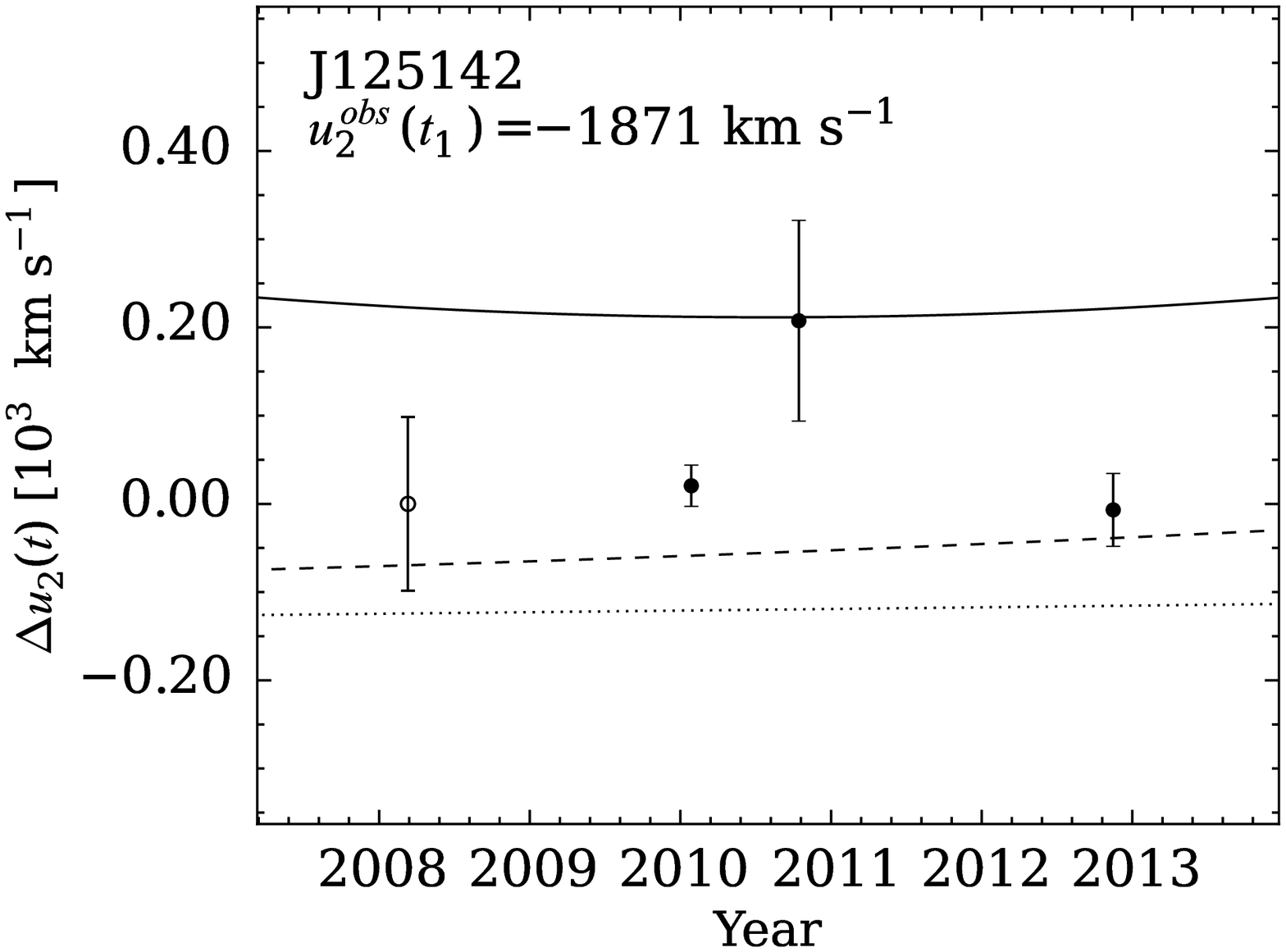} 
  \includegraphics[width=8.9cm]{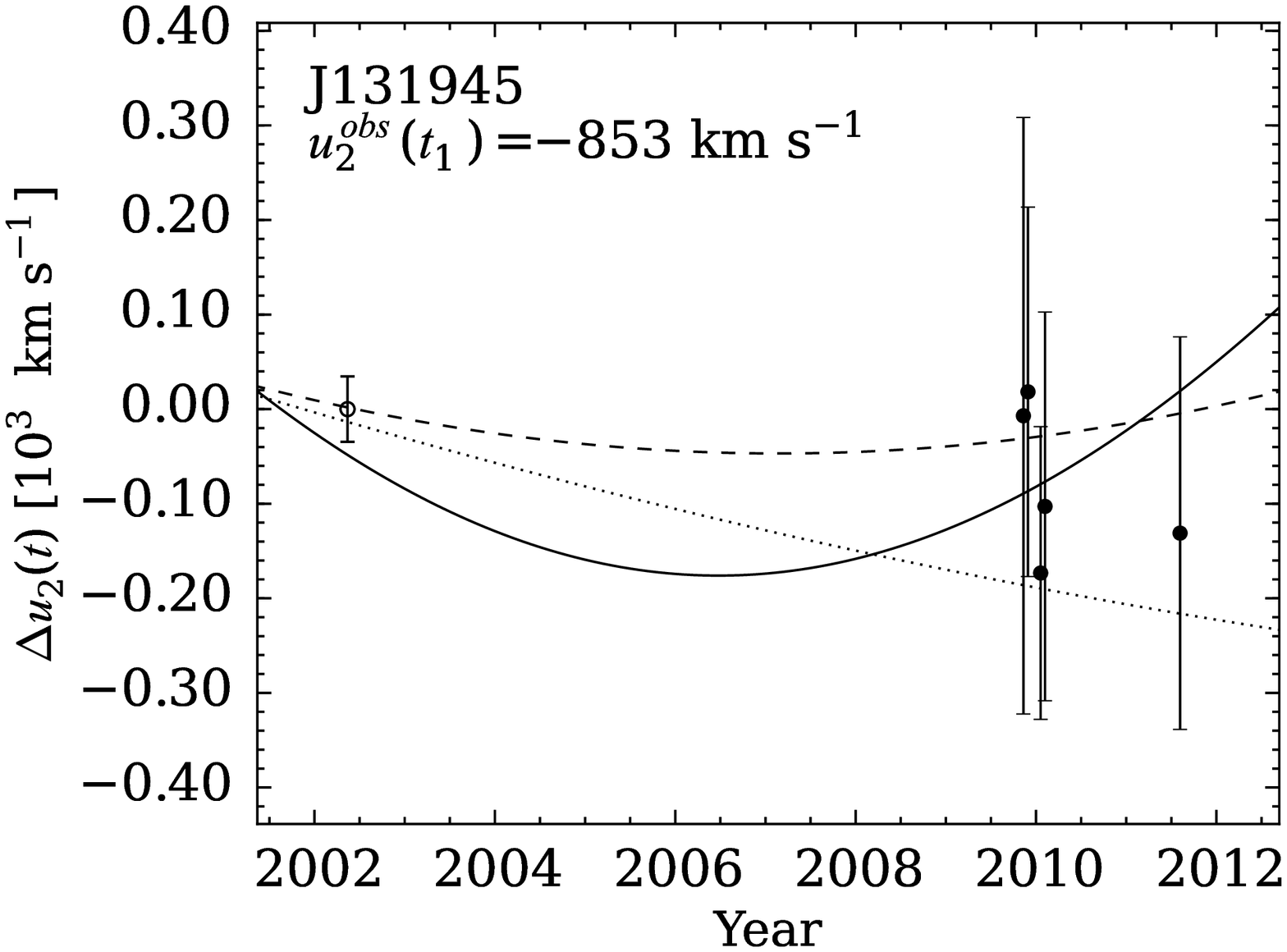} 
}
\centerline{
  \includegraphics[width=8.9cm]{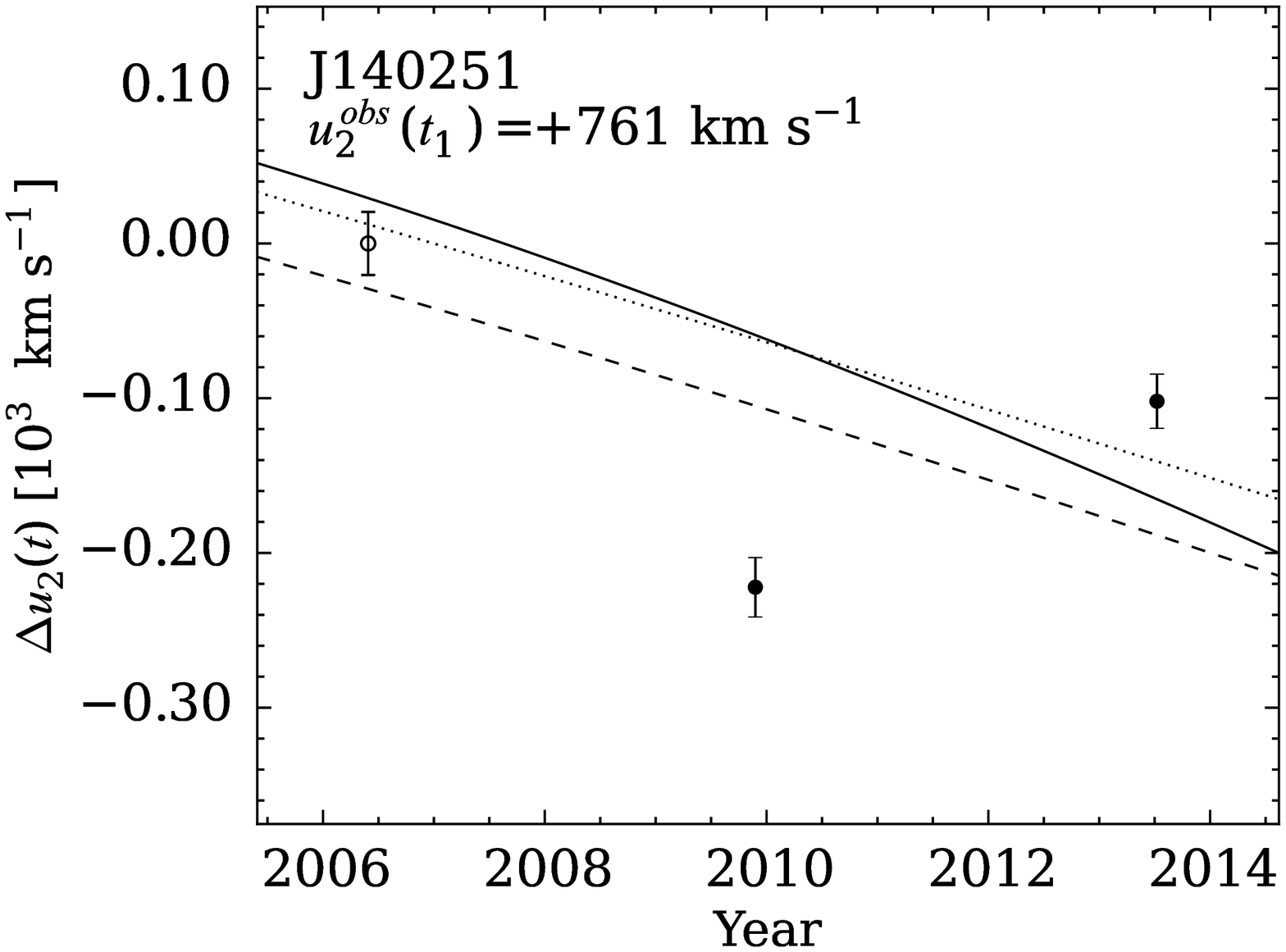} 
  \includegraphics[width=8.9cm]{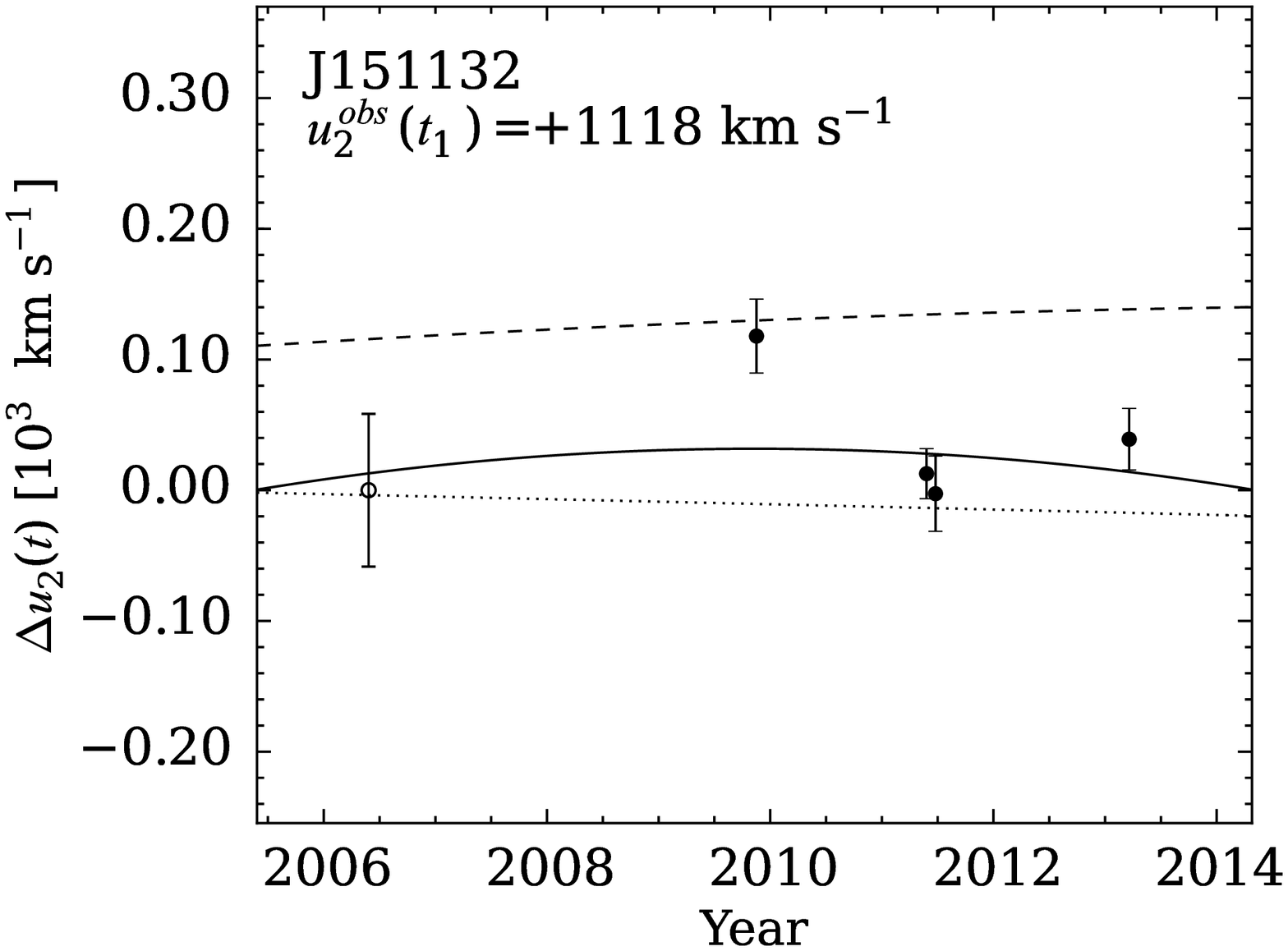} 
}
\centerline{Figure~\ref{fig:zbrv}. -- Continued.}
\end{figure*}

\begin{figure*}
\centerline{
  \includegraphics[width=8.9cm]{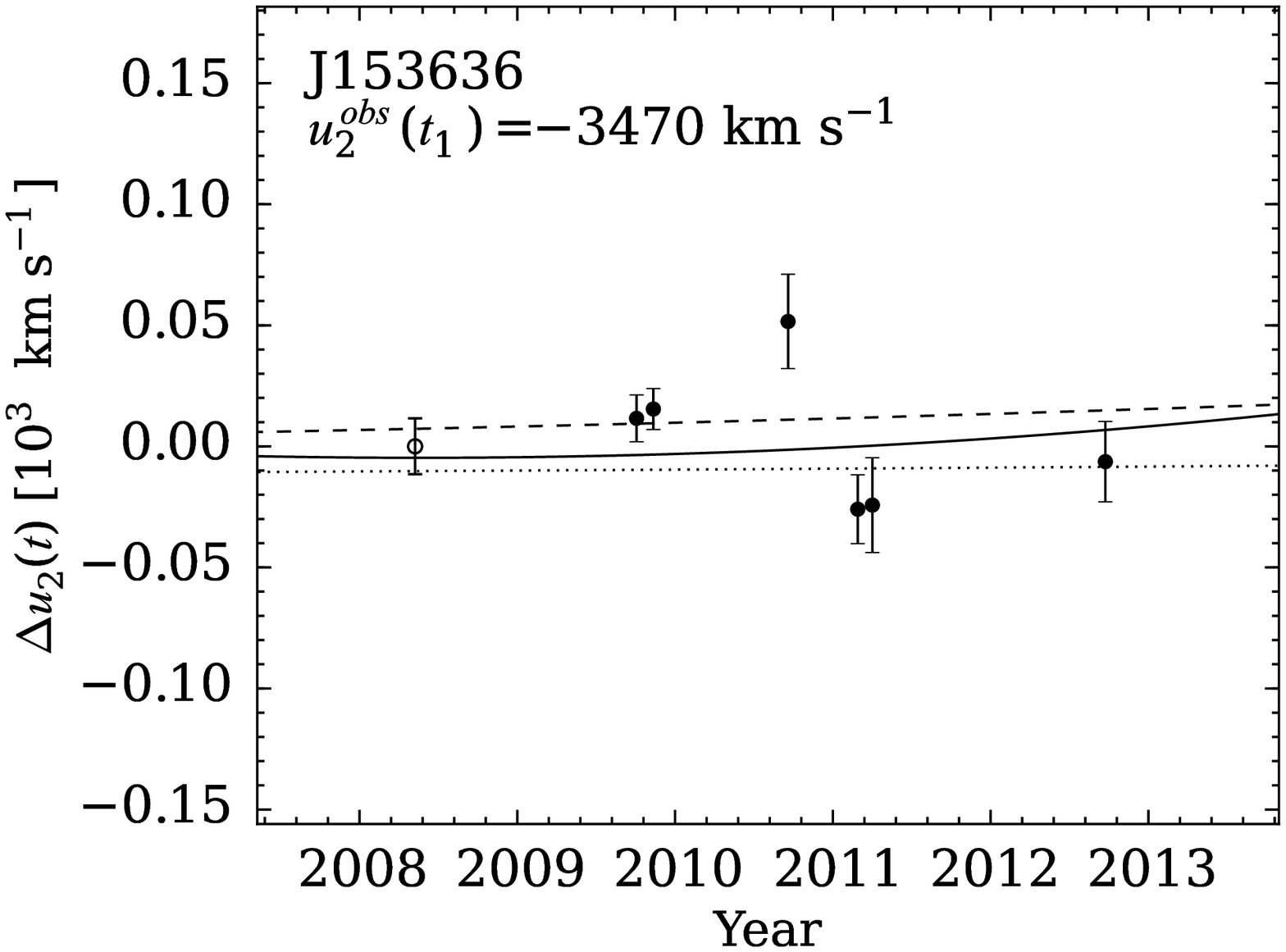} 
  \includegraphics[width=8.9cm]{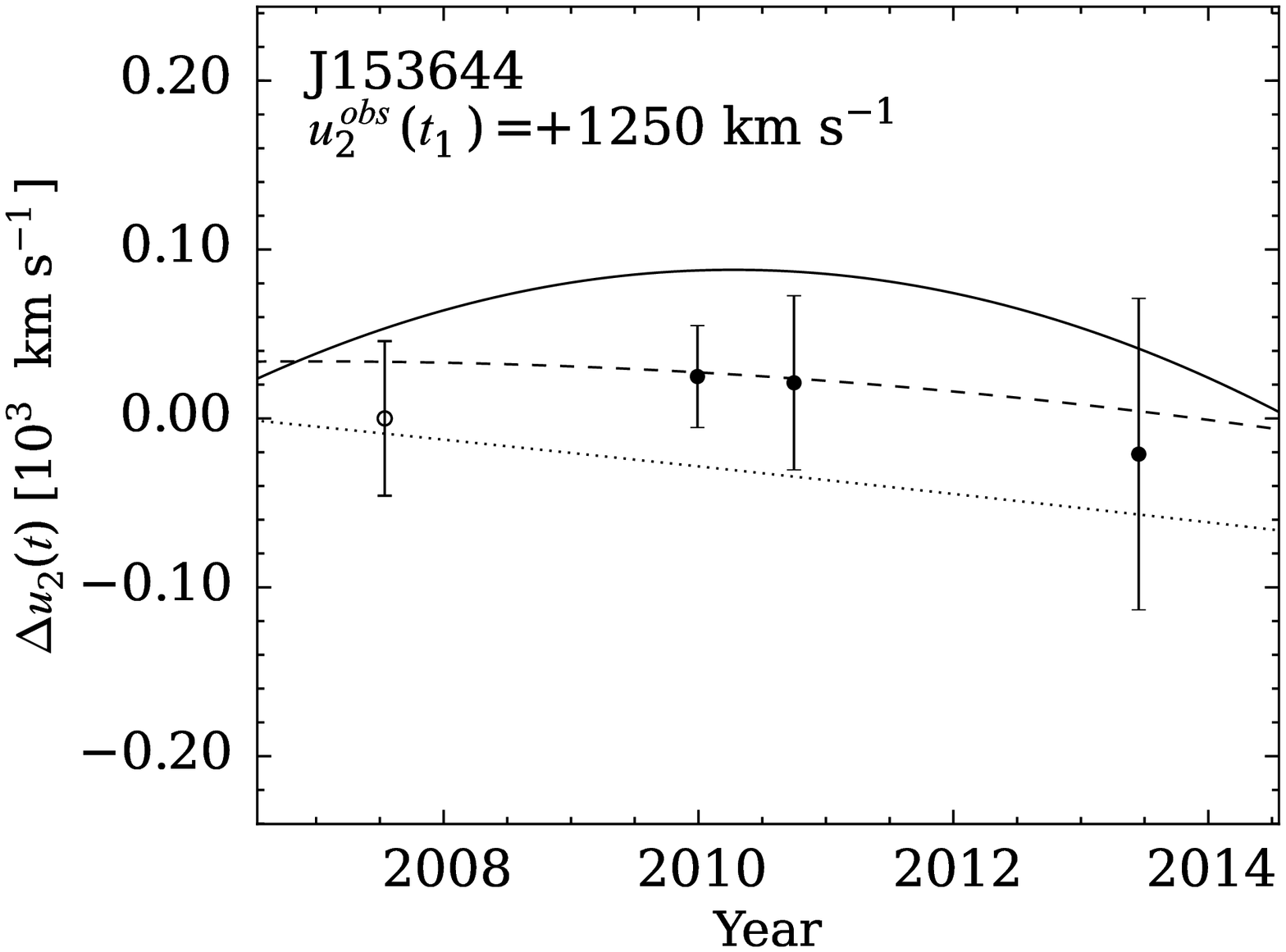} 
}
\centerline{
  \includegraphics[width=8.9cm]{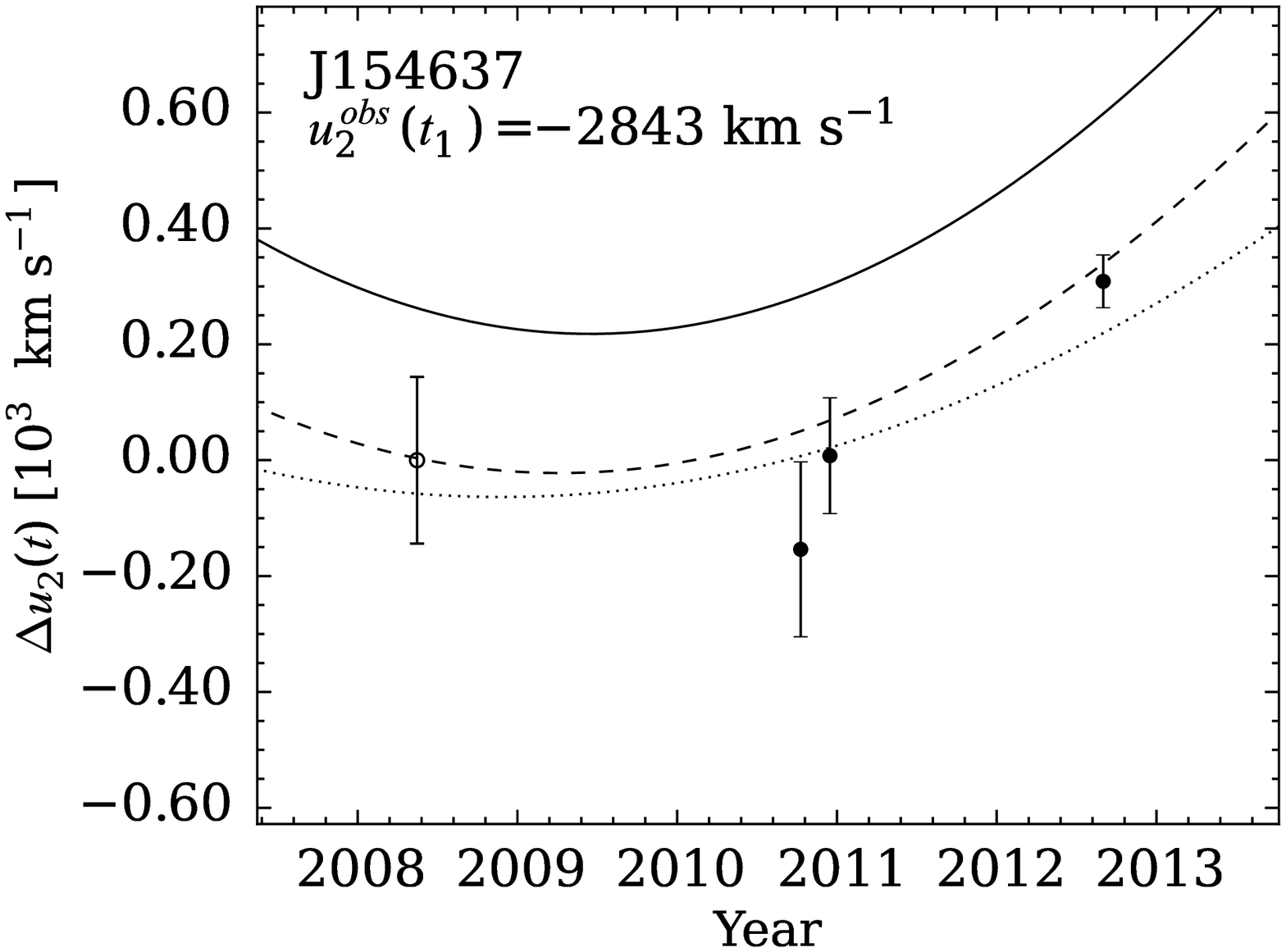} 
  \includegraphics[width=8.9cm]{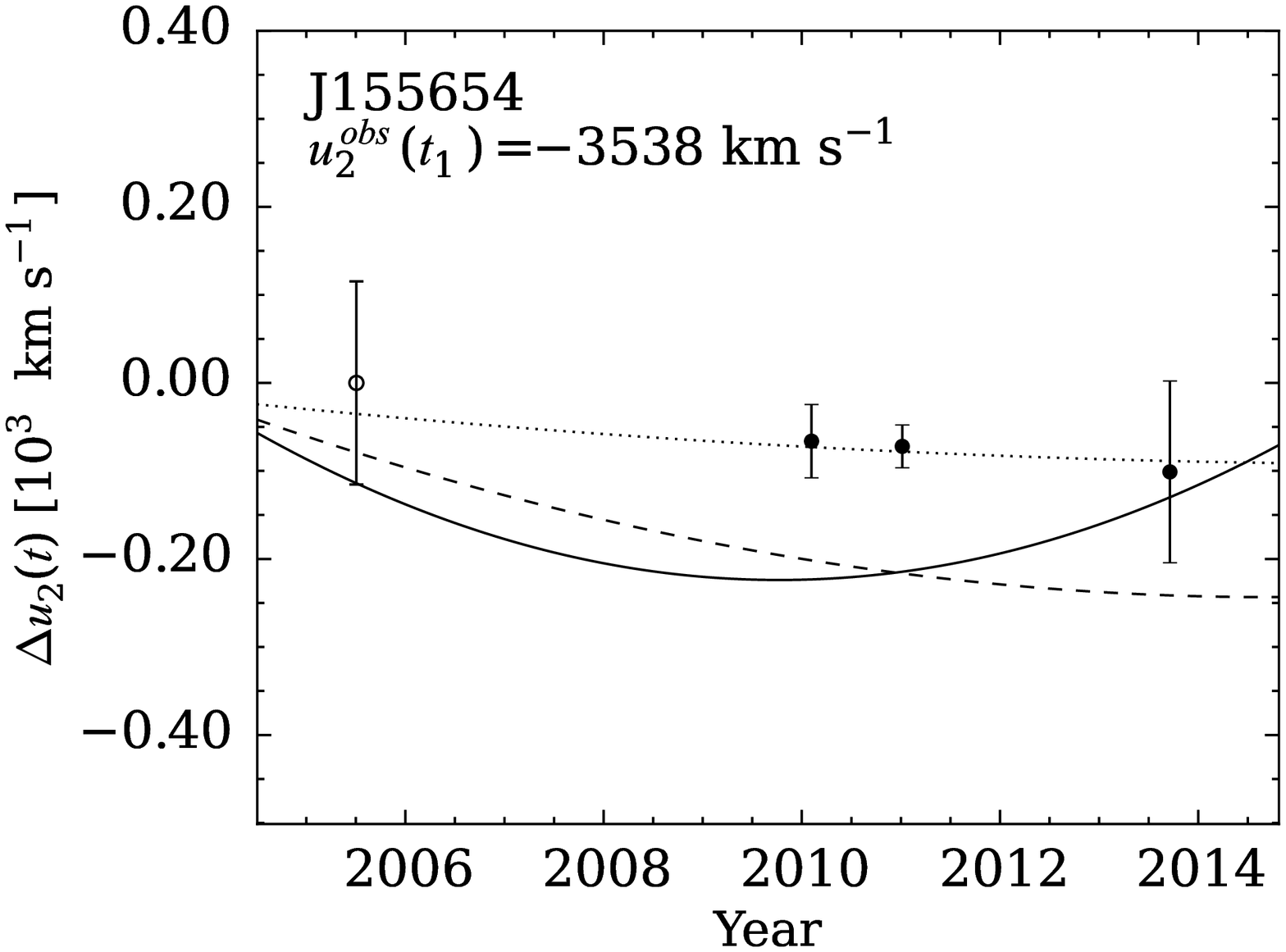} 
}
\centerline{
  \includegraphics[width=8.9cm]{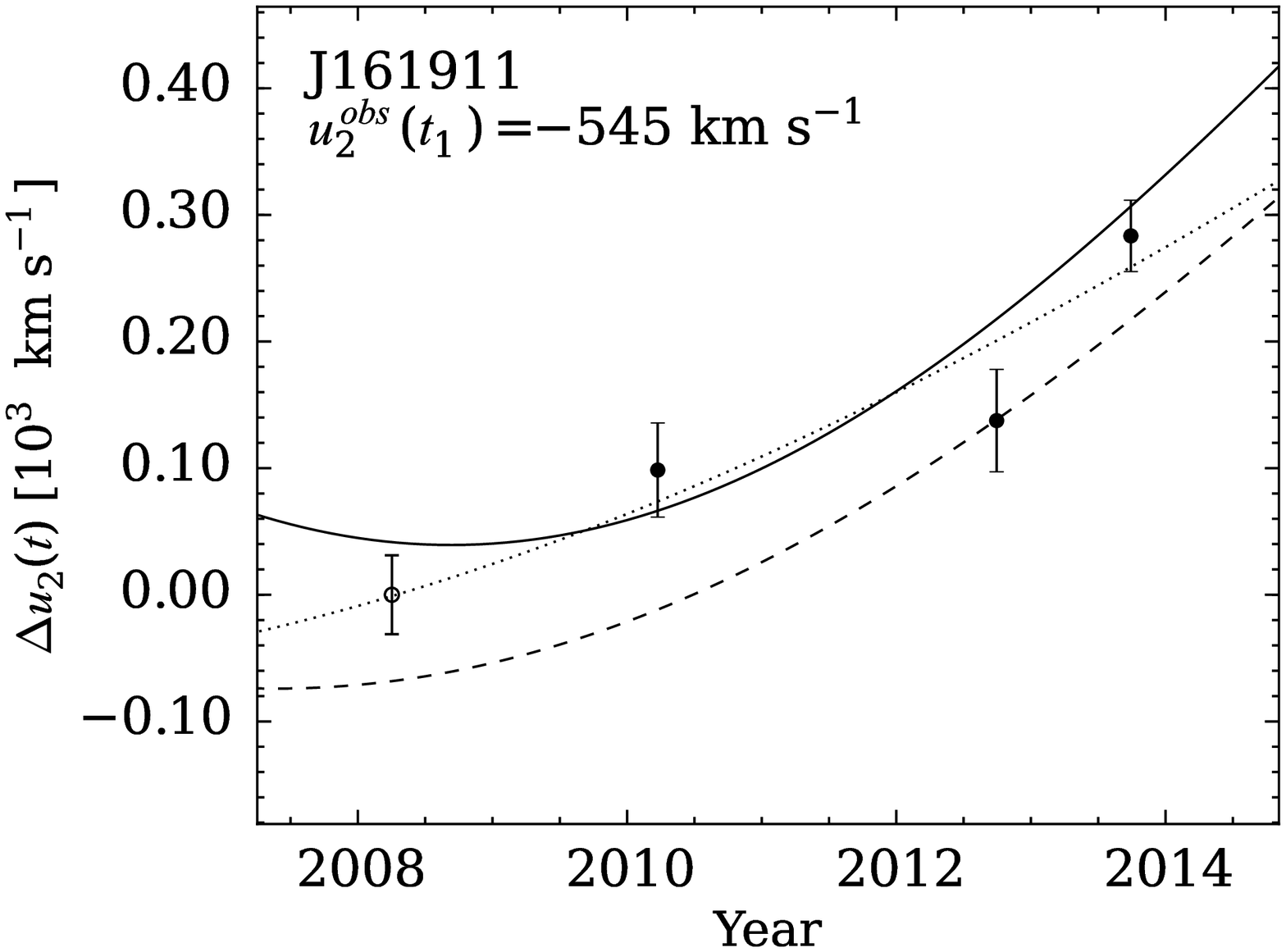} 
  \includegraphics[width=8.9cm]{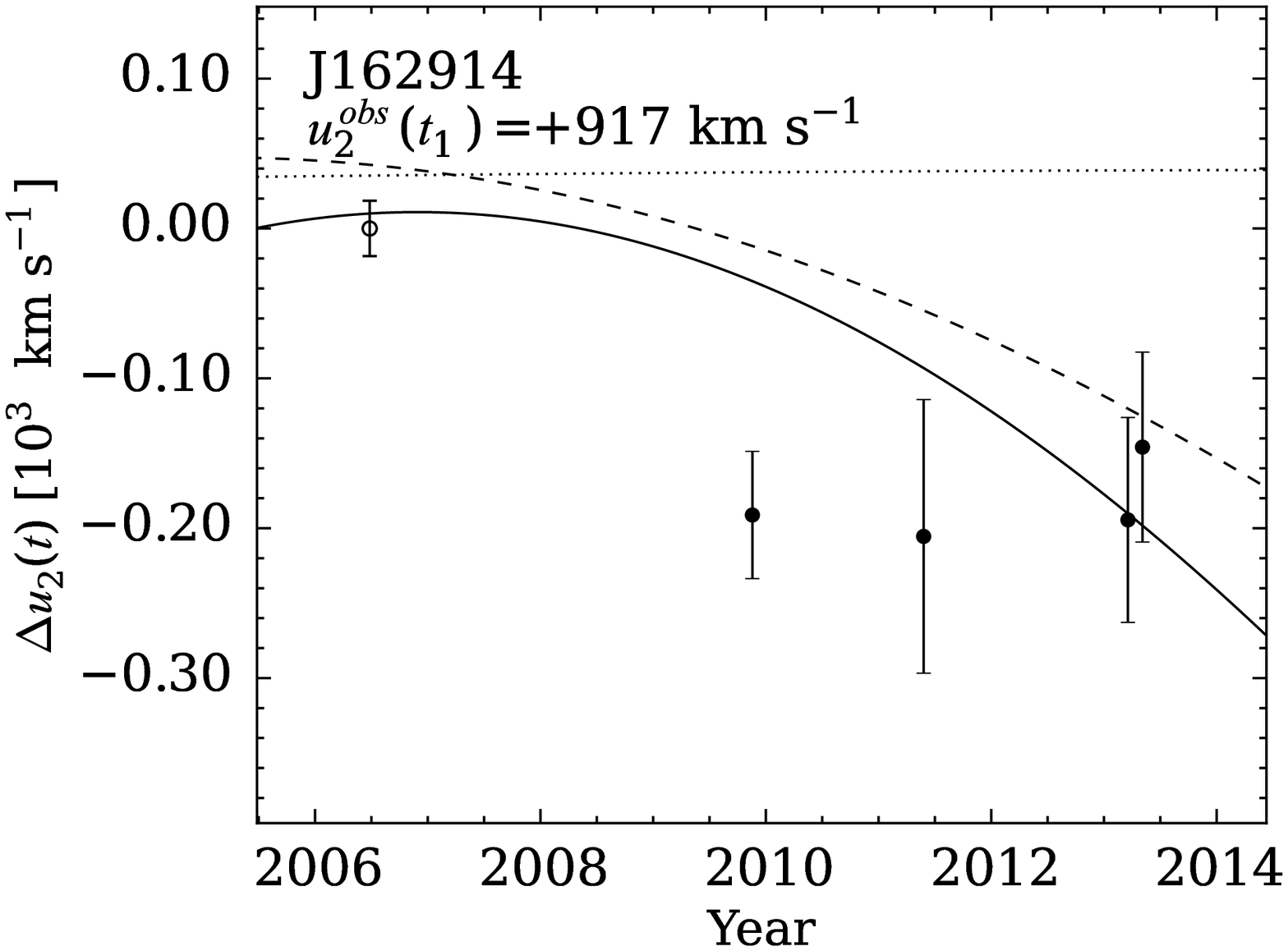} 
}
\centerline{Figure~\ref{fig:zbrv}. -- Continued.}
\end{figure*}

\subsection{Caveats and Uncertainties}
\label{sec:caveats}

The most significant caveat to this method of searching for SBHBs is that variability of the broad H$\beta$ profiles may mimic the radial velocity shifts that we hope to measure. Indeed, based on our visual inspection, 44 per cent (39/88) of the sample showed substantial profile shape variability over the course of our campaign. Profile shape variability at smaller amplitudes would not be apparent in a visual inspection but could be mistaken for a velocity shift. There are multiple mechanisms that can produce profile shape variability on time-scales of the order of the span of our observations, which we describe below.

On time-scales comparable to the variability time-scale of the continuum source, the broad-line centroids or peaks can display velocity shifts as a result of changes in the illumination of the BLR that are similar to those we expect from orbital motion. \citet{barth15} highlight this issue in the context of searches for SBHBs and demonstrate the danger based on high-quality reverberation mapping data for several well-studied AGN. If one side of the broad-line profile responds to variations in the continuum source before the other, the centroid of the line profile can shift by up to a few hundred km~s$^{-1}$, even when measured using cross-correlation techniques. Such an asymmetric response can result from an non-axisymmetric distribution of matter or the presence of radial motions in the BLR. Examples of BLRs with such properties have been found through reverberation mapping campaigns with sufficiently high-S/N spectra \citep[e.g.,][]{bentz10,grier13,skielboe15}. Searches for SBHBs based on radial velocity changes of line profiles that are close to their systemic velocity are particularly susceptible to this effect. We cannot rule out the possibility that this effect is also at work in our targets since we do not monitor the continuum variations of our targets. Even in a SBHB, this type of variability might be expected and will produce jitter on top of a systematic velocity change. Thus, long-term monitoring is crucial in order to measure long-term radial velocity variations.

On time-scales comparable to the dynamical time-scale in the BLR, of order years or longer, the broad-line profile may vary independently of the continuum source as a result of structural changes in the BLR (e.g., redistribution of the gas in position and/or velocity space). The variations of the line profiles resemble a ``see saw'' pattern and can be quantified by following the magnitude and sign of the skewness. This behavior has been observed in both quasars \citep[e.g.,][]{kassebaum97,romano98,dediego98} and Seyfert galaxies \citep[e.g.,][]{peterson99conf,kollatschny00,wangli11,doroshenko12,schimoia12}. The same behavior is also fairly common in the profiles of broad, double-peaked Balmer lines found in some radio-loud AGNs \citep[e.g.,][]{eracleous97,sergeev07,gezari07,lewis10,popovic14}. Thus, we performed simulations to investigate whether profile shape variability can masquerade as the radial velocity shifts we hope to measure. We describe the simulations in detail in Appendix~\ref{app:prosim}. In summary, we induced both shape variability and a shift of the broad \Hb\ profile in the observed spectra of the SBHB candidates, and applied the same cross-correlation method to measure shifts. We found that it is much easier to mistake profile shape variability for a bulk shift of the broad line in broad boxy profiles, while in cuspy profiles it is easier to notice in a visual inspection that the profile has changed. However, we were able to find examples of both boxy and cuspy profiles where the profile shape variability and bulk shift produced visually indistinguishable shift measurements, as we illustrate in Appendix~\ref{app:prosim}. We conclude that it is possible to be misled by this type of profile shape variability into thinking that we are measuring radial velocity shifts. Again, long-term spectroscopic monitoring is the best way to distinguish between profile shape variability and real shifts. Moreover, independent tests of the nature of the SBHB candidates can bolster the radial velocity results.

\section{Interpretation of radial velocity variations in the context of the binary hypothesis}
\label{sec:interpretation}

It is useful to consider whether the apparent radial velocity variations that we have measured in our campaign so far are consistent with those expected from orbital motion. To this end, we present here a methodology for testing whether the observed radial velocity curves are consistent with sinusoids and obtaining constraints on the orbital parameters. In view of the long periods that we expect, the main attainable goal of this exercise is to set restrictive lower limits on the BH masses that make the SBHB hypothesis untenable for some systems. Even objects in which no significant variations are observed can lead to interesting limits on the properties of the hypothesized SBHBs so we consider all objects with reliable radial velocity curves, even if their radial velocity variations are consistent with zero within uncertainties. We follow a variant of the methodology described by \citet{eracleous97} \citep[see also][]{halpern88, halpern92, liu16} to identify the minimum period that is consistent with the observed radial velocity curve. We assume that both members of the binary follow circular orbits around the centre of mass and only the secondary BH is active.

Our assumption of circular orbits is motivated by both practical and theoretical considerations. From a practical point of view, the radial velocity curves shown in Figure~\ref{fig:zbrv} often include three or four measurements, which are not enough to constrain the free parameters of an elliptical orbit. In some cases the radial velocity curves include five to seven points but in these cases there are closely separated pairs of measurements, which hampers our ability to get meaningful constraints on the parameters of an elliptical orbit. From a theoretical point of view, predictions for the orbital eccentricity of SBHBs vary substantially and depend on the formation channel of the system and the mass of the circumbinary disc. Simulations of the evolution of a SBHB through interactions with stars suggest that the eccentricity is damped very quickly \citep{khan13,holley15}. However, the question of eccentricity evolution in these models is not settled because it depends sensitively on the initial conditions, the SBHB mass ratio, and other properties of the system \citep[e.g.,][]{quinlan96,sesana06,wang14,vasiliev15}, and can be high \citep{vecchio94,aarseth03}. Eccentricity evolution in models where the SBHB interacts with a gaseous reservoir is also uncertain. In models where the SBHB initially evolves in a large, {\it smooth}, gaseous disc \citep[e.g.][]{dotti06}, the eccentricity is damped very quickly. However, if the disc includes massive clumps, the interaction of the BHs with these clumps imparts a substantial eccentricity to their orbits \citep[e.g.][]{fiacconi13,delvalle15}. At the late stages of the evolution of the SBHB, after it has opened a gap in the gaseous disc, the evolution of the eccentricity depends on the mass of the disc exterior to the binary orbit. If the disc is sufficiently massive, the combination of torques from the disc and accretion streams will give rise to an appreciable eccentricity, even if the orbit is initially circular \citep[see][and references therein]{roedig11}. Otherwise, the eccentricity will not be affected by interactions of the SBHB with the circumbinary disc \citep[e.g.][]{hayasaki07,farris14}. Thus, there is a wide variety of physical scenarios in which SBHBs have circular orbits.
 
Under the assumption of circular orbits, we write the model for the radial velocity curve of the secondary as
\begin{equation}
\label{eqn:sine}
u_{2}(t) = (V_2\,\sin\,i)\,\sin\left[ \frac{2\pi}{P} (t-t_0)\right],
\end{equation}
where $u_{2}$ is the observed velocity of the secondary, $V_2$ is the (true, tangential) orbital velocity of the secondary, $i$ is the inclination angle between the normal to the orbital plane and the line of sight, $P$ is the orbital period, and $t_0$ is the time of the next inferior conjunction of the secondary (i.e., the time at which the secondary starts to recede from the observer and its projected position is between the primary and the observer; if the orbital plane were viewed edge on, this would be the time of eclipse of the primary by the secondary). Moreover, we denote the time of the $n^{th}$ observation in our time series as $t_n$ and the observed radial velocity at that time as $u_2^{\rm obs}(t_n)$ (the superscript ``obs'' identifies {\it measured} radial velocities). In this convention, $u_2^{\rm obs}(t_1)$ is the radial velocity measured from the first spectrum in the series, taken as part of SDSS.

We have to treat the uncertainty in the first radial velocity measurement differently from the uncertainties in subsequent measurements because the first measurement is an absolute one while subsequent measurements are differential (relative to the first one). Therefore, we re-cast the model for the radial velocity curve as 
\begin{eqnarray}
\label{eqn:delsine}
\nonumber \Delta u_{2}(t) &\equiv& u_{2}(t)  - u_2(t_1) \\
&=& (V_2\,\sin\,i)\,\sin\left[ \frac{2\pi}{P} (t-t_0)\right] - u_2(t_1),
\end{eqnarray}
where $u_2(t_1)$ is the initial velocity offset in the time series and $\Delta u_{2}(t)$ represents a velocity change {\it relative to the initial value.} Following the convention of the previous paragraph, we use $\Delta u_2^{\rm obs}(t_n)$ to identify {\it differential} velocity measurements. By writing the model in terms of $\Delta u_{2}(t)$, we can use the uncertainties in relative velocity changes determined as described in \S\ref{sec:xc}. 

\begin{figure*}
\centerline{
  \includegraphics[width=8.9cm]{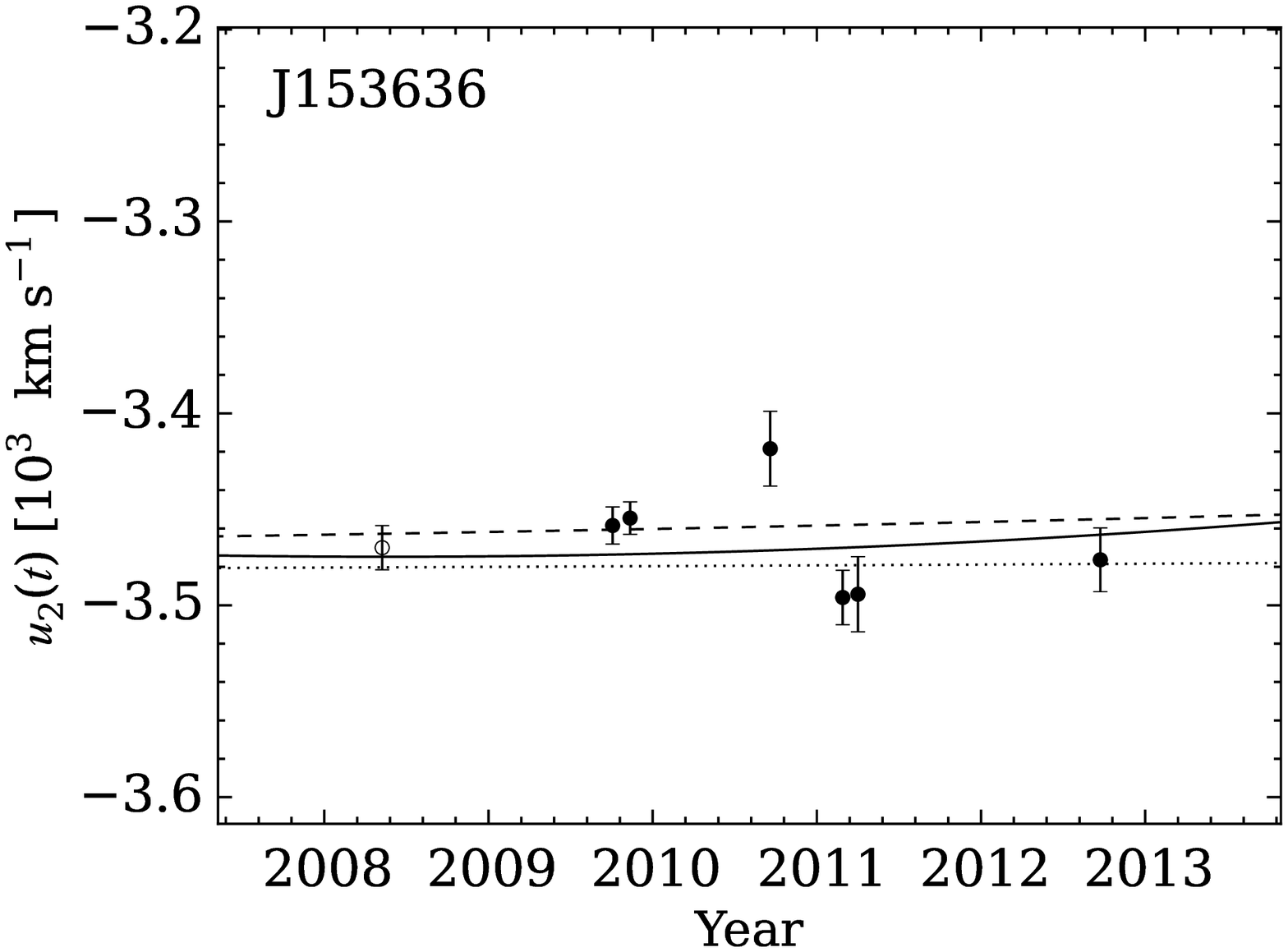}
  \includegraphics[width=8.9cm]{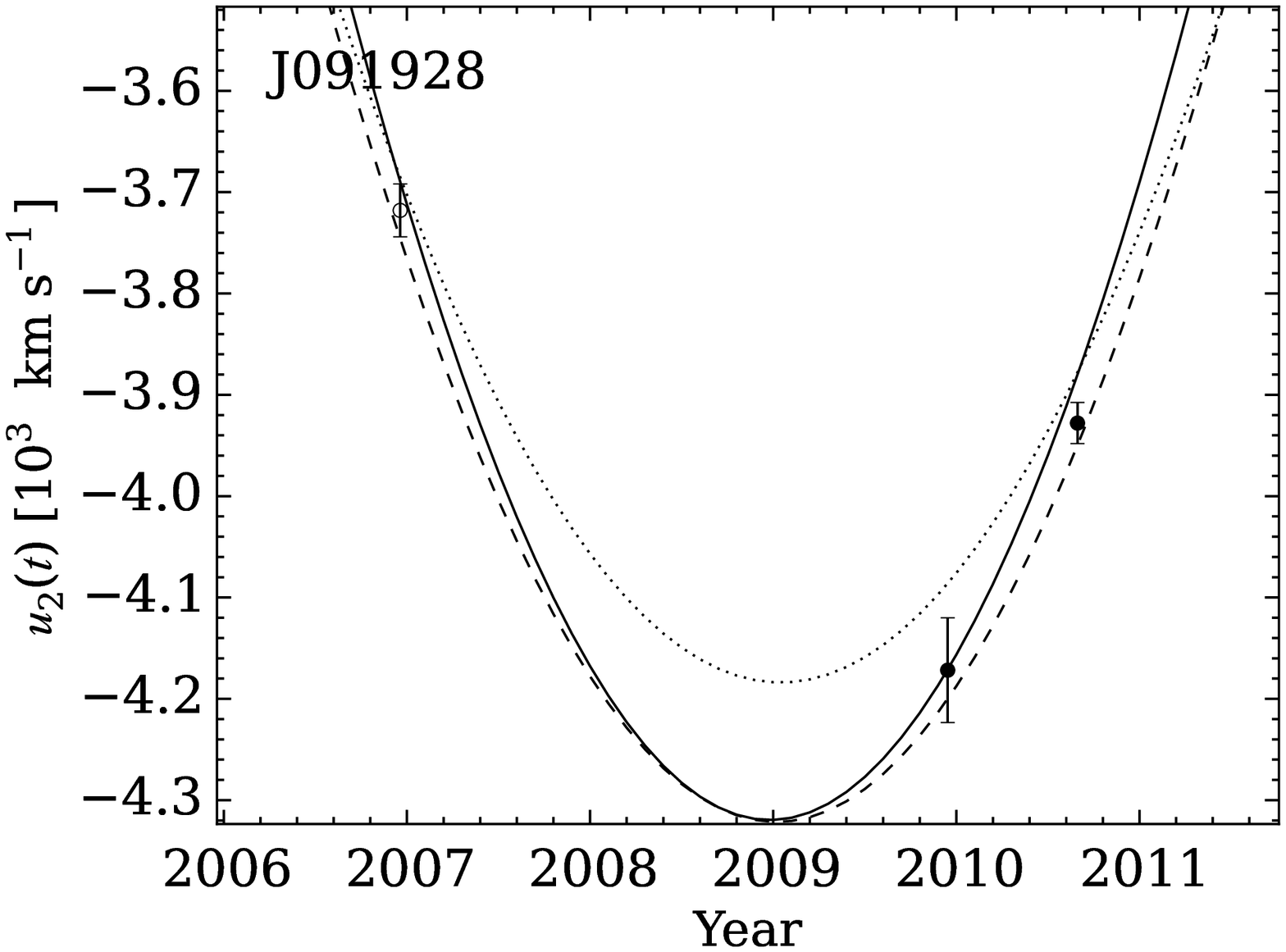} 
}
\centerline{
  \includegraphics[width=8.9cm]{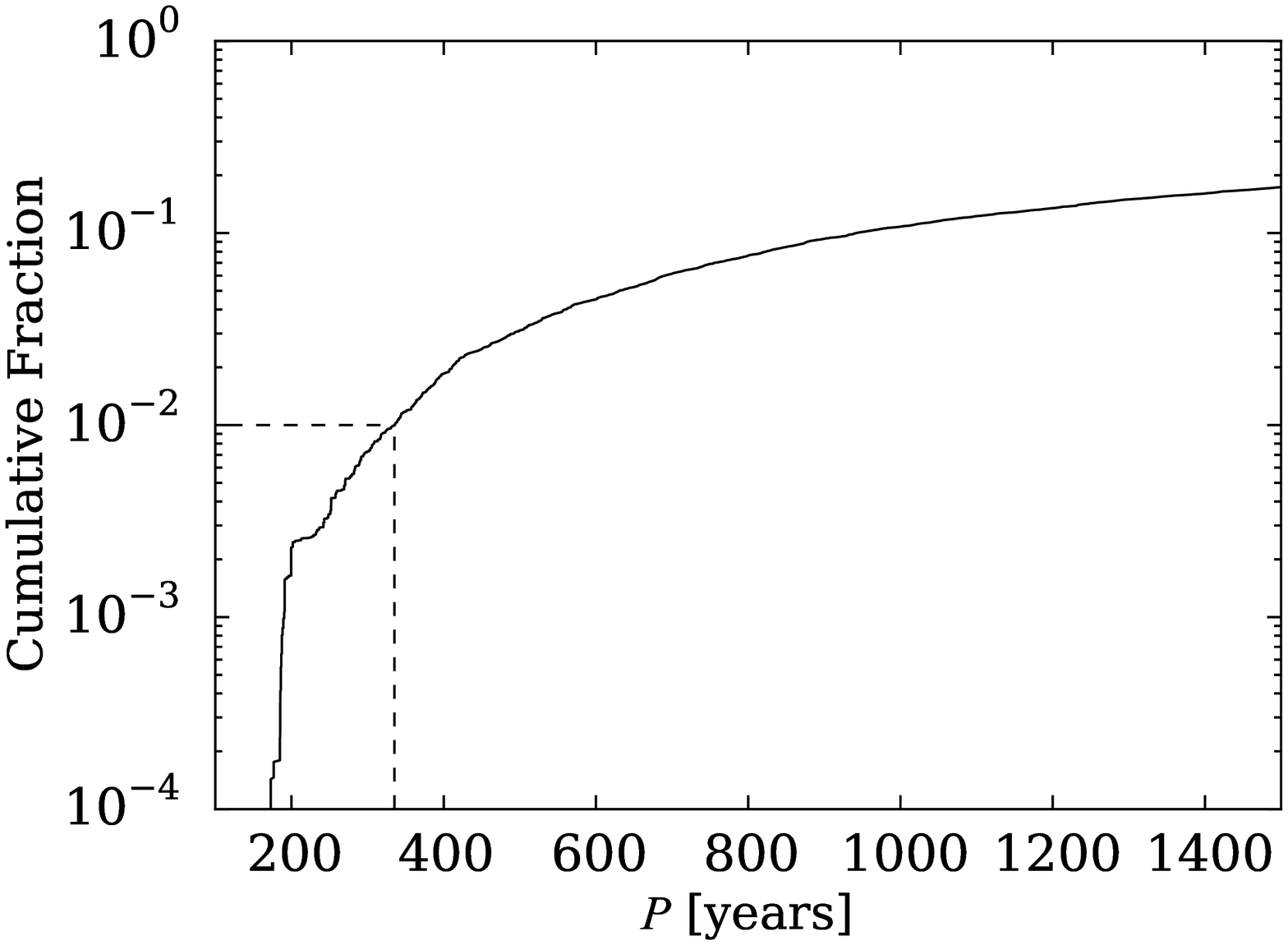} 
  \includegraphics[width=8.9cm]{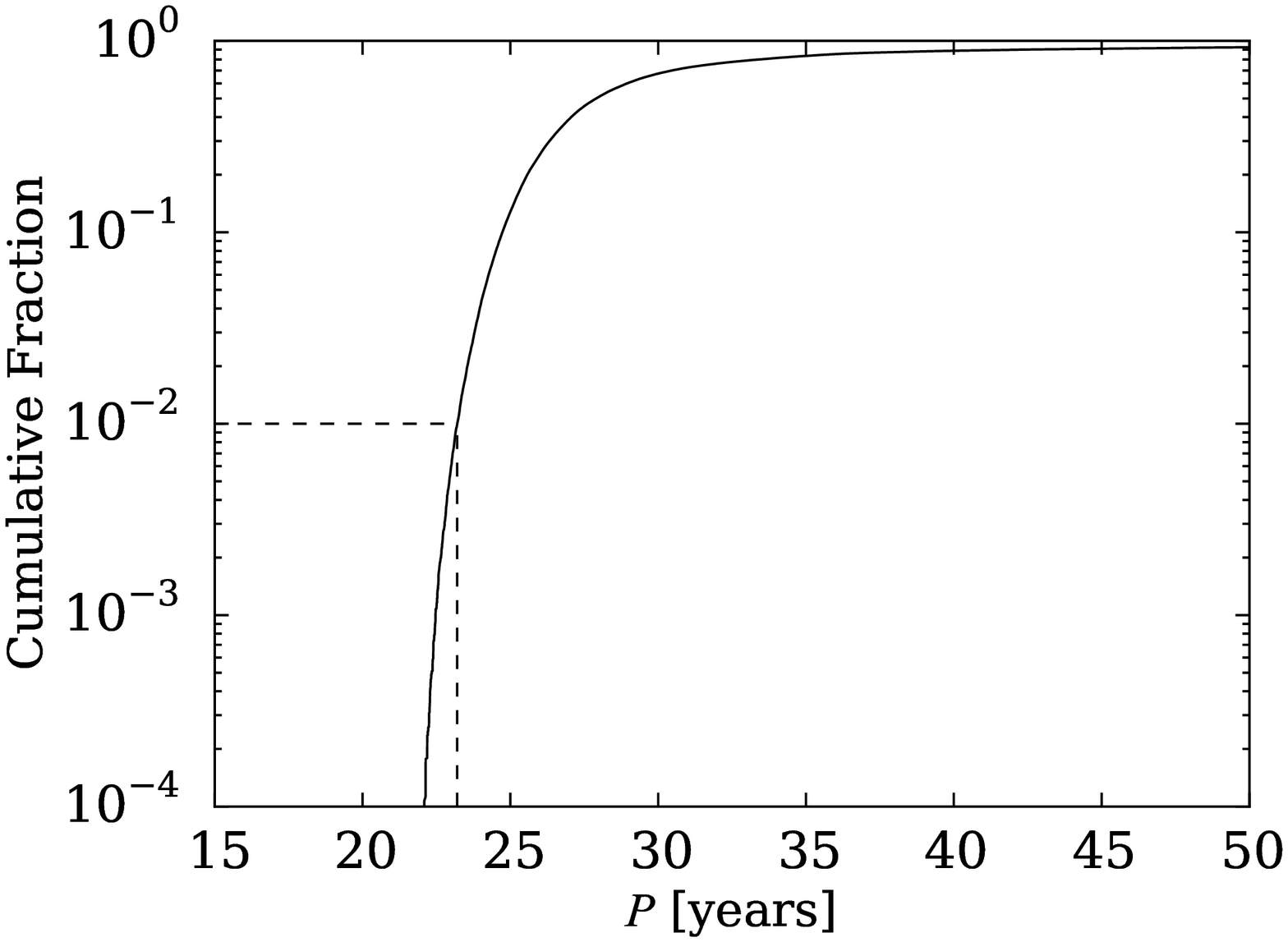} 
}
  \caption{Results from the MCMC simulations to determine the minimum period consistent with the observed radial velocity curves. The top panels show the measured radial velocity curve with ``$1\sigma$'' (68 per cent confidence) error bars. Dotted, dashed, and solid lines show the sinusoids corresponding to the 68th, 90th, and 99th minimum percentile periods, respectively. The bottom panels show the corresponding cumulative distributions, $\psi(>\!P)$, from the combined high- and low-resolution MCMC simulations (see Appendix~\ref{app:mcmc}). The dashed line indicates the 99th percentile minimum period in the observed frame. The left column shows J153636 ($z=0.3889$), where the flat radial velocity curve pushes the minimum period to large values, yielding a lower limit on the rest-frame mass of $M > 3.8\times10^{8}$~\Msun. The right column shows J094603 ($z=0.2203$), where differences in the radial velocity measurements likely due to jitter are interpreted as systematic trends in the radial velocity curve due to the small number of measurements. The sinusoids corresponding to the 68th, 90th, and 99th percentile periods from these distributions can be viewed for all objects in Figure~\ref{fig:zbrv}. The 99th percentile periods determined from these cumulative distributions for all objects are listed in Table~\ref{tab:Pmin}.} \label{fig:mcmc}
\end{figure*}

In order to test this model, we ran the Markov Chain Monte Carlo (MCMC) sampler {\tt emcee} \citep{foreman13}, which we describe in detail in Appendix~\ref{app:mcmc}. In summary, the simulations determine the likelihood of our data given the above model, and produce posterior distributions for all model parameters and an accurate accounting of their covariance. We then generate the cumulative posterior distribution of $P$, $\psi(>\!P)$, from which we can determine the 68th, 90th, and 99th percentile minimum periods and calculate the corresponding orbital models. In Figure~\ref{fig:mcmc} we show two examples of the ultimate output of the MCMC simulations, one for an object with a flat radial velocity curve and one where jitter has likely been interpreted as orbital motion. The output consists of the $\psi(>\!P)$ distributions and the radial velocity curve with the orbital models corresponding to the 68th, 90th, and 99th percentile minimum periods superposed. These models are illustrated for all objects in Figure~\ref{fig:zbrv}.

We take the period corresponding to the 99th percentile of $\psi(>\!P)$ to be the minimum period that is consistent with the observations, $P_{\rm min}$. We combine this period with the corresponding velocity normalization of the sinusoid to place a lower limit on the total mass and orbital separation of the SBHB. In the framework described in \S{2} of \pone, 
\begin{equation}
\label{eqn:Mmin}
M > 3.8\times10^{5}\,(1+q)^3\,\left(\frac{P_{\rm min}}{10\,\textrm{yr}}\right)\,\left(\frac{V_2\,\textrm{sin}\,i}{10^3\,\textrm{km s}^{-1}}\right)^{3}\;{\rm M}_{\sun}\; ,
\end{equation}
where we have defined $M\equiv M_1+M_2$ as the total mass of the SBHB, $q\equiv M_2/M_1\le 1$ as the mass ratio, and we note that $(1+q)^3>1$. If, instead, we were observing the radial velocity variations of the primary, we would replace $V_2$ by $V_1$ and $(1+q)^3\ge 1$ with $(1+q)^3/q^3\ge 8$ in Equation~(\ref{eqn:Mmin}). Thus, any limits on the total mass would be multiplied by 8. The limits on the periods and total masses, derived using Equation~(\ref{eqn:Mmin}), for objects with reliable radial velocity curves are given in Table~\ref{tab:Pmin}. The physical models corresponding to the 68th, 90th, and 99th percentile periods are shown in Figure~\ref{fig:zbrv}.

\begin{table*}
\begin{minipage}{0.7\linewidth}
\renewcommand{\thefootnote}{\alph{footnote}}
\caption[]{Lower limits on binary orbital parameters \footnotemark[1] \label{tab:Pmin}}
\begin{tabular}{lccccc}
            	& Minimum 			&                                                    & Minimum                       & Minimum & Minimum \\
 Object 	& Period\footnotemark[2]   &  $V_2\,\sin\, i$\footnotemark[3] & $t_{0}$\footnotemark[3] & Total Mass\footnotemark[4] & Orbital Separation\footnotemark[4] \\
 (SDSS J) & (years)  				& (km s$^{-1}$)  			   & (year)      			  & (\Msun)			        & ($10^{-2}$ pc) \\
\hline
001224	& \phantom{1}80	& 2000 			& 2033 	&   $2.5\times10^7$ & \phantom{1}2.6 \\
015530	& \phantom{1}50  	& 1400 			& 2050 	&   $5.3\times10^6$ & \phantom{1}1.1 \\
031715	& \phantom{1}70  	& 1500 			& 2073 	&   $9.5\times10^6$ & \phantom{1}1.7 \\
074007	& \phantom{1}90	&  \phantom{1}700  	& 2084 	&   $1.1\times10^6$ & \phantom{1}1.0 \\
082150	& \phantom{1}50  	& 1300 			& 2061 	&   $4.7\times10^6$ & \phantom{1}1.1 \\
091928	& \phantom{1}20  	& 4300 			& 2015 	&   $5.9\times10^7$ & \phantom{1}1.3 \\
092712	& \phantom{1}30  	& 2600 			& 2022 	&   $2.2\times10^7$ & \phantom{1}1.3 \\
093844	& 110			& \phantom{1}910	& 2044 	&   $3.2\times10^6$ & \phantom{1}1.6 \\
094603	& \phantom{1}50  	& 2000 			& 2050 	&   $1.5\times10^7$ & \phantom{1}1.5 \\
094620	& \phantom{1}50	& 1800 			& 2026 	&   $1.1\times10^7$ & \phantom{1}1.5 \\
095036	& \phantom{1}20  	& 1700 			& 2012 	&   $3.7\times10^6$ & \phantom{1}0.6 \\
095539	& \phantom{1}90	&  \phantom{1}640  	& 2079 	&   $9.2\times10^5$ & \phantom{1}0.9 \\
105041	& 140			& 3600 			& 2052 	&   $2.5\times10^8$ & \phantom{1}8.1 \\
110556	& \phantom{1}70  	& 1200 			& 2028 	&   $4.2\times10^6$ & \phantom{1}1.3 \\
111537	& 100			& 2200 			& 2044 	&   $3.7\times10^7$ & \phantom{1}3.3 \\
113330	& \phantom{1}90	& 1200 			& 2052 	&   $6.0\times10^6$ & \phantom{1}1.7 \\
113904	& 130			& 1500 			& 2156 	&   $1.8\times10^7$ & \phantom{1}3.2 \\
115158	& \phantom{1}20  	&  \phantom{1}390 	& 2025 	&   $4.7\times10^4$ & \phantom{1}0.1 \\
120924	& \phantom{1}50  	& 1800 			& 2024 	&   $1.1\times10^7$ & \phantom{1}1.4 \\
125142	& 110			& 1700 			& 2043 	&   $1.9\times10^7$ & \phantom{1}2.9 \\
131945	& \phantom{1}40  	& 1000		   	& 2019 	&   $1.5\times10^6$ & \phantom{1}0.6 \\
140251	& 130			& \phantom{1}980   	& 2109 	&   $4.8\times10^6$ & \phantom{1}2.1 \\
151132	& \phantom{1}90	& 1100			& 2100 	&   $5.4\times10^6$ & \phantom{1}1.7 \\
153636	& 240			& 3500	 		& 2092 	&   $3.8\times10^8$ & 13.4 \\
153644	& \phantom{1}50  	& 1300 			& 2067 	&   $4.9\times10^6$ & \phantom{1}1.2 \\ 
154637	& \phantom{1}30  	& 2600 			& 2019 	&   $1.8\times10^7$ & \phantom{1}1.1 \\ 
155654	& 100			& 3900 			& 2037 	&   $1.9\times10^8$ & \phantom{1}5.7 \\ 
161911	& \phantom{1}20  	& \phantom{1}510   	& 2016 	&   $1.2\times10^5$ & \phantom{1}0.2 \\ 
162914	& \phantom{1}40  	& \phantom{1}930   	& 2051 	&   $1.4\times10^6$ & \phantom{1}0.7 \\ 
\hline
\end{tabular}
\footnotetext[1]{Orbital parameters are determined assuming that the observed radial velocities refer to the secondary BH, which is in a circular orbit.  See Equation~(\ref{eqn:Mmin}) in Section~\ref{sec:interpretation} and associated discussion.}
\footnotetext[2]{Expressed in the rest frame.}
\footnotetext[3]{The time of the next inferior conjunction of the secondary corresponding to the minimum period.}
\footnotetext[4]{Computed using the the values of the rest-frame minimum period and $V_2\,\sin\, i$ reported in this table.}
\end{minipage}
\end{table*}

With the data available so far, none of the limits on the total mass of the SBHB are sufficiently stringent to rule out the binary hypothesis. According to Equation~(\ref{eqn:Mmin}), for velocity offsets of order a few$\times10^{3}$~km~s$^{-1}$ we would need a very high minimum period, $\sim 10^{4}$~yrs, to obtain a limiting mass $\sim 10^{10}$~\Msun. With only 3--4 observations, it is rare that the radial velocity curve is actually flat over the duration of our campaign. In other words, velocity jitter and finite measurement uncertainties can make a flat radial velocity curve appear curved if only a few measurements are available. Specifically, in 10/88 cases, we measured statistically significant radial velocity variations, which introduce curvature into the radial velocity curves and lead to less restrictive constraints on the periods. In 19/88 cases we found that all of our reliable radial velocity measurements were consistent with the original SDSS measurement at the 99 per cent confidence level, however the sampling pattern and measurement errors allow for substantial curvature in the orbital models. For example J153636, which has a stable broad line profile observed over six years but substantial jitter, yields the largest mass limit at $M>3.8\times10^{8}$~\Msun. In the case of J105041, the 2012 measurement with its large uncertainty does not constrain the model as much as the 2010--2011 measurements, with the result that shorter periods are possible. From these results, we conclude that five or more radial velocity measurements with small measurement uncertainties are needed to get better period constraints than what we have gotten so far.

\section{Discussion}
\label{sec:discussion}

Our main observational result is the measurement of radial velocities of the SBHB candidates in our sample at several epochs. We are able to produce reliable radial velocity curves for 29/88 candidates. As we noted in \S\ref{sec:caveats}, variability of the line profiles is the most serious limitation in our ability to put together long time series of radial velocity measurements. In approximately half of the objects in our sample we have observed substantial changes in the profiles of the broad H$\beta$ lines that make the measurements of radial velocity variations impossible. In objects where the broad H$\beta$ profiles appear stable, there is still a danger that profile shape variability causes radial velocity jitter beyond the measurement errors and complicates the interpretation of the radial velocity curves. The potential effects of profile shape variability were pointed out and discussed by \citet{decarli13}, \citet{shen13a}, and \citet{liu14} who also noted that broad lines with larger velocity offsets are more likely to display profile shape variability. Because the dynamical time of the broad-line region of any one of the two BHs is shorter than the orbital period of the SBHB, we cannot expect that the profiles of the broad lines will remain stable over the course of an orbital cycle. This means that we will typically be able to measure only segments of the radial velocity curve during which the line profiles are stable. In order to put together long-term time series, we will need to devise methods of combining such short segments together and overcoming the problem caused by the changing velocity zero point that is brought about by profile variations.

The above cautions notwithstanding, we would like to draw attention to three objects in our sample, J093844, J095036, and J161911, which have shown systematic and monotonic velocity changes by several hundred kms~s$^{-1}$. These represent the most promising cases in our sample and deserve further attention. They are high-priority targets for continued spectroscopic monitoring as well as additional observational tests, such as high-resolution imaging of the host galaxies in search of morphological features associated with a merger.

In addition to making new measurements, we have also developed the methodology for setting a lower limit on the period, hence the mass, of a hypothesized SBHB using a radial velocity curve of finite duration. The method uses an MCMC algorithm to explore the space of parameters of circular orbits and allows one to determine the shortest orbital period consistent with the observed radial velocities at the 99 per cent confidence level. Applying this method to the available data, we find that we are not yet able to put interesting constraints on the SBHB candidates. The limits we give in Table~\ref{tab:Pmin} reflect the properties of our radial velocity curves (duration, uncertainties, jitter, etc.) rather than physically interesting constraints on the properties of the hypothesized binaries. With more measurements over a longer baseline, we can obtain more stringent constraints and also attempt to constrain models of elliptical orbits, which is not possible with the data available so far. Another noteworthy point is that the value of $V_2\sin i$, derived from the model fits, influences the mass constraints considerably because it enters in the third power in Equation~(\ref{eqn:Mmin}). For objects with small observed radial velocities, $V_2\sin i$ is likely to be small, therefore a combination of a longer temporal baseline and smaller error bars is needed in order to obtain interesting constraints on the mass of the hypothesized SBHB.

\begin{figure*}
\centerline{
  \includegraphics[width=8.9cm]{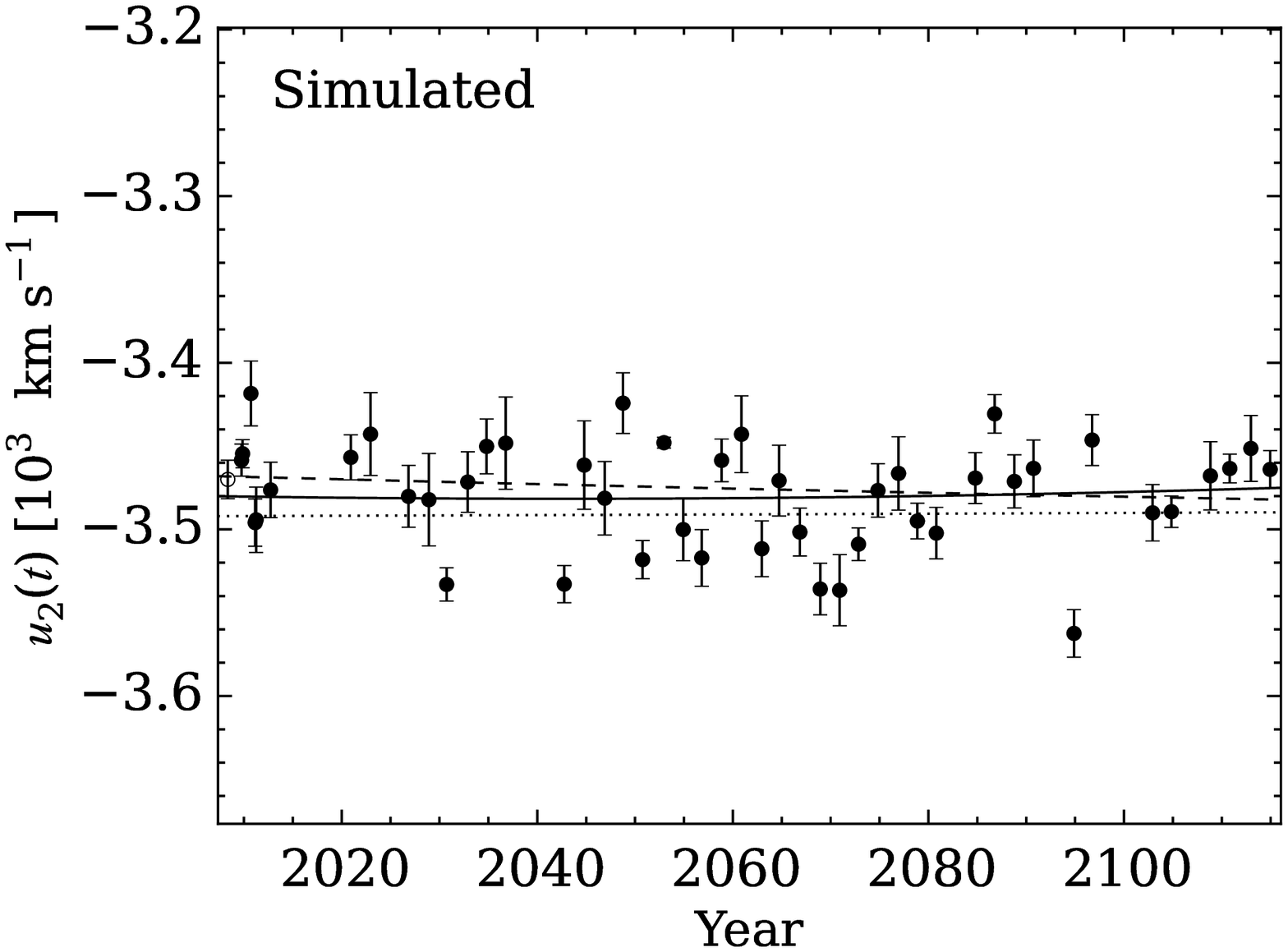}
  \includegraphics[width=8.9cm]{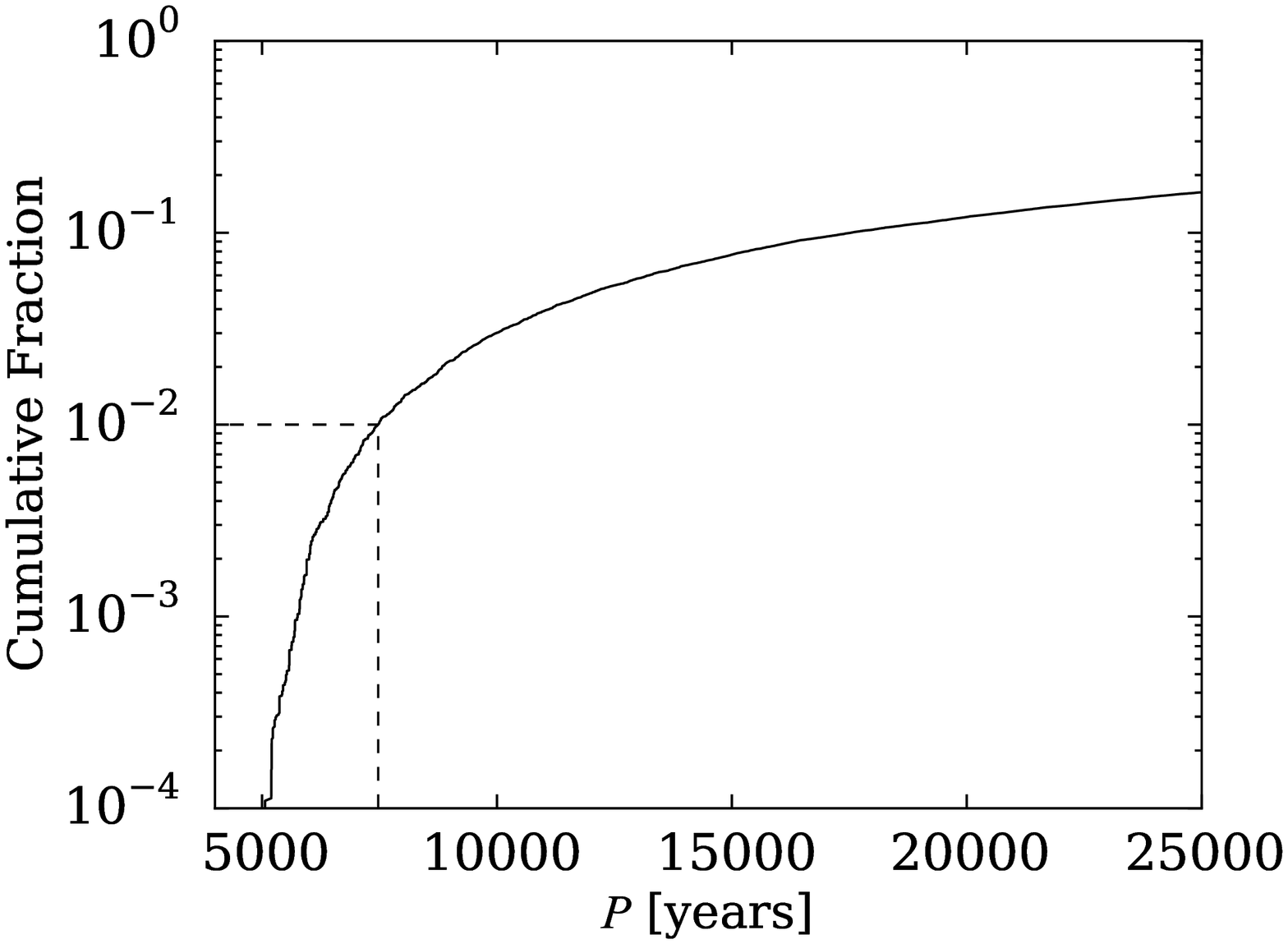} 
}
  \caption{Results from the MCMC simulation for a synthetic radial velocity curve of J153636. Starting with the observed radial velocity curve for this object, new radial velocity measurements are generated every two years (with breaks to emulate bad weather) assuming no evolution in the offset and the noise properties of the observed measurements. The left panel shows the simulated radial velocity curve with 68 per cent error bars. Dotted, dashed, and solid lines show the sinusoids corresponding to the 68th, 90th, and 99th percentile minimum periods, respectively. The right panel shows the corresponding cumulative observed-frame $P$ distributions from the combined high- and low-resolution MCMC simulations (see Appendix~\ref{app:mcmc}). The dashed line indicates the 99th percentile period. The minimum period yields a lower limit on the mass of $M > 9\times10^{9}$~\Msun. Despite the long time baseline, jitter in the radial velocity measurements allows curvature in the sinusoids.} \label{fig:fakemcmc}
\end{figure*}

If the orbital periods are of order centuries, we cannot hope to confirm in our lifetimes that any specific objects in our sample are SBHBs based on radial velocity variations. However, we will be able to place more stringent constraints on the properties of the hypothesized binary orbits with continued monitoring. In the spirit of \citet{decarli13} we ask how much longer we need to monitor the SBHB candidates before we obtain interesting constraints on their orbital parameters. With simulations, we explore how long it will take to obtain a restrictive mass constraint in the event that the observed radial velocity curve is constant, within the uncertainties and allowing for jitter. We use the observed radial velocity curve of J153636 as the starting point for this experiment, because it has a relatively large velocity amplitude and has remained flat over our monitoring period. We simulate 50 additional observations using the measurement uncertainties and jitter properties determined from the available observations. The simulated observations are made every two years with several gaps to simulate the effects of bad weather. The simulated data for a century of monitoring are shown in Figure~\ref{fig:fakemcmc}. After analyzing these data with the procedure of \S\ref{sec:interpretation}, we find that a monitoring period of of $\sim 100$ years yields an interesting mass limit of $M>9\times10^9$~\Msun (see Figure~\ref{fig:fakemcmc}), provided that the time series is not sparsely sampled, i.e., as long as we have an observation every couple of years. For reference, we note that the largest BH masses measured so far \citep[in brightest cluster galaxies, e.g.,][and references therein]{mcconnell12,thomas16} are approximately $2\times 10^{10}\;{\rm M}_{\sun}$ while arguments have been made that BH with masses sevaral times higher than this value could exist \citep[see][and references therein]{inayoshi16}. The velocity jitter, which is evident in the observed radial velocity curve for J153636 and which we reproduce in our simulations, significantly hinders our ability to get an interesting mass limit by allowing orbital models with short orbital periods to describe the data adequately. Carrying out an analogous simulation with measurement errors only and no jitter yields a more restrictive mass constraint of $M>1\times10^{10}$~\Msun. It may be possible to obtain such interesting constraints within a few decades by introducing a model for the jitter in the MCMC simulation. This is beyond the scope of this work and, notably, would require several more observations of radial velocity changes before a reliable model for the jitter can be derived. Thus, patience and continued monitoring are required for the radial velocity test to achieve its maximum discriminating power.

With additional monitoring we can possibly carry out additional tests of the SBHB hypothesis by looking for radial velocity variations that are inconsistent with orbital motion. The double-peaked emitter 3C~390.3 is a good case in point: the double-line spectroscopic binary scenario was tested and rejected for this object after three decades of monitoring \citep[][and references therein]{liu16}. The velocity offsets of the peaks of the broad Balmer lines abruptly departed from their long-term monotonic trend (positive acceleration) and remained steady for more than a decade before the acceleration changed sign without the velocity passing through zero (e.g., see Fig.~5 of \citealt{eracleous97} and Fig.~10 of \citealt{shapovalova01}). This behavior is inconsistent with orbital motion. Additional monitoring can also constrain the properties of the BLR, if we assume that our candidates are true SBHBs. As we noted in \ptwo\ and illustrated in Figure~14 therein, constraints on the orbital period of an SBHB translate into constraints on the extent of the BLR around the secondary (active) BH. Thus, if further monitoring increases the lower limit on the orbital periods to values of 500~years or greater, the scenario of a tidally truncated BLR will be favoured over the scenario of a BLR that fills the Roche lobe of the secondary BH.

The shifted broad lines that were used to select the SBHB candidates may also be a signpost of recoiling BHs. As discussed in detail in \pone, this scenario seems unlikely to produce the observed properties of the objects in our sample. Theoretical models of recoiling BHs suggest that velocity offsets in excess of 400~km~s$^{-1}$ are rare \citep[e.g.,][]{dotti10}. Additionally, the illumination of the narrow-line region and the obscuring torus from outside the host galaxy should leave its imprint on the relative strengths and widths of the line profiles and the shape of the spectral energy distribution in the infrared \citep{bonning07,shields09a,shields09b,lusso14}, which is not observed in these objects. In addition to these tests, which were carried out after the initial selection of the sample, recoiling BHs are expected to exhibit stable radial velocity curves. The 19 objects with reliable, non-variable radial velocity curves are therefore the best candidates for this interpretation, notwithstanding the  disagreement of the high velocity offsets, narrow-line ratios, and IR spectral energy distributions with expectations. Of these, J092712 is particularly noteworthy because recent observations by \citet{decarli14} find molecular gas in the host galaxy at the same redshift as the broad emission lines from the quasar. This result strongly disfavors the recoiling BH interpretation for this object. The 10 objects with variable radial velocity curves are unlikely to represent recoiling BHs unless the radial velocity variations can be associated with variability of a single BLR. Based on our spectroscopic monitoring, of these 10 objects there are three (J093844, J095036, J161911) in which the radial velocity curves show monotonic variations by many hundreds of kms~s$^{-1}$, incompatible with a single BLR bound to a recoiling BH. It is difficult to evaluate the remaining 7/10 of the objects with reliable variable radial velocity curves because their radial velocities are erratic and consistent with jitter.

\section{Summary and Conclusions}
\label{sec:summary}

In \pone\ we identified a sample of 88 SBHB candidates selected based on their broad \Hb\ emission lines, which are offset by thousands of kilometers per second relative to the frame of the narrow lines. As part of a long-term effort we have been monitoring the candidates spectroscopically since 2009, with the latest installment of data from our campaign, typically 3--4 spectra per object spanning an interval of 12 years in the observer's frame, reported in \ptwo. The main observational result of this paper, the third of the series, is the measurement of radial velocity variations of the broad \Hb\ lines from these spectra and the presentation of the radial velocity curves of the SBHB candidates. Furthermore, we establish a methodology for evaluating the binary hypothesis by setting a lower limit on the period, and thus the mass, of the hypothetical SBHBs. Our results are:

\begin{enumerate}
\item We identify 29/88 of SBHB candidates where, based on visual inspection, no profile shape variability has occurred and we can make reliable measurements of the radial velocities. This number includes cases where we detect statistically significant radial velocity changes as well as those where the radial velocities are reliably steady. For these objects, we construct and present radial velocity curves.

\item We identify 3 objects (J093844, J095036, and J161911) in the sample that show systematic and monotonic velocity changes consistent with the binary hypothesis. These sources are high-priority targets for further spectroscopic monitoring and additional observational tests.

\item We observe substantial profile shape variability in at least one spectrum in 39/88 of the SBHB candidates. This highlights the main caveat associated with searching for SBHBs with our method: variability of the broad-line profile shape can mimic radial velocity variations. Through simulations of profile shape variability, we have demonstrated that subtle profile shape variability may elude visual inspection and masquerade as a radial velocity change. Broad lines with wide, boxy profiles are more susceptible to this effect than cuspy profiles, where it is easier to discern that the profile shape has changed. As a result, it is critical to characterize the long-term variability properties of the broad emission line profiles of regular quasars, and to continue to observe the candidates on time-scales where orbital motion can be distinguished. 

\item Under the assumption that we can interpret the radial velocity variations of the broad \Hb\ emission line as orbital motion, we develop a methodology for finding the minimum period that is consistent with the observed radial velocity curves at the 99 per cent level. This allows us to calculate the minimum total mass of the SBHB. A sufficiently high mass limit can lead to a rejection of the SBHB hypothesis. The short duration of our campaign so far leads to mass limits between a few$\times 10^6$ and a few$\times 10^8\;{\rm M}_\odot$, which are not restrictive enough. 
\end{enumerate}  

Continued spectroscopic monitoring will enable substantial gains in several ways. The addition of points to the radial velocity curves will allow us to test more complex physical models, such as eccentric orbits. By characterizing the jitter and noise properties in the radial velocity curves we may be able to uncover the underlying trends and obtain more restrictive mass limits within a few decades of observing. With extended monitoring we also open the door to observing radial velocity variations that are inconsistent with orbital motion. Given that the periods of the SBHBs are expected to be long, abrupt departures from established monotonic changes, as has been observed in double-peaked emitters, are not expected. We therefore conclude that continued spectroscopic monitoring is required to further evaluate the credentials of the SBHB candidates.

\section{Acknowledgements}
We thank Anh N. D. Doan for his help in testing our local implementation of the {\tt emcee} code. JCR would like to thank Becky Nevin and Alex Hagen for helpful discussions during the preparation of this work. The authors also wish to thank the referee, Roberto Decarli, for a careful reading of the manuscript and suggestions that improved the paper.

This work was supported by grant AST-1211756 from the National Science Foundation and an associated REU supplement. ME thanks the members of the Center for Relativistic Astrophysics at Georgia Tech and the Department of Astronomy at the University of Washington, where he was based during some phases of this work, for their warm hospitality. SS thanks the Aspen Center for Physics for hospitality and the ACP is supported by NSF grant PHY-1066293. TB acknowledges support from the National Aeronautics and Space Administration under Grant No. NNX15AK84G issued through the Astrophysics Theory Program and support from the Research Corporation for Science Advancement through a Cottrell Scholar Award. We thank the staff at Kitt Peak National Observatory, Apache Point Observatory, and the Hobby-Eberly Telescope for their expert help in carrying out the observations. 

This work is based on observations obtained with the Apache Point Observatory 3.5-meter telescope, which is owned and operated by the Astrophysical Research Consortium. 

This work is based on observations at Kitt Peak National Observatory, National Optical Astronomy Observatory (NOAO Prop. ID: 2014A-0098; PI: Runnoe), which is operated by the Association of Universities for Research in Astronomy (AURA) under cooperative agreement with the National Science Foundation.

The Hobby-Eberly Telescope (HET) is a joint project of the University of Texas at Austin, the Pennsylvania State University, Stanford University, Ludwig-`Maximillians-Universit\"at M\"unchen, and Georg-August-Universit\"at G\"ottingen. The HET is named in honor of its principal benefactors, William P. Hobby and Robert E. Eberly.

The Marcario Low-Resolution Spectrograph is named for Mike Marcario of High Lonesome Optics, who fabricated several optics for the instrument but died before its completion; it is a joint project of the Hobby-Eberly Telescope partnership and the Instituto de Astronom\'{\i}a de la Universidad Nacional Aut\'onoma de M\'exico.

This research has made use of the NASA/IPAC Extragalactic Database (NED) which is operated by the Jet Propulsion Laboratory, California Institute of Technology, under contract with the National Aeronautics and Space Administration.


\appendix
\section{Simulated profile shape variability}
\label{app:prosim}
Variability of the profile shape is an important impediment in our ability to measure velocity shifts of the broad \Hb\ emission line. In making the radial velocity measurements, profile shape variability was diagnosed by eye and with the help of the $\chi^2$ test so that shifts were only measured in cases where the profile was stable. Here, we investigate whether changes in the profile shape can cause us to measure radial velocity shifts and whether we can identify these cases based on visual inspection. That is, we ask if we can actually distinguish between true velocity shifts in the broad-line profile and subtle changes in the profile shape that may occur over the course of several years. In order to address this question, we simulate both profile shape variability and real shifts in the observed spectra of the SBHB candidates and consider the results of our shift measurements.

We specifically simulate different types of ``see saw'' variability of the line profiles (i.e., the shapes) because this type of variability is often observed (see discussion in \S\ref{sec:caveats}) and because it has the potential of mimicking the effects of a velocity shift. We take advantage of the decomposition of the spectra into continuum and broad and narrow lines, carried out in \ptwo, in order to isolate the profile of the broad H$\beta$ line. Thus, we extract the smooth parametrization of the broad H$\beta$ profile, perturb it in a variety of ways, and then put it back in the spectrum.

\begin{figure}
\begin{center}
\includegraphics[width=8.5 truecm]{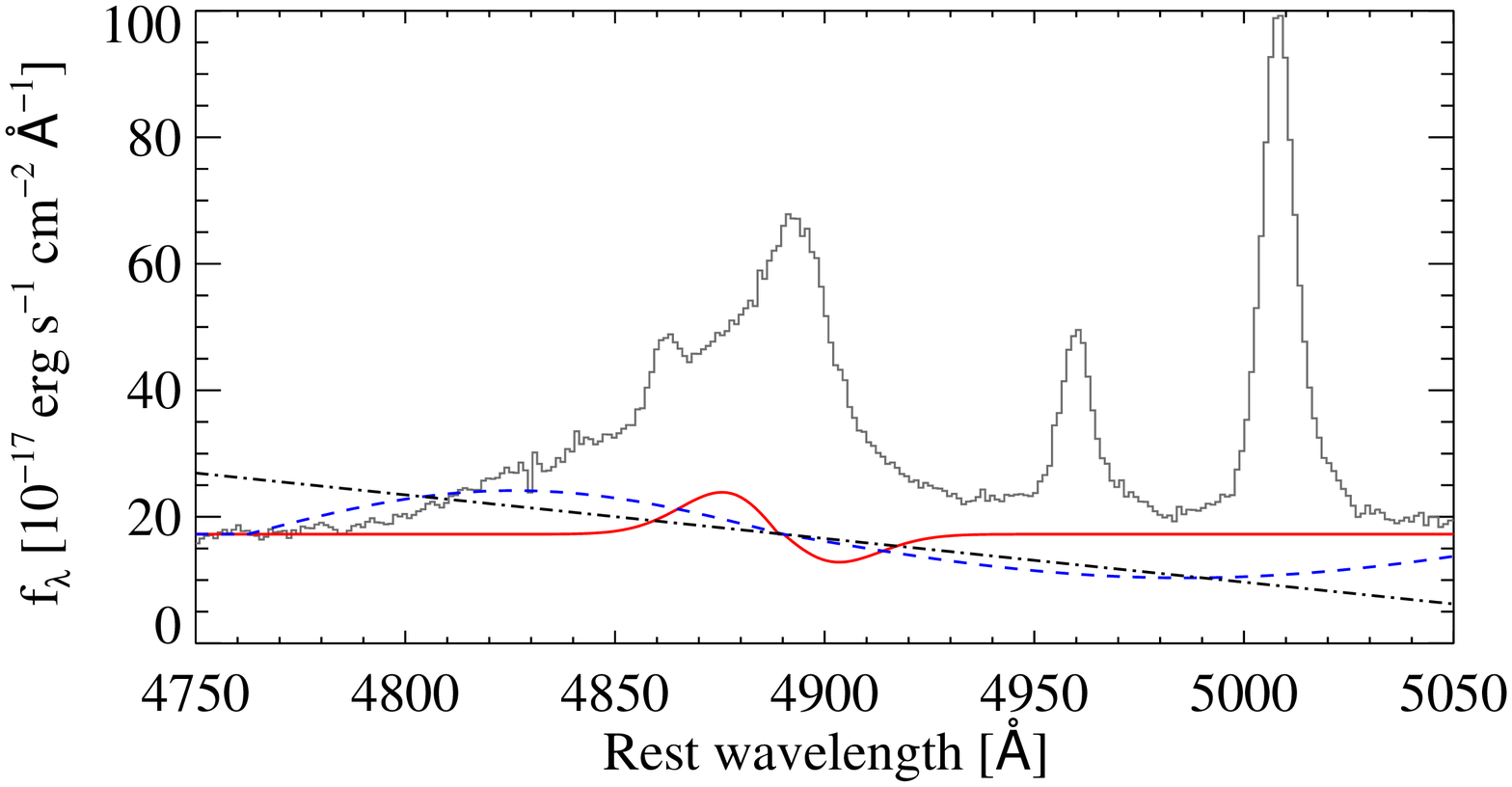}
\end{center}
\caption{An illustration of the variability functions used to induce profile shape variability in the observed spectra of the SBHB candidates. The grey histogram shows the observed spectrum of one of the SBHB candidates. The tilted line (black dot-dashed), sinusoid (blue dashed), and modified Gaussian (red) are overplotted in the real wavelength domain and their parameters are adjusted according to the constraints discussed in \S\ref{sec:caveats} and Appendix~\ref{app:prosim}.}
\label{fig:vfunc}
\end{figure}

The perturbations we use take the form of a suppression of one shoulder/wing of the broad H$\beta$ line and a boosting of the other. In practice, we multiply the parametric model for the broad H$\beta$ profile with three families of odd functions: a tilted line of the form $f(x)=1+ax$, a sinusoid of the form $f(x)=1+a\,\sin(b\,x)$, and a modified Gaussian of the form $f(x) = 1 + a\, x\,e^{-b\,x^2}$. In all cases $x=\lambda-\lambda_{peak}\propto \Delta v$ with $\lambda_{peak}$ the wavelength of the peak of the line profile and $\Delta v$ the velocity separation relative to the peak of the profile. The parameter $a$ controls the amplitude of the perturbations and the parameter $b$ controls how the perturbation evolves with velocity relative to the peak of the profile. These functions are illustrated in Figure~\ref{fig:vfunc}. In the case of the tilted line, the perturbation increases steadily with increasing $\Delta v$. In the case of the sinusoid, the parameter $b$ has different values on the blue and red sides of the peak so that the variability reaches zero where the line profile falls below 1 per cent of the peak value. In the case of the modified Gaussian, the parameter $b$ also has different values on the blue and red sides of the peak so that the maximal perturbation occurs at 40 per cent of the full-width at half-maximum (see references in \S\ref{sec:caveats}). We find that the modified Gaussian creates the most realistic profile shape variability, particularly for boxy line profiles. The tilted line and sinusoid induce unrealistically large variability in the far wings of broad line profiles.

\begin{figure*}
\centerline{
  \includegraphics[width=8.9cm]{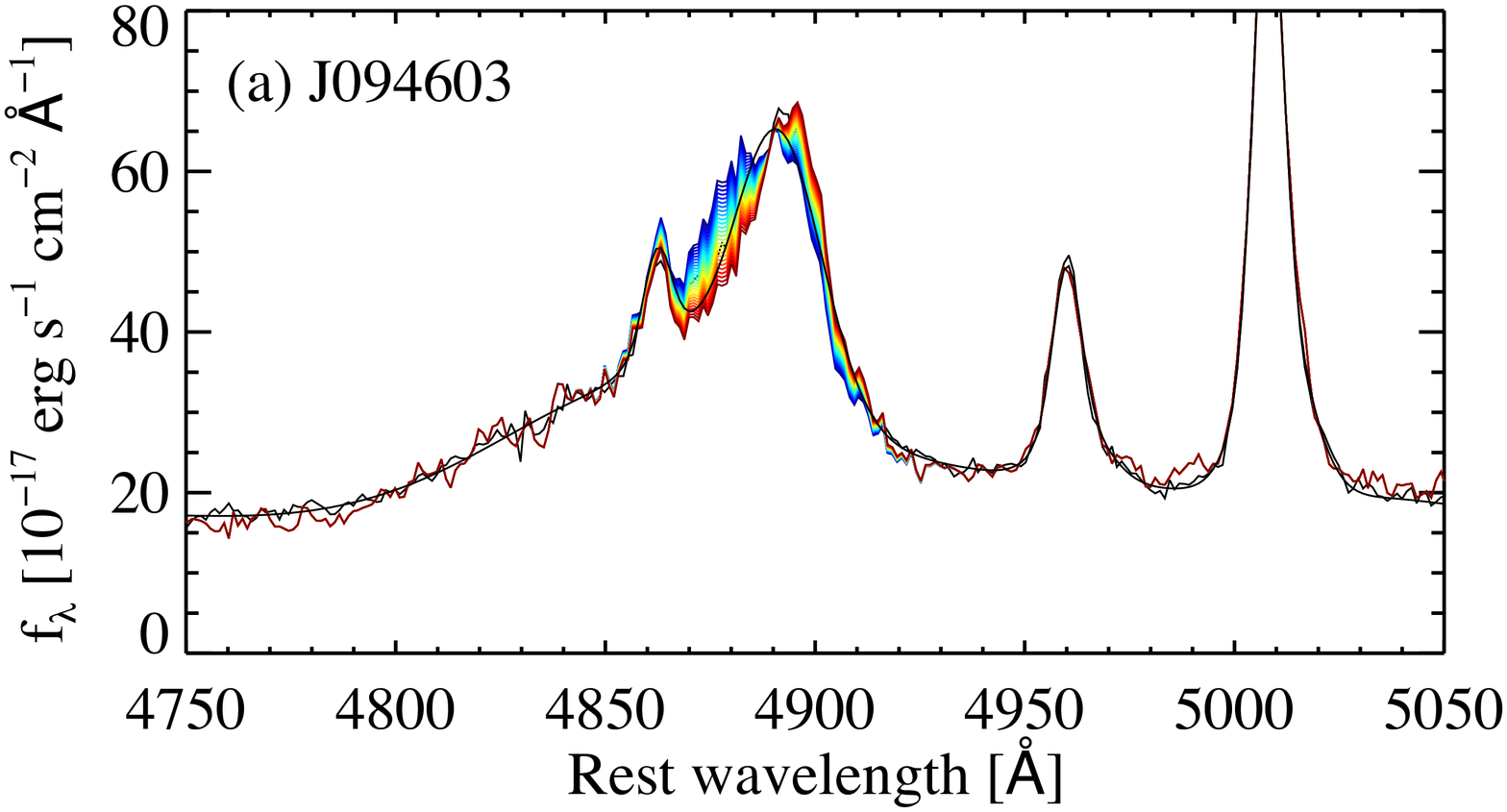}
  \includegraphics[width=8.9cm]{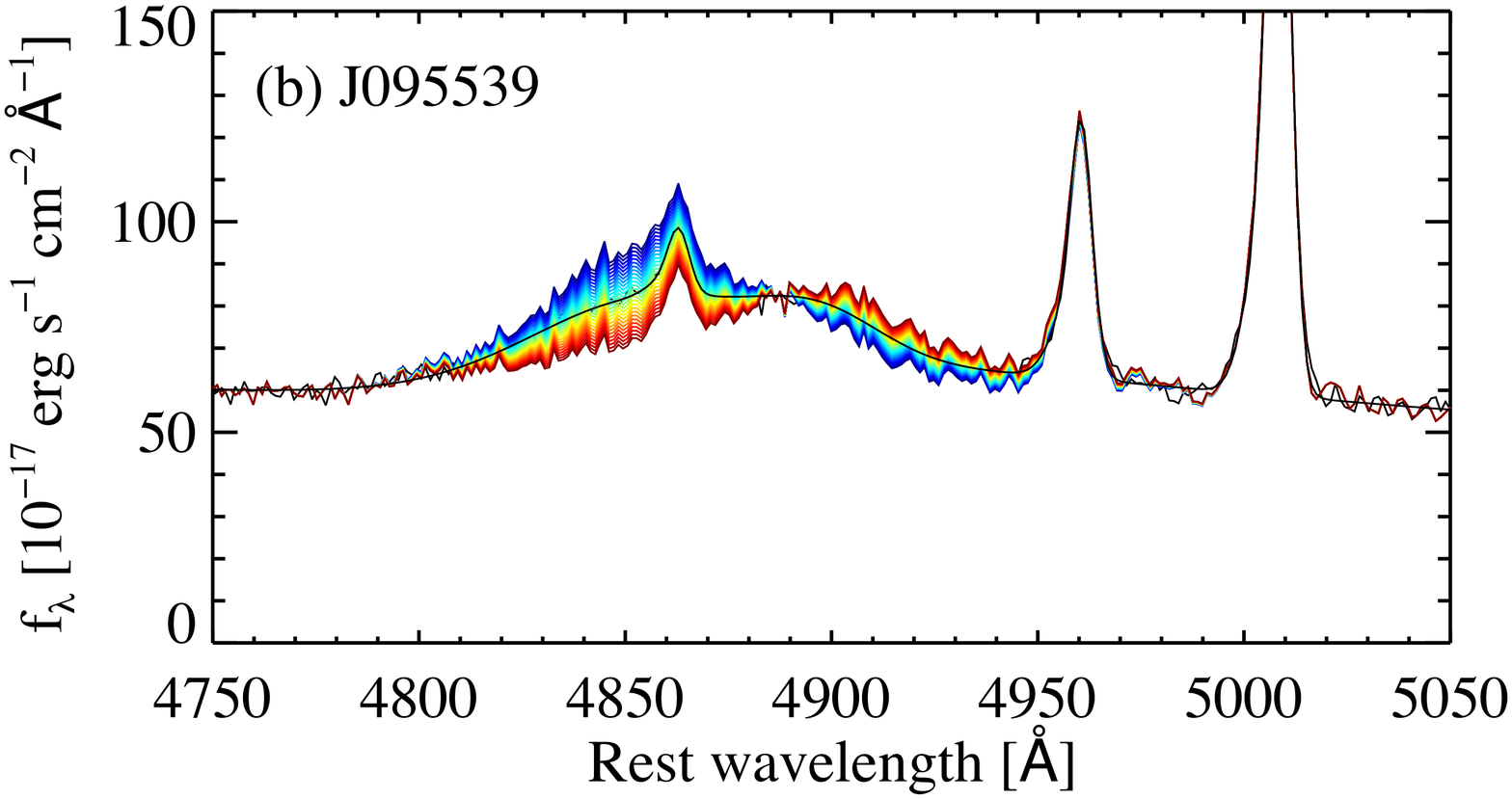} 
}
  \caption{The simulated profile shape variability of (a) J094603, an object with a cuspy \Hb\ profile and (b) J095539, an object with a boxy \Hb\ profile. The black line is the best-fiting model from the decomposition of the observed spectrum. Each colour traces a synthetic spectrum where the parametric model of the broad \Hb\ line has been perturbed by multiplying it with the function $f(x) = 1+a\,x\,e^{-b\,x^2}$ with a different value of the parameter $a$.} \label{fig:varspec}
\end{figure*}

To explore the effect of the simulated perturbations we apply them to the SDSS spectrum of each SBHB candidate in our sample as described above. For each object and perturbation prescription we generate 50 synthetic spectra with a range in  the perturbation amplitude. Two examples of synthetic spectra, generated using the modified Gaussian perturbation, are shown in Figure~\ref{fig:varspec}, one for a ``cuspy'' and one for a ``boxy'' line profile. In a similar fashion, we also generated spectra where the broad \Hb\ line was actually shifted in order to carry out a visual comparison with the spectra where we induced profile shape variability.

\begin{figure*}
  \centerline{
    \includegraphics[width=8cm]{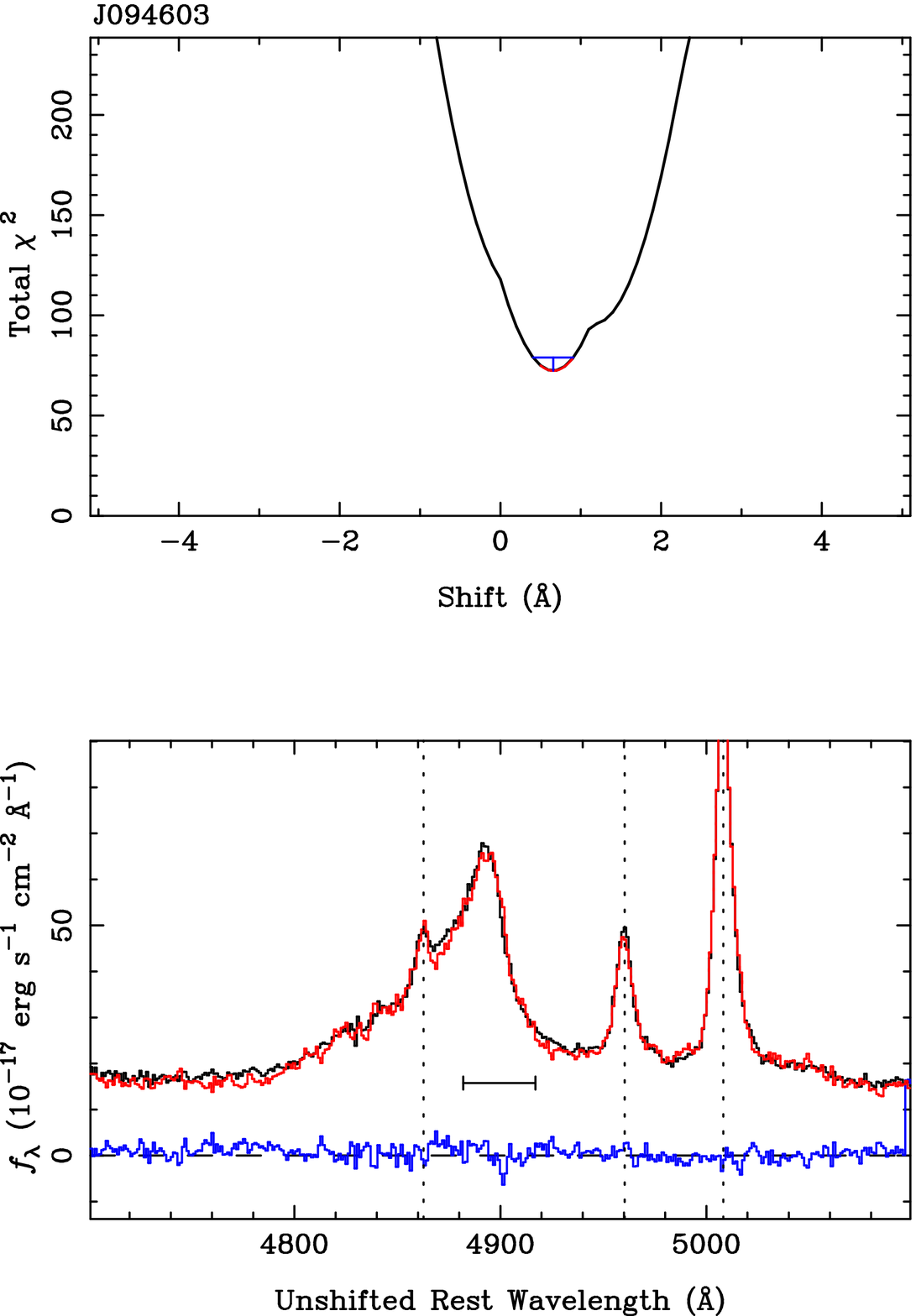}
    \hskip 1cm
    \includegraphics[width=8cm]{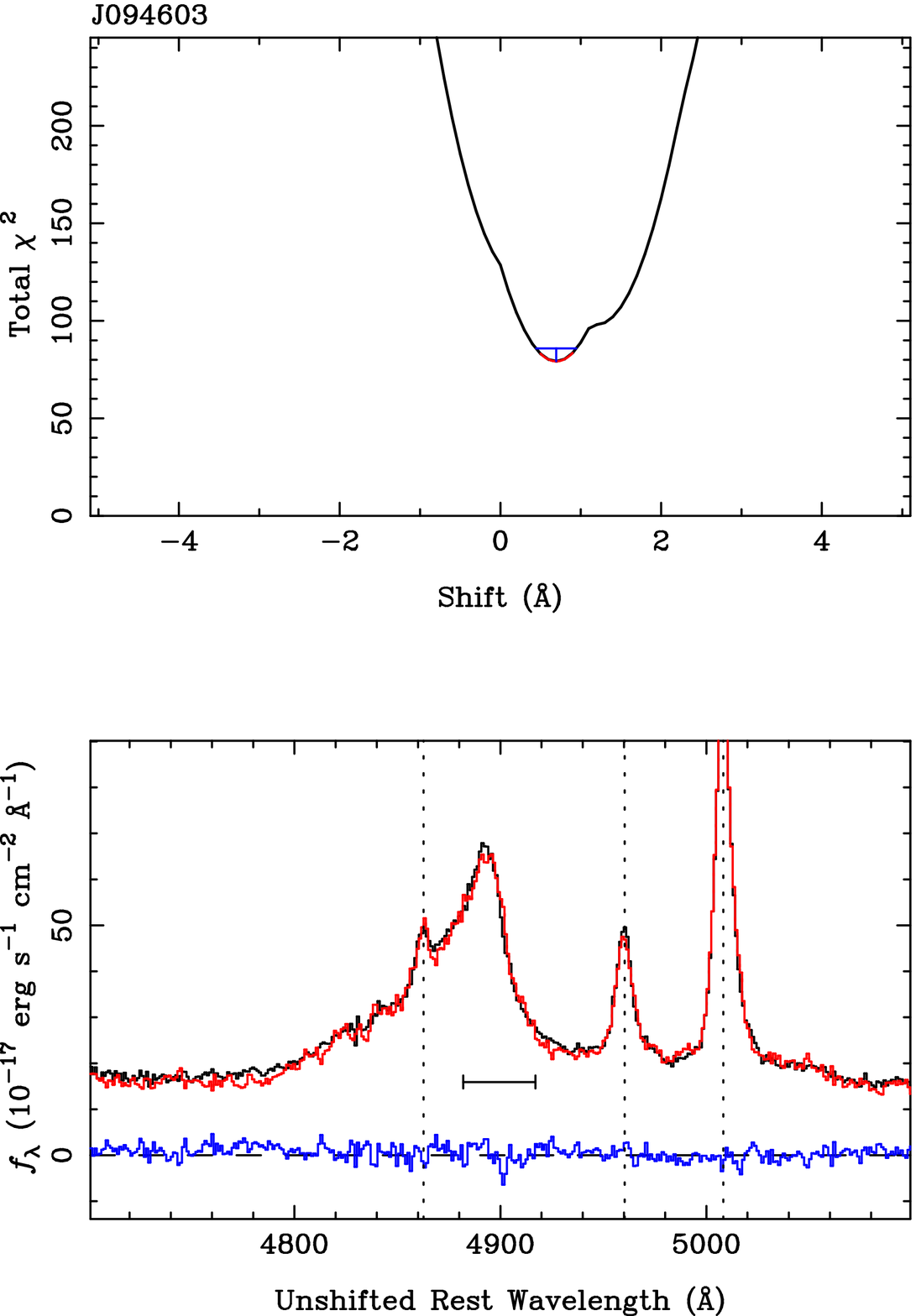}
  }
\caption{A comparison of the shift measurement for a cuspy profile whose shape has been perturbed (left) and a cuspy profile that has actually been shifted (right). As in Figure~\ref{fig:xc}, the upper panels show the $\chi^{2}$  curve as a function of shift of the synthetic (red) spectrum. The blue horizontal bars mark the 99 per cent  (or $2.6\sigma$) confidence intervals. The lower panels show the observed profile (black) and synthetic profile (red), scaled and superposed, and the blue line is their difference. The cases where varied the profile shape (left) and actually shifted the broad profile (right) are visually indistinguishable.} \label{fig:xcpvarc}
\end{figure*}

\begin{figure*}
  \centerline{
    \includegraphics[width=8cm]{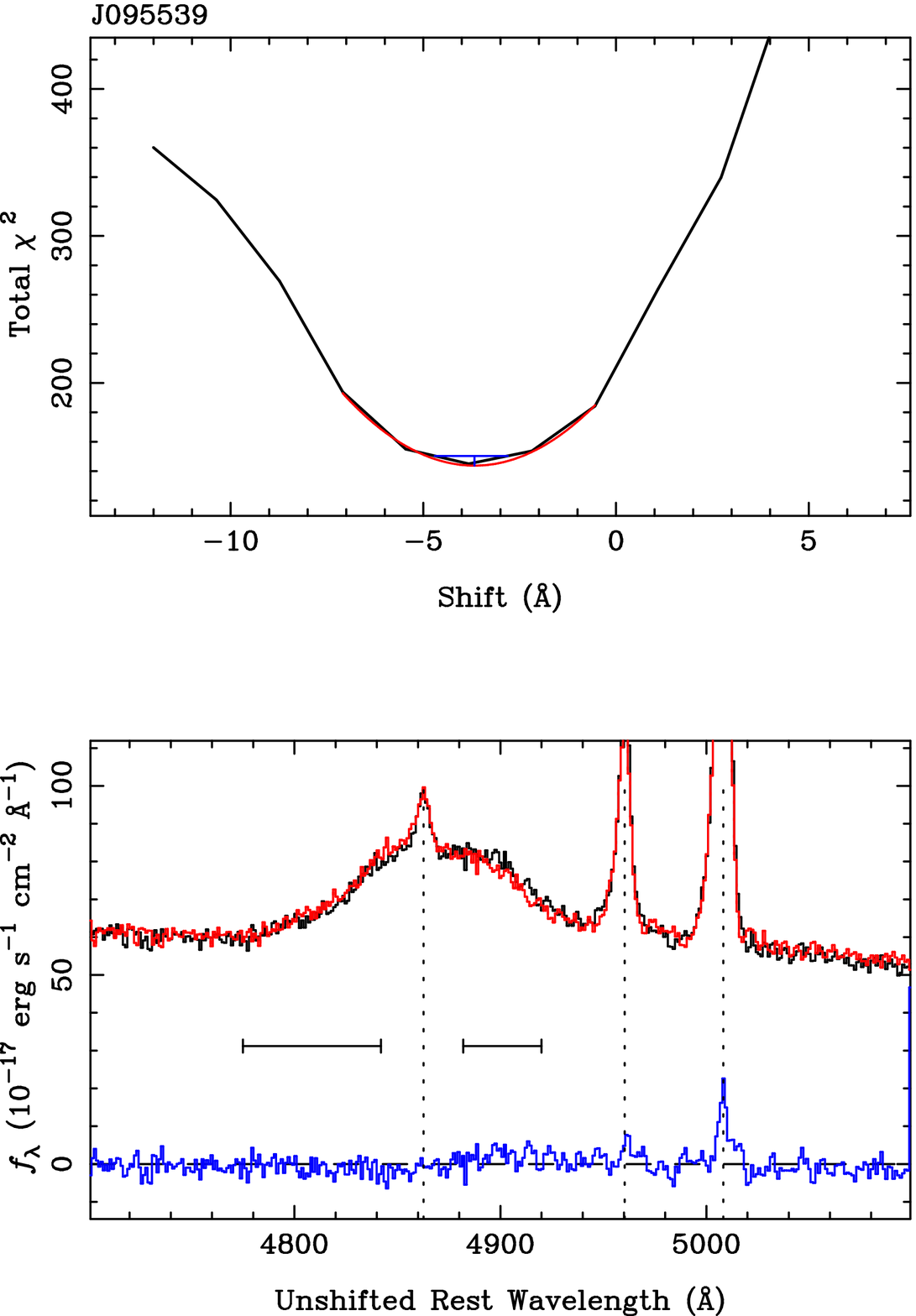}
    \hskip 1cm
    \includegraphics[width=8cm]{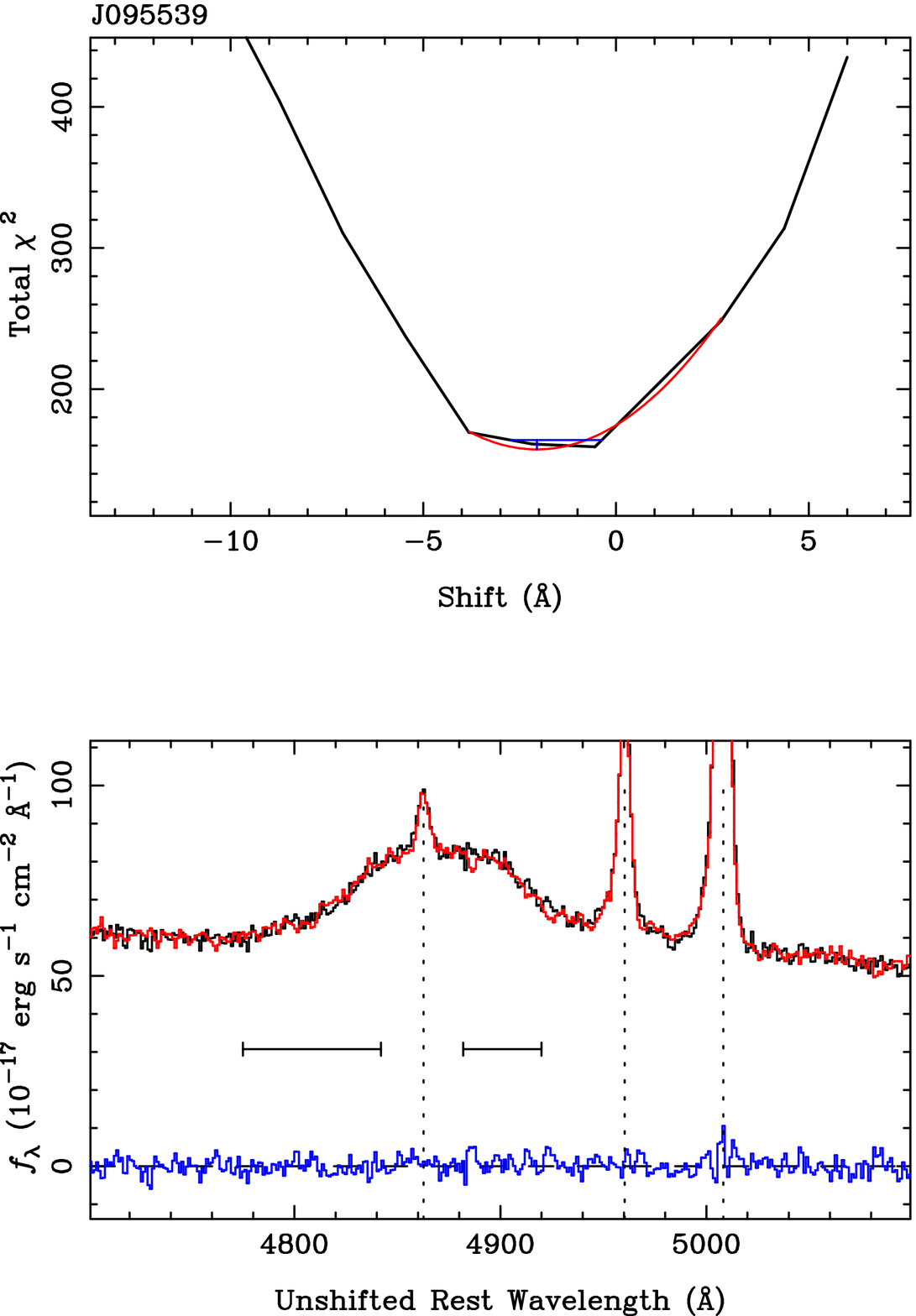}
    }
  \vskip 1cm
\caption{Same as Figure~\ref{fig:xcpvarc} but for a boxy line profile.} \label{fig:xcpvarb}
\end{figure*}

We measure the shifts of the broad \Hb\ lines in the synthetic spectra relative to the observed spectrum using the methodology of \S\ref{sec:xc}. Examples of the results of this exercise are shown in Figures~\ref{fig:xcpvarc} and \ref{fig:xcpvarb} for the same two objects shown Figure~\ref{fig:varspec}. We have chosen these examples to illustrate that shifted and perturbed profiles produce an effectively indistinguishable signal. Visual inspection in such cases does not help us distinguish between shifts and profile variations because the magnitude of the perturbations is small. Such cases are common in the tests that we have carried out; we encounter them in 24 per cent of the objects in our sample. Moreover, we find that genuine shifts are more difficult to distinguish from profile shape variability if the broad line profile is ``boxy'' e.g., J095539 in Figure~\ref{fig:varspec}b and Figure~\ref{fig:xcpvarb}. Of course, our ability to distinguish profile shape variability from real shifts using visual inspection and a quantitative comparison of the line profiles via the $\chi^2$ test depends critically not only on the shape of the profile but also on the type of perturbation and the S/N of the spectra. Therefore, we cannot make a reliable estimate of the likelihood that any of the shifts that we measure may be false. Nor can we formulate specific criteria for identifying cases that are particularly susceptible to this effect. None the less, the simulations we have carried out do indicate that there is cause for concern. The best way to guard against false positive signals of this type is continued spectroscopic monitoring. By extending the time series of spectra we can see if perturbations in the line profiles grow. Also, the long-term radial velocity curves will show different patterns in cases of shifts and profile variations.

\section{Markov Chain Monte Carlo simulations of the radial velocity curves}
\label{app:mcmc}
In order to determine the minimum period consistent with the observed radial velocity curves at the 99\% confidence level, we adopt the physical model and measurements described in Section~\ref{sec:interpretation}, and run an MCMC simulation which we describe here in detail. We use the code {\tt emcee}\footnote{{\tt emcee} is written in Python and made publicly available via {\tt http://dan.iel.fm/emcee/current/}} \citep{foreman13}, which implements an affine invariant MCMC sampler. This code determines the likelihood of the data given the model in Equation~\ref{eqn:delsine} and produces posterior probability distributions of all model parameters. As inputs the code takes the data, prior distributions for all model parameters, and an initial distribution of ``walkers,'' i.e., starting points of trajectories for exploring the model parameter space. In practice, a ``burn-in phase'' is required before the simulation will yield meaningful results. That is, initial, throw-away iterations of the simulation are needed in order to transform the input parameter and walker distributions to a distribution of walkers in parameter space that reflects the desired posterior distributions. The simulation itself is then started with the walker distribution reached at the end of the burn-in phase and all other input distributions as supplied initially by the user.

This code also allows us to properly treat the uncertainty in the first radial velocity measurement, as follows. We allow $u_{2}(t_1)$ in Equation~(\ref{eqn:delsine}) to be a free parameter but we constrain it to follow a Gaussian distribution with a mean equal to the measured value of $u_{2}^{\rm obs}(t_1)$ and a standard deviation equal to the `$1\sigma$' uncertainty on the measurement. For the first measurement in the relative radial velocity curve, we adopt $\Delta u_2^{\rm obs}(t_1)=0\;{\rm km\; s}^{-1}$ and an uncertainty of 1~km~s$^{-1}$, much smaller than the other points, effectively forcing the model to follow the $u_{2}(t)$ model parameter. Other inputs to this code (i.e., ``priors'') include the distributions of the orbital period, $P$, the time of the next conjunction, $t_0$, and the amplitude of the radial velocity variations, $V_2\,\sin\, i$. We choose the orbital period distribution to be $\psi(P) \propto 1/P$ (i.e., a Jeffreys prior, flat in $\log P$) and ranging from $2T$ (where $T$ is the duration of the monitoring period for a particular object) to a maximum value, $P_{\rm max}$, chosen as follows. We use equation~(1) of \pone, which relates the orbital period to the measured velocity offset and the mass, phase angle, and inclination angle of the binary orbit. We set the total SBHB mass to $10^{11}\;\rm M_\odot$ (see \S\ref{sec:discussion}) and select the values of other parameters so as to maximize the period ($q\ll 1$, $\sin i =\sin\phi=1$). Thus we get the maximum orbital period corresponding to a mass of $10^{11}\;\rm M_\odot$ and compatible with the measured value of $u_2^{\rm obs}(t_1)$, $P_{\rm max} = 2.652\times 10^6\,[u_2^{\rm obs}(t_1)/10^3\;{\rm km\;s}^{-1}]^{-1}\;$yr. With this scheme, the code explores orbital models corresponding to SBHB masses that are at least as large as $10^{11}\;M_\odot$. Since the SBHB mass is a derived parameter that depends on the values of $P$ and $V_2\,\sin\, i$, the code can actually explore SBHB masses that are larger than $10^{11}\;{\rm M}_\odot$ because the value of $V_2\,\sin\, i$ obtained from the model fit can be larger than the value of $u_2^{\rm obs}(t_1)$ used to set $P_{\rm max}$. For $t_0$ we choose a Jeffreys distribution ranging from $t_1$ to $t_1+T+P$ (where the upper limit is set during the simulation depending on $P$). We also explore the effect of adopting a flat distribution for $t_{0}$ and find that it yields systematically higher periods by 1 per cent with a scatter of 10 per cent, a negligible difference. For $V_2\,\sin\, i$ we choose a modified Jeffreys prior (which is flat below 1~km~s$^{-1}$) ranging from 0 to the speed of light. We summarize our selection of prior distributions in Table~\ref{tab:priors}.

\begin{table*}
\begin{minipage}{10.5cm}
\renewcommand{\thefootnote}{\alph{footnote}}
\caption{Parameters of prior distributions$^a$ \label{tab:priors}}
\renewcommand{\thefootnote}{\alph{footnote}}
\begin{tabular}{lccc}
Parameter & Distribution & Lower Bound & Upper Bound \\
\hline
$P$ (years)...............							& Jeffreys							&  $2T$ 		  	& $2.652\times 10^6\,/\,u_{2,3}^{\rm obs}(t_1)\;$ \\
$V_2\,\sin\,i$ (km s$^{-1}$)...					& Modified Jeffreys					&  0 				& $c$ \\
$t_{0}$ (years)...............						& Jeffreys/Uniform\footnotemark[2]		&  $t_{1}$ 		  	& $t_{1}+T+P$ \\
$u_{2}(t_{1})\,$(km s$^{-1}$)\footnotemark[3]....  	& Gaussian 						& All reals			& \nodata \\
\hline
\end{tabular}
\footnotetext[1]{See \S\ref{sec:interpretation} of the text and Appendix~\ref{app:mcmc} for a detailed discussion.}
\footnotetext[2]{We do not have strong physical constraints on the prior distribution of the time of the next conjunction, $t_{0}$. Therefore, we try two fairly different distributions.}
\footnotetext[3]{This parameter is effectively not free because the prior distribution is constrained by measurements of the SDSS data point.}
\end{minipage}
\end{table*}

We run the {\tt emcee} code in two stages so as to explore the large parameter space efficiently but also sample the shortest periods fully. First, we perform a high-resolution simulation of the shortest periods. We initialize 500 walkers clustered very close together around an initial guess for each model parameter ($\delta Q/Q\sim 10^{-4}$, where $Q$ is any model parameter). We follow a three-stage burn-in process, where for each stage we run the simulation for $10^3$ iterations and then reset the walkers in a cluster around their average value for each parameter. Then a final simulation is run for another $10^3$ iterations without resetting the walkers after the last burn-in phase. Second, we perform a lower-resolution simulation that samples the longest periods. Based on the results of the high-resolution simulation, we identify a high-likelihood ridge in the $P-t_{0}$ parameter space. The 500 walkers for the low-resolution simulation are initialized along this ridge according to the priors on $P$ and $t_{0}$, while the walkers for $V_2\,\sin\, i$ are distributed in a cluster around the high-resolution average value with dispersion equal to the measured standard deviation. After a burn-in phase of $3\times10^3$ iterations, the low-resolution simulation is run for $10^3$ iterations.

Each simulation produces posterior distributions for all model parameters and an accurate accounting of their covariance. We inspect the posterior distributions of $u_2(t_1)$ to confirm that, indeed, the prior constraints were obeyed in all cases. Using the posterior $\psi(P)$ distributions from the high- and low-resolution simulations, we can create a combined, cumulative posterior distribution of $P$, $\psi(>\!P)$. We identify a range of periods where the posterior distributions from the high- and low-resolution simulations overlap and are well populated and we normalize them so that they yield equal probability over this range and then combine them. From the final $\psi(>\!P)$ we can determine the 68th, 90th, and 99th percentile periods and calculate the corresponding orbital models.

\newpage
\bibliographystyle{mnras}
\bibliography{all.021716}

\begin{thebibliography}{}
\makeatletter
\relax
\def\mn@urlcharsother{\let\do\@makeother \do\$\do\&\do\#\do\^\do\_\do\%\do\~}
\def\mn@doi{\begingroup\mn@urlcharsother \@ifnextchar [ {\mn@doi@}
  {\mn@doi@[]}}
\def\mn@doi@[#1]#2{\def\@tempa{#1}\ifx\@tempa\@empty \href
  {http://dx.doi.org/#2} {doi:#2}\else \href {http://dx.doi.org/#2} {#1}\fi
  \endgroup}
\def\mn@eprint#1#2{\mn@eprint@#1:#2::\@nil}
\def\mn@eprint@arXiv#1{\href {http://arxiv.org/abs/#1} {{\tt arXiv:#1}}}
\def\mn@eprint@dblp#1{\href {http://dblp.uni-trier.de/rec/bibtex/#1.xml}
  {dblp:#1}}
\def\mn@eprint@#1:#2:#3:#4\@nil{\def\@tempa {#1}\def\@tempb {#2}\def\@tempc
  {#3}\ifx \@tempc \@empty \let \@tempc \@tempb \let \@tempb \@tempa \fi \ifx
  \@tempb \@empty \def\@tempb {arXiv}\fi \@ifundefined
  {mn@eprint@\@tempb}{\@tempb:\@tempc}{\expandafter \expandafter \csname
  mn@eprint@\@tempb\endcsname \expandafter{\@tempc}}}

\bibitem[\protect\citeauthoryear{{Aarseth}}{{Aarseth}}{2003}]{aarseth03}
{Aarseth} S.~J.,  2003, \mn@doi [\apss] {10.1023/A:1025492510715}, \href
  {http://adsabs.harvard.edu/abs/2003Ap%26SS.285..367A} {285, 367}

\bibitem[\protect\citeauthoryear{{Amaro-Seoane} et~al.,}{{Amaro-Seoane}
  et~al.}{2012}]{amaro-seoane12}
{Amaro-Seoane} P.,  et~al., 2012, \mn@doi [Classical and Quantum Gravity]
  {10.1088/0264-9381/29/12/124016}, \href
  {http://adsabs.harvard.edu/abs/2012CQGra..29l4016A} {29, 124016}

\bibitem[\protect\citeauthoryear{{Armitage} \& {Natarajan}}{{Armitage} \&
  {Natarajan}}{2002}]{armitage02}
{Armitage} P.~J.,  {Natarajan} P.,  2002, \mn@doi [\apjl] {10.1086/339770},
  \href {http://adsabs.harvard.edu/abs/2002ApJ...567L...9A} {567, L9}

\bibitem[\protect\citeauthoryear{{Artymowicz} \& {Lubow}}{{Artymowicz} \&
  {Lubow}}{1996}]{artymowicz96}
{Artymowicz} P.,  {Lubow} S.~H.,  1996, \mn@doi [\apjl] {10.1086/310200}, \href
  {http://adsabs.harvard.edu/abs/1996ApJ...467L..77A} {467, L77}

\bibitem[\protect\citeauthoryear{{Arzoumanian} et~al.,}{{Arzoumanian}
  et~al.}{2014}]{arzoumanian14}
{Arzoumanian} Z.,  et~al., 2014, \mn@doi [\apj] {10.1088/0004-637X/794/2/141},
  \href {http://adsabs.harvard.edu/abs/2014ApJ...794..141A} {794, 141}

\bibitem[\protect\citeauthoryear{{Baker}, {Boggs}, {Centrella}, {Kelly},
  {McWilliams}, {Miller}  \& {van Meter}}{{Baker} et~al.}{2008}]{baker08}
{Baker} J.~G.,  {Boggs} W.~D.,  {Centrella} J.,  {Kelly} B.~J.,  {McWilliams}
  S.~T.,  {Miller} M.~C.,   {van Meter} J.~R.,  2008, \mn@doi [\apjl]
  {10.1086/590927}, \href {http://adsabs.harvard.edu/abs/2008ApJ...682L..29B}
  {682, L29}

\bibitem[\protect\citeauthoryear{{Barrows} et~al.,}{{Barrows}
  et~al.}{2012}]{barrows12}
{Barrows} R.~S.,  et~al., 2012, \mn@doi [\apj] {10.1088/0004-637X/744/1/7},
  \href {http://adsabs.harvard.edu/abs/2012ApJ...744....7B} {744, 7}

\bibitem[\protect\citeauthoryear{{Barth} et~al.,}{{Barth}
  et~al.}{2015}]{barth15}
{Barth} A.~J.,  et~al., 2015, \mn@doi [\apjs] {10.1088/0067-0049/217/2/26},
  \href {http://adsabs.harvard.edu/abs/2015ApJS..217...26B} {217, 26}

\bibitem[\protect\citeauthoryear{{Begelman}, {Blandford}  \& {Rees}}{{Begelman}
  et~al.}{1980}]{begelman80}
{Begelman} M.~C.,  {Blandford} R.~D.,   {Rees} M.~J.,  1980, \mn@doi [\nat]
  {10.1038/287307a0}, \href {http://adsabs.harvard.edu/abs/1980Natur.287..307B}
  {287, 307}

\bibitem[\protect\citeauthoryear{{Bentz} et~al.,}{{Bentz}
  et~al.}{2010}]{bentz10}
{Bentz} M.~C.,  et~al., 2010, \mn@doi [\apjl] {10.1088/2041-8205/720/1/L46},
  \href {http://adsabs.harvard.edu/abs/2010ApJ...720L..46B} {720, L46}

\bibitem[\protect\citeauthoryear{{Berczik}, {Merritt}, {Spurzem}  \&
  {Bischof}}{{Berczik} et~al.}{2006}]{berczik06}
{Berczik} P.,  {Merritt} D.,  {Spurzem} R.,   {Bischof} H.-P.,  2006, \mn@doi
  [\apjl] {10.1086/504426}, \href
  {http://adsabs.harvard.edu/abs/2006ApJ...642L..21B} {642, L21}

\bibitem[\protect\citeauthoryear{{Blecha}, {Cox}, {Loeb}  \&
  {Hernquist}}{{Blecha} et~al.}{2011}]{blecha11}
{Blecha} L.,  {Cox} T.~J.,  {Loeb} A.,   {Hernquist} L.,  2011, \mn@doi
  [\mnras] {10.1111/j.1365-2966.2010.18042.x}, \href
  {http://adsabs.harvard.edu/abs/2011MNRAS.412.2154B} {412, 2154}

\bibitem[\protect\citeauthoryear{{Blecha} et~al.,}{{Blecha}
  et~al.}{2016}]{blecha16}
{Blecha} L.,  et~al., 2016, \mn@doi [\mnras] {10.1093/mnras/stv2646}, \href
  {http://adsabs.harvard.edu/abs/2016MNRAS.456..961B} {456, 961}

\bibitem[\protect\citeauthoryear{{Bon} et~al.,}{{Bon} et~al.}{2012}]{bon12}
{Bon} E.,  et~al., 2012, \mn@doi [\apj] {10.1088/0004-637X/759/2/118}, \href
  {http://adsabs.harvard.edu/abs/2012ApJ...759..118B} {759, 118}

\bibitem[\protect\citeauthoryear{{Bon} et~al.,}{{Bon} et~al.}{2016}]{bon16}
{Bon} E.,  et~al., 2016, preprint, \href
  {http://adsabs.harvard.edu/abs/2016arXiv160604606B} {} (\mn@eprint {arXiv}
  {1606.04606})

\bibitem[\protect\citeauthoryear{{Bonning}, {Shields}  \&
  {Salviander}}{{Bonning} et~al.}{2007}]{bonning07}
{Bonning} E.~W.,  {Shields} G.~A.,   {Salviander} S.,  2007, \mn@doi [\apjl]
  {10.1086/521674}, \href {http://adsabs.harvard.edu/abs/2007ApJ...666L..13B}
  {666, L13}

\bibitem[\protect\citeauthoryear{{Boroson} \& {Lauer}}{{Boroson} \&
  {Lauer}}{2010}]{boroson10}
{Boroson} T.~A.,  {Lauer} T.~R.,  2010, \mn@doi [\aj]
  {10.1088/0004-6256/140/2/390}, \href
  {http://adsabs.harvard.edu/abs/2010AJ....140..390B} {140, 390}

\bibitem[\protect\citeauthoryear{{Campanelli}, {Lousto}, {Zlochower}  \&
  {Merritt}}{{Campanelli} et~al.}{2007a}]{campanelli07a}
{Campanelli} M.,  {Lousto} C.~O.,  {Zlochower} Y.,   {Merritt} D.,  2007a,
  \mn@doi [Physical Review Letters] {10.1103/PhysRevLett.98.231102}, \href
  {http://adsabs.harvard.edu/abs/2007PhRvL..98w1102C} {98, 231102}

\bibitem[\protect\citeauthoryear{{Campanelli}, {Lousto}, {Zlochower}  \&
  {Merritt}}{{Campanelli} et~al.}{2007b}]{campanelli07b}
{Campanelli} M.,  {Lousto} C.,  {Zlochower} Y.,   {Merritt} D.,  2007b, \mn@doi
  [\apjl] {10.1086/516712}, \href
  {http://adsabs.harvard.edu/abs/2007ApJ...659L...5C} {659, L5}

\bibitem[\protect\citeauthoryear{{Charisi}, {Bartos}, {Haiman}, {Price-Whelan},
  {Graham}, {Bellm}, {Laher}  \& {Marka}}{{Charisi} et~al.}{2016}]{charisi16}
{Charisi} M.,  {Bartos} I.,  {Haiman} Z.,  {Price-Whelan} A.~M.,  {Graham}
  M.~J.,  {Bellm} E.~C.,  {Laher} R.~R.,   {Marka} S.,  2016, preprint, \href
  {http://adsabs.harvard.edu/abs/2016arXiv160401020C} {} (\mn@eprint {arXiv}
  {1604.01020})

\bibitem[\protect\citeauthoryear{{Comerford}, {Griffith}, {Gerke}, {Cooper},
  {Newman}, {Davis}  \& {Stern}}{{Comerford} et~al.}{2009}]{comerford09b}
{Comerford} J.~M.,  {Griffith} R.~L.,  {Gerke} B.~F.,  {Cooper} M.~C.,
  {Newman} J.~A.,  {Davis} M.,   {Stern} D.,  2009, \mn@doi [\apjl]
  {10.1088/0004-637X/702/1/L82}, \href
  {http://adsabs.harvard.edu/abs/2009ApJ...702L..82C} {702, L82}

\bibitem[\protect\citeauthoryear{{Comerford}, {Gerke}, {Stern}, {Cooper},
  {Weiner}, {Newman}, {Madsen}  \& {Barrows}}{{Comerford}
  et~al.}{2012}]{comerford12}
{Comerford} J.~M.,  {Gerke} B.~F.,  {Stern} D.,  {Cooper} M.~C.,  {Weiner}
  B.~J.,  {Newman} J.~A.,  {Madsen} K.,   {Barrows} R.~S.,  2012, \mn@doi
  [\apj] {10.1088/0004-637X/753/1/42}, \href
  {http://adsabs.harvard.edu/abs/2012ApJ...753...42C} {753, 42}

\bibitem[\protect\citeauthoryear{{Comerford}, {Schluns}, {Greene}  \&
  {Cool}}{{Comerford} et~al.}{2013}]{comerford13}
{Comerford} J.~M.,  {Schluns} K.,  {Greene} J.~E.,   {Cool} R.~J.,  2013,
  \mn@doi [\apj] {10.1088/0004-637X/777/1/64}, \href
  {http://adsabs.harvard.edu/abs/2013ApJ...777...64C} {777, 64}

\bibitem[\protect\citeauthoryear{{Comerford}, {Pooley}, {Barrows}, {Greene},
  {Zakamska}, {Madejski}  \& {Cooper}}{{Comerford} et~al.}{2015}]{comerford15}
{Comerford} J.~M.,  {Pooley} D.,  {Barrows} R.~S.,  {Greene} J.~E.,  {Zakamska}
  N.~L.,  {Madejski} G.~M.,   {Cooper} M.~C.,  2015, \mn@doi [\apj]
  {10.1088/0004-637X/806/2/219}, \href
  {http://adsabs.harvard.edu/abs/2015ApJ...806..219C} {806, 219}

\bibitem[\protect\citeauthoryear{{Cuadra}, {Armitage}, {Alexander}  \&
  {Begelman}}{{Cuadra} et~al.}{2009}]{cuadra09}
{Cuadra} J.,  {Armitage} P.~J.,  {Alexander} R.~D.,   {Begelman} M.~C.,  2009,
  \mn@doi [\mnras] {10.1111/j.1365-2966.2008.14147.x}, \href
  {http://adsabs.harvard.edu/abs/2009MNRAS.393.1423C} {393, 1423}

\bibitem[\protect\citeauthoryear{{Decarli}, {Dotti}, {Fumagalli}, {Tsalmantza},
  {Montuori}, {Lusso}, {Hogg}  \& {Prochaska}}{{Decarli}
  et~al.}{2013}]{decarli13}
{Decarli} R.,  {Dotti} M.,  {Fumagalli} M.,  {Tsalmantza} P.,  {Montuori} C.,
  {Lusso} E.,  {Hogg} D.~W.,   {Prochaska} J.~X.,  2013, \mn@doi [\mnras]
  {10.1093/mnras/stt831}, \href
  {http://adsabs.harvard.edu/abs/2013MNRAS.433.1492D} {433, 1492}

\bibitem[\protect\citeauthoryear{{Decarli}, {Dotti}, {Mazzucchelli}, {Montuori}
   \& {Volonteri}}{{Decarli} et~al.}{2014}]{decarli14}
{Decarli} R.,  {Dotti} M.,  {Mazzucchelli} C.,  {Montuori} C.,   {Volonteri}
  M.,  2014, \mn@doi [\mnras] {10.1093/mnras/stu1810}, \href
  {http://adsabs.harvard.edu/abs/2014MNRAS.445.1558D} {445, 1558}

\bibitem[\protect\citeauthoryear{{Doroshenko}, {Sergeev}, {Klimanov}, {Pronik}
  \& {Efimov}}{{Doroshenko} et~al.}{2012}]{doroshenko12}
{Doroshenko} V.~T.,  {Sergeev} S.~G.,  {Klimanov} S.~A.,  {Pronik} V.~I.,
  {Efimov} Y.~S.,  2012, \mn@doi [\mnras] {10.1111/j.1365-2966.2012.20843.x},
  \href {http://adsabs.harvard.edu/abs/2012MNRAS.426..416D} {426, 416}

\bibitem[\protect\citeauthoryear{{Dotti}, {Colpi}  \& {Haardt}}{{Dotti}
  et~al.}{2006}]{dotti06}
{Dotti} M.,  {Colpi} M.,   {Haardt} F.,  2006, \mn@doi [\mnras]
  {10.1111/j.1365-2966.2005.09956.x}, \href
  {http://adsabs.harvard.edu/abs/2006MNRAS.367..103D} {367, 103}

\bibitem[\protect\citeauthoryear{{Dotti}, {Colpi}, {Haardt}  \&
  {Mayer}}{{Dotti} et~al.}{2007}]{dotti07}
{Dotti} M.,  {Colpi} M.,  {Haardt} F.,   {Mayer} L.,  2007, \mn@doi [\mnras]
  {10.1111/j.1365-2966.2007.12010.x}, \href
  {http://adsabs.harvard.edu/abs/2007MNRAS.379..956D} {379, 956}

\bibitem[\protect\citeauthoryear{{Dotti}, {Montuori}, {Decarli}, {Volonteri},
  {Colpi}  \& {Haardt}}{{Dotti} et~al.}{2009}]{dotti09b}
{Dotti} M.,  {Montuori} C.,  {Decarli} R.,  {Volonteri} M.,  {Colpi} M.,
  {Haardt} F.,  2009, \mn@doi [\mnras] {10.1111/j.1745-3933.2009.00714.x},
  \href {http://adsabs.harvard.edu/abs/2009MNRAS.398L..73D} {398, L73}

\bibitem[\protect\citeauthoryear{{Dotti}, {Volonteri}, {Perego}, {Colpi},
  {Ruszkowski}  \& {Haardt}}{{Dotti} et~al.}{2010}]{dotti10}
{Dotti} M.,  {Volonteri} M.,  {Perego} A.,  {Colpi} M.,  {Ruszkowski} M.,
  {Haardt} F.,  2010, \mn@doi [\mnras] {10.1111/j.1365-2966.2009.15922.x},
  \href {http://adsabs.harvard.edu/abs/2010MNRAS.402..682D} {402, 682}

\bibitem[\protect\citeauthoryear{{Eracleous}, {Halpern}, {M.~Gilbert}, {Newman}
   \& {Filippenko}}{{Eracleous} et~al.}{1997}]{eracleous97}
{Eracleous} M.,  {Halpern} J.~P.,  {M.~Gilbert} A.,  {Newman} J.~A.,
  {Filippenko} A.~V.,  1997, \apj, \href
  {http://adsabs.harvard.edu/abs/1997ApJ...490..216E} {490, 216}

\bibitem[\protect\citeauthoryear{{Eracleous}, {Boroson}, {Halpern}  \&
  {Liu}}{{Eracleous} et~al.}{2012}]{eracleous12}
{Eracleous} M.,  {Boroson} T.~A.,  {Halpern} J.~P.,   {Liu} J.,  2012, \mn@doi
  [\apjs] {10.1088/0067-0049/201/2/23}, \href
  {http://adsabs.harvard.edu/abs/2012ApJS..201...23E} {201, 23}

\bibitem[\protect\citeauthoryear{{Escala}, {Larson}, {Coppi}  \&
  {Mardones}}{{Escala} et~al.}{2004}]{escala04}
{Escala} A.,  {Larson} R.~B.,  {Coppi} P.~S.,   {Mardones} D.,  2004, \mn@doi
  [\apj] {10.1086/386278}, \href
  {http://adsabs.harvard.edu/abs/2004ApJ...607..765E} {607, 765}

\bibitem[\protect\citeauthoryear{{Farris}, {Duffell}, {MacFadyen}  \&
  {Haiman}}{{Farris} et~al.}{2014}]{farris14}
{Farris} B.~D.,  {Duffell} P.,  {MacFadyen} A.~I.,   {Haiman} Z.,  2014,
  \mn@doi [\apj] {10.1088/0004-637X/783/2/134}, \href
  {http://adsabs.harvard.edu/abs/2014ApJ...783..134F} {783, 134}

\bibitem[\protect\citeauthoryear{{Fiacconi}, {Mayer}, {Ro{\v s}kar}  \&
  {Colpi}}{{Fiacconi} et~al.}{2013}]{fiacconi13}
{Fiacconi} D.,  {Mayer} L.,  {Ro{\v s}kar} R.,   {Colpi} M.,  2013, \mn@doi
  [\apjl] {10.1088/2041-8205/777/1/L14}, \href
  {http://adsabs.harvard.edu/abs/2013ApJ...777L..14F} {777, L14}

\bibitem[\protect\citeauthoryear{{Foreman-Mackey}, {Hogg}, {Lang}  \&
  {Goodman}}{{Foreman-Mackey} et~al.}{2013}]{foreman13}
{Foreman-Mackey} D.,  {Hogg} D.~W.,  {Lang} D.,   {Goodman} J.,  2013, \mn@doi
  [\pasp] {10.1086/670067}, \href
  {http://adsabs.harvard.edu/abs/2013PASP..125..306F} {125, 306}

\bibitem[\protect\citeauthoryear{{Fu}, {Myers}, {Djorgovski}  \& {Yan}}{{Fu}
  et~al.}{2011}]{fu11a}
{Fu} H.,  {Myers} A.~D.,  {Djorgovski} S.~G.,   {Yan} L.,  2011, \mn@doi [\apj]
  {10.1088/0004-637X/733/2/103}, \href
  {http://adsabs.harvard.edu/abs/2011ApJ...733..103F} {733, 103}

\bibitem[\protect\citeauthoryear{{Fu}, {Yan}, {Myers}, {Stockton},
  {Djorgovski}, {Aldering}  \& {Rich}}{{Fu} et~al.}{2012}]{fu12}
{Fu} H.,  {Yan} L.,  {Myers} A.~D.,  {Stockton} A.,  {Djorgovski} S.~G.,
  {Aldering} G.,   {Rich} J.~A.,  2012, \mn@doi [\apj]
  {10.1088/0004-637X/745/1/67}, \href
  {http://adsabs.harvard.edu/abs/2012ApJ...745...67F} {745, 67}

\bibitem[\protect\citeauthoryear{{Gaskell}}{{Gaskell}}{1996}]{gaskell96a}
{Gaskell} C.~M.,  1996, in {Kundt} W.,  ed.,  Lecture Notes in Physics, Berlin
  Springer Verlag Vol. 471, Jets from Stars and Galactic Nuclei. p.~165
  (\mn@eprint {} {astro-ph/9605175}), \mn@doi{10.1007/BFb0102607}

\bibitem[\protect\citeauthoryear{{Gezari}, {Halpern}  \& {Eracleous}}{{Gezari}
  et~al.}{2007}]{gezari07}
{Gezari} S.,  {Halpern} J.~P.,   {Eracleous} M.,  2007, \mn@doi [\apjs]
  {10.1086/511032}, \href {http://adsabs.harvard.edu/abs/2007ApJS..169..167G}
  {169, 167}

\bibitem[\protect\citeauthoryear{{Graham} et~al.,}{{Graham}
  et~al.}{2015a}]{graham15b}
{Graham} M.~J.,  et~al., 2015a, \mn@doi [\mnras] {10.1093/mnras/stv1726}, \href
  {http://adsabs.harvard.edu/abs/2015MNRAS.453.1562G} {453, 1562}

\bibitem[\protect\citeauthoryear{{Graham} et~al.,}{{Graham}
  et~al.}{2015b}]{graham15a}
{Graham} M.~J.,  et~al., 2015b, \mn@doi [\nat] {10.1038/nature14143}, \href
  {http://adsabs.harvard.edu/abs/2015Natur.518...74G} {518, 74}

\bibitem[\protect\citeauthoryear{{Grier} et~al.,}{{Grier}
  et~al.}{2013}]{grier13}
{Grier} C.~J.,  et~al., 2013, \mn@doi [\apj] {10.1088/0004-637X/764/1/47},
  \href {http://adsabs.harvard.edu/abs/2013ApJ...764...47G} {764, 47}

\bibitem[\protect\citeauthoryear{{Guedes}, {Madau}, {Mayer}  \&
  {Callegari}}{{Guedes} et~al.}{2011}]{guedes11}
{Guedes} J.,  {Madau} P.,  {Mayer} L.,   {Callegari} S.,  2011, \mn@doi [\apj]
  {10.1088/0004-637X/729/2/125}, \href
  {http://adsabs.harvard.edu/abs/2011ApJ...729..125G} {729, 125}

\bibitem[\protect\citeauthoryear{{Halpern} \& {Filippenko}}{{Halpern} \&
  {Filippenko}}{1988}]{halpern88}
{Halpern} J.~P.,  {Filippenko} A.~V.,  1988, \mn@doi [\nat] {10.1038/331046a0},
  \href {http://adsabs.harvard.edu/abs/1988Natur.331...46H} {331, 46}

\bibitem[\protect\citeauthoryear{{Halpern} \& {Filippenko}}{{Halpern} \&
  {Filippenko}}{1992}]{halpern92}
{Halpern} J.,  {Filippenko} A.,  1992, in {Holt} S.~S.,  {Neff} S.~G.,   {Urry}
  C.~M.,  eds,  American Institute of Physics Conference Series Vol. 254,
  American Institute of Physics Conference Series. pp 57--60,
  \mn@doi{10.1063/1.42243}

\bibitem[\protect\citeauthoryear{{Hayasaki}, {Mineshige}  \&
  {Sudou}}{{Hayasaki} et~al.}{2007}]{hayasaki07}
{Hayasaki} K.,  {Mineshige} S.,   {Sudou} H.,  2007, \mn@doi [\pasj]
  {10.1093/pasj/59.2.427}, \href
  {http://adsabs.harvard.edu/abs/2007PASJ...59..427H} {59, 427}

\bibitem[\protect\citeauthoryear{{Holley-Bockelmann} \&
  {Khan}}{{Holley-Bockelmann} \& {Khan}}{2015}]{holley15}
{Holley-Bockelmann} K.,  {Khan} F.~M.,  2015, \mn@doi [\apj]
  {10.1088/0004-637X/810/2/139}, \href
  {http://adsabs.harvard.edu/abs/2015ApJ...810..139H} {810, 139}

\bibitem[\protect\citeauthoryear{{Hopkins}, {Hernquist}, {Cox}, {Di Matteo},
  {Robertson}  \& {Springel}}{{Hopkins} et~al.}{2006}]{hopkins06}
{Hopkins} P.~F.,  {Hernquist} L.,  {Cox} T.~J.,  {Di Matteo} T.,  {Robertson}
  B.,   {Springel} V.,  2006, \mn@doi [\apjs] {10.1086/499298}, \href
  {http://adsabs.harvard.edu/abs/2006ApJS..163....1H} {163, 1}

\bibitem[\protect\citeauthoryear{{Inayoshi} \& {Haiman}}{{Inayoshi} \&
  {Haiman}}{2016}]{inayoshi16}
{Inayoshi} K.,  {Haiman} Z.,  2016, preprint, \href
  {http://adsabs.harvard.edu/abs/2016arXiv160102611I} {} (\mn@eprint {arXiv}
  {1601.02611})

\bibitem[\protect\citeauthoryear{{Ju}, {Greene}, {Rafikov}, {Bickerton}  \&
  {Badenes}}{{Ju} et~al.}{2013}]{ju13}
{Ju} W.,  {Greene} J.~E.,  {Rafikov} R.~R.,  {Bickerton} S.~J.,   {Badenes} C.,
   2013, \mn@doi [\apj] {10.1088/0004-637X/777/1/44}, \href
  {http://adsabs.harvard.edu/abs/2013ApJ...777...44J} {777, 44}

\bibitem[\protect\citeauthoryear{{Kassebaum}, {Peterson}, {Wanders}, {Pogge},
  {Bertram}  \& {Wagner}}{{Kassebaum} et~al.}{1997}]{kassebaum97}
{Kassebaum} T.~M.,  {Peterson} B.~M.,  {Wanders} I.,  {Pogge} R.~W.,  {Bertram}
  R.,   {Wagner} R.~M.,  1997, \apj, \href
  {http://adsabs.harvard.edu/abs/1997ApJ...475..106K} {475, 106}

\bibitem[\protect\citeauthoryear{{Khan}, {Holley-Bockelmann}, {Berczik}  \&
  {Just}}{{Khan} et~al.}{2013}]{khan13}
{Khan} F.~M.,  {Holley-Bockelmann} K.,  {Berczik} P.,   {Just} A.,  2013,
  \mn@doi [\apj] {10.1088/0004-637X/773/2/100}, \href
  {http://adsabs.harvard.edu/abs/2013ApJ...773..100K} {773, 100}

\bibitem[\protect\citeauthoryear{{Kollatschny}, {Bischoff}  \&
  {Dietrich}}{{Kollatschny} et~al.}{2000}]{kollatschny00}
{Kollatschny} W.,  {Bischoff} K.,   {Dietrich} M.,  2000, \aap, \href
  {http://adsabs.harvard.edu/abs/2000A%26A...361..901K} {361, 901}

\bibitem[\protect\citeauthoryear{{Kormendy} \& {Richstone}}{{Kormendy} \&
  {Richstone}}{1995}]{kormendy95}
{Kormendy} J.,  {Richstone} D.,  1995, \mn@doi [\araa]
  {10.1146/annurev.aa.33.090195.003053}, \href
  {http://adsabs.harvard.edu/abs/1995ARA%26A..33..581K} {33, 581}

\bibitem[\protect\citeauthoryear{{Koss} et~al.,}{{Koss} et~al.}{2016}]{koss16}
{Koss} M.~J.,  et~al., 2016, \mn@doi [\apjl] {10.3847/2041-8205/824/1/L4},
  \href {http://adsabs.harvard.edu/abs/2016ApJ...824L...4K} {824, L4}

\bibitem[\protect\citeauthoryear{{Lewis}, {Eracleous}  \&
  {Storchi-Bergmann}}{{Lewis} et~al.}{2010}]{lewis10}
{Lewis} K.~T.,  {Eracleous} M.,   {Storchi-Bergmann} T.,  2010, \mn@doi [\apjs]
  {10.1088/0067-0049/187/2/416}, \href
  {http://adsabs.harvard.edu/abs/2010ApJS..187..416L} {187, 416}

\bibitem[\protect\citeauthoryear{{Li} et~al.,}{{Li} et~al.}{2016}]{li16}
{Li} Y.-R.,  et~al., 2016, preprint, \href
  {http://adsabs.harvard.edu/abs/2016arXiv160205005L} {} (\mn@eprint {arXiv}
  {1602.05005})

\bibitem[\protect\citeauthoryear{{Liu}, {Shen}, {Strauss}  \& {Greene}}{{Liu}
  et~al.}{2010a}]{liu10a}
{Liu} X.,  {Shen} Y.,  {Strauss} M.~A.,   {Greene} J.~E.,  2010a, \mn@doi
  [\apj] {10.1088/0004-637X/708/1/427}, \href
  {http://adsabs.harvard.edu/abs/2010ApJ...708..427L} {708, 427}

\bibitem[\protect\citeauthoryear{{Liu}, {Greene}, {Shen}  \& {Strauss}}{{Liu}
  et~al.}{2010b}]{liu10b}
{Liu} X.,  {Greene} J.~E.,  {Shen} Y.,   {Strauss} M.~A.,  2010b, \mn@doi
  [\apjl] {10.1088/2041-8205/715/1/L30}, \href
  {http://adsabs.harvard.edu/abs/2010ApJ...715L..30L} {715, L30}

\bibitem[\protect\citeauthoryear{{Liu}, {Shen}, {Bian}, {Loeb}  \&
  {Tremaine}}{{Liu} et~al.}{2014}]{liu14}
{Liu} X.,  {Shen} Y.,  {Bian} F.,  {Loeb} A.,   {Tremaine} S.,  2014, \mn@doi
  [\apj] {10.1088/0004-637X/789/2/140}, \href
  {http://adsabs.harvard.edu/abs/2014ApJ...789..140L} {789, 140}

\bibitem[\protect\citeauthoryear{{Liu} et~al.,}{{Liu} et~al.}{2015}]{liu15}
{Liu} T.,  et~al., 2015, \mn@doi [\apjl] {10.1088/2041-8205/803/2/L16}, \href
  {http://adsabs.harvard.edu/abs/2015ApJ...803L..16L} {803, L16}

\bibitem[\protect\citeauthoryear{{Liu}, {Eracleous}  \& {Halpern}}{{Liu}
  et~al.}{2016}]{liu16}
{Liu} J.,  {Eracleous} M.,   {Halpern} J.~P.,  2016, \mn@doi [\apj]
  {10.3847/0004-637X/817/1/42}, \href
  {http://adsabs.harvard.edu/abs/2016ApJ...817...42L} {817, 42}

\bibitem[\protect\citeauthoryear{{Lodato}, {Nayakshin}, {King}  \&
  {Pringle}}{{Lodato} et~al.}{2009}]{lodato09}
{Lodato} G.,  {Nayakshin} S.,  {King} A.~R.,   {Pringle} J.~E.,  2009, \mn@doi
  [\mnras] {10.1111/j.1365-2966.2009.15179.x}, \href
  {http://adsabs.harvard.edu/abs/2009MNRAS.398.1392L} {398, 1392}

\bibitem[\protect\citeauthoryear{{Loeb}}{{Loeb}}{2010}]{loeb10}
{Loeb} A.,  2010, \mn@doi [\prd] {10.1103/PhysRevD.81.047503}, \href
  {http://adsabs.harvard.edu/abs/2010PhRvD..81d7503L} {81, 047503}

\bibitem[\protect\citeauthoryear{{Lusso}, {Decarli}, {Dotti}, {Montuori},
  {Hogg}, {Tsalmantza}, {Fumagalli}  \& {Prochaska}}{{Lusso}
  et~al.}{2014}]{lusso14}
{Lusso} E.,  {Decarli} R.,  {Dotti} M.,  {Montuori} C.,  {Hogg} D.~W.,
  {Tsalmantza} P.,  {Fumagalli} M.,   {Prochaska} J.~X.,  2014, \mn@doi
  [\mnras] {10.1093/mnras/stu572}, \href
  {http://adsabs.harvard.edu/abs/2014MNRAS.441..316L} {441, 316}

\bibitem[\protect\citeauthoryear{{MacLeod} et~al.,}{{MacLeod}
  et~al.}{2010}]{macleod10}
{MacLeod} C.~L.,  et~al., 2010, \mn@doi [\apj] {10.1088/0004-637X/721/2/1014},
  \href {http://adsabs.harvard.edu/abs/2010ApJ...721.1014M} {721, 1014}

\bibitem[\protect\citeauthoryear{{McConnell}, {Ma}, {Murphy}, {Gebhardt},
  {Lauer}, {Graham}, {Wright}  \& {Richstone}}{{McConnell}
  et~al.}{2012}]{mcconnell12}
{McConnell} N.~J.,  {Ma} C.-P.,  {Murphy} J.~D.,  {Gebhardt} K.,  {Lauer}
  T.~R.,  {Graham} J.~R.,  {Wright} S.~A.,   {Richstone} D.~O.,  2012, \mn@doi
  [\apj] {10.1088/0004-637X/756/2/179}, \href
  {http://adsabs.harvard.edu/abs/2012ApJ...756..179M} {756, 179}

\bibitem[\protect\citeauthoryear{{McGurk}, {Max}, {Medling}, {Shields}  \&
  {Comerford}}{{McGurk} et~al.}{2015}]{mcgurk15}
{McGurk} R.~C.,  {Max} C.~E.,  {Medling} A.~M.,  {Shields} G.~A.,   {Comerford}
  J.~M.,  2015, \mn@doi [\apj] {10.1088/0004-637X/811/1/14}, \href
  {http://adsabs.harvard.edu/abs/2015ApJ...811...14M} {811, 14}

\bibitem[\protect\citeauthoryear{{Menou}, {Haiman}  \& {Narayanan}}{{Menou}
  et~al.}{2001}]{menou01}
{Menou} K.,  {Haiman} Z.,   {Narayanan} V.~K.,  2001, \mn@doi [\apj]
  {10.1086/322310}, \href {http://adsabs.harvard.edu/abs/2001ApJ...558..535M}
  {558, 535}

\bibitem[\protect\citeauthoryear{{Merritt} \& {Poon}}{{Merritt} \&
  {Poon}}{2004}]{merritt04a}
{Merritt} D.,  {Poon} M.~Y.,  2004, \mn@doi [\apj] {10.1086/382497}, \href
  {http://adsabs.harvard.edu/abs/2004ApJ...606..788M} {606, 788}

\bibitem[\protect\citeauthoryear{{M{\"u}ller-S{\'a}nchez}, {Comerford},
  {Nevin}, {Barrows}, {Cooper}  \& {Greene}}{{M{\"u}ller-S{\'a}nchez}
  et~al.}{2015}]{muller-sanchez15}
{M{\"u}ller-S{\'a}nchez} F.,  {Comerford} J.~M.,  {Nevin} R.,  {Barrows} R.~S.,
   {Cooper} M.~C.,   {Greene} J.~E.,  2015, \mn@doi [\apj]
  {10.1088/0004-637X/813/2/103}, \href
  {http://adsabs.harvard.edu/abs/2015ApJ...813..103M} {813, 103}

\bibitem[\protect\citeauthoryear{{Peterson}, {Pogge}  \& {Wanders}}{{Peterson}
  et~al.}{1999}]{peterson99conf}
{Peterson} B.~M.,  {Pogge} R.~W.,   {Wanders} I.,  1999, in {Gaskell} C.~M.,
  {Brandt} W.~N.,  {Dietrich} M.,  {Dultzin-Hacyan} D.,   {Eracleous} M.,  eds,
   Astronomical Society of the Pacific Conference Series Vol. 175, Structure
  and Kinematics of Quasar Broad Line Regions. p.~41

\bibitem[\protect\citeauthoryear{{Popovi{\'c}} et~al.,}{{Popovi{\'c}}
  et~al.}{2014}]{popovic14}
{Popovi{\'c}} L.~{\v C}.,  et~al., 2014, \mn@doi [\aap]
  {10.1051/0004-6361/201423555}, \href
  {http://adsabs.harvard.edu/abs/2014A%26A...572A..66P} {572, A66}

\bibitem[\protect\citeauthoryear{{Quinlan}}{{Quinlan}}{1996}]{quinlan96}
{Quinlan} G.~D.,  1996, \mn@doi [\na] {10.1016/S1384-1076(96)00018-8}, \href
  {http://adsabs.harvard.edu/abs/1996NewA....1..255Q} {1, 255}

\bibitem[\protect\citeauthoryear{{Roedig}, {Dotti}, {Sesana}, {Cuadra}  \&
  {Colpi}}{{Roedig} et~al.}{2011}]{roedig11}
{Roedig} C.,  {Dotti} M.,  {Sesana} A.,  {Cuadra} J.,   {Colpi} M.,  2011,
  \mn@doi [\mnras] {10.1111/j.1365-2966.2011.18927.x}, \href
  {http://adsabs.harvard.edu/abs/2011MNRAS.415.3033R} {415, 3033}

\bibitem[\protect\citeauthoryear{{Romano}, {Marziani}  \&
  {Dultzin-Hacyan}}{{Romano} et~al.}{1998}]{romano98}
{Romano} P.,  {Marziani} P.,   {Dultzin-Hacyan} D.,  1998, \mn@doi [\apj]
  {10.1086/305293}, \href {http://adsabs.harvard.edu/abs/1998ApJ...495..222R}
  {495, 222}

\bibitem[\protect\citeauthoryear{{Runnoe} et~al.,}{{Runnoe}
  et~al.}{2015}]{runnoe15}
{Runnoe} J.~C.,  et~al., 2015, \mn@doi [\apjs] {10.1088/0067-0049/221/1/7},
  \href {http://adsabs.harvard.edu/abs/2015ApJS..221....7R} {221, 7}

\bibitem[\protect\citeauthoryear{{Schimoia}, {Storchi-Bergmann}, {Nemmen},
  {Winge}  \& {Eracleous}}{{Schimoia} et~al.}{2012}]{schimoia12}
{Schimoia} J.~S.,  {Storchi-Bergmann} T.,  {Nemmen} R.~S.,  {Winge} C.,
  {Eracleous} M.,  2012, \mn@doi [\apj] {10.1088/0004-637X/748/2/145}, \href
  {http://adsabs.harvard.edu/abs/2012ApJ...748..145S} {748, 145}

\bibitem[\protect\citeauthoryear{{Sergeev}, {Doroshenko}, {Dzyuba}, {Peterson},
  {Pogge}  \& {Pronik}}{{Sergeev} et~al.}{2007}]{sergeev07}
{Sergeev} S.~G.,  {Doroshenko} V.~T.,  {Dzyuba} S.~A.,  {Peterson} B.~M.,
  {Pogge} R.~W.,   {Pronik} V.~I.,  2007, \mn@doi [\apj] {10.1086/520697},
  \href {http://adsabs.harvard.edu/abs/2007ApJ...668..708S} {668, 708}

\bibitem[\protect\citeauthoryear{{Sesana}}{{Sesana}}{2015}]{sesana15}
{Sesana} A.,  2015, \mn@doi [Astrophysics and Space Science Proceedings]
  {10.1007/978-3-319-10488-1_13}, \href
  {http://adsabs.harvard.edu/abs/2015ASSP...40..147S} {40, 147}

\bibitem[\protect\citeauthoryear{{Sesana}, {Haardt}  \& {Madau}}{{Sesana}
  et~al.}{2006}]{sesana06}
{Sesana} A.,  {Haardt} F.,   {Madau} P.,  2006, \mn@doi [\apj]
  {10.1086/507596}, \href {http://adsabs.harvard.edu/abs/2006ApJ...651..392S}
  {651, 392}

\bibitem[\protect\citeauthoryear{{Shapovalova} et~al.,}{{Shapovalova}
  et~al.}{2001}]{shapovalova01}
{Shapovalova} A.~I.,  et~al., 2001, \mn@doi [\aap]
  {10.1051/0004-6361:20011011}, \href
  {http://adsabs.harvard.edu/abs/2001A%26A...376..775S} {376, 775}

\bibitem[\protect\citeauthoryear{{Shapovalova} et~al.,}{{Shapovalova}
  et~al.}{2016}]{shapovalova16}
{Shapovalova} A.~I.,  et~al., 2016, \mn@doi [\apjs]
  {10.3847/0067-0049/222/2/25}, \href
  {http://adsabs.harvard.edu/abs/2016ApJS..222...25S} {222, 25}

\bibitem[\protect\citeauthoryear{{Shen}, {Liu}, {Greene}  \& {Strauss}}{{Shen}
  et~al.}{2011}]{shen11b}
{Shen} Y.,  {Liu} X.,  {Greene} J.~E.,   {Strauss} M.~A.,  2011, \mn@doi [\apj]
  {10.1088/0004-637X/735/1/48}, \href
  {http://adsabs.harvard.edu/abs/2011ApJ...735...48S} {735, 48}

\bibitem[\protect\citeauthoryear{{Shen}, {Liu}, {Loeb}  \& {Tremaine}}{{Shen}
  et~al.}{2013}]{shen13a}
{Shen} Y.,  {Liu} X.,  {Loeb} A.,   {Tremaine} S.,  2013, \mn@doi [\apj]
  {10.1088/0004-637X/775/1/49}, \href
  {http://adsabs.harvard.edu/abs/2013ApJ...775...49S} {775, 49}

\bibitem[\protect\citeauthoryear{{Shields}, {Bonning}  \&
  {Salviander}}{{Shields} et~al.}{2009a}]{shields09a}
{Shields} G.~A.,  {Bonning} E.~W.,   {Salviander} S.,  2009a, \mn@doi [\apj]
  {10.1088/0004-637X/696/2/1367}, \href
  {http://adsabs.harvard.edu/abs/2009ApJ...696.1367S} {696, 1367}

\bibitem[\protect\citeauthoryear{{Shields} et~al.,}{{Shields}
  et~al.}{2009b}]{shields09b}
{Shields} G.~A.,  et~al., 2009b, \mn@doi [\apj] {10.1088/0004-637X/707/2/936},
  \href {http://adsabs.harvard.edu/abs/2009ApJ...707..936S} {707, 936}

\bibitem[\protect\citeauthoryear{{Skielboe}, {Pancoast}, {Treu}, {Park},
  {Barth}  \& {Bentz}}{{Skielboe} et~al.}{2015}]{skielboe15}
{Skielboe} A.,  {Pancoast} A.,  {Treu} T.,  {Park} D.,  {Barth} A.~J.,
  {Bentz} M.~C.,  2015, \mn@doi [\mnras] {10.1093/mnras/stv1917}, \href
  {http://adsabs.harvard.edu/abs/2015MNRAS.454..144S} {454, 144}

\bibitem[\protect\citeauthoryear{{Smith}, {Shields}, {Bonning}, {McMullen},
  {Rosario}  \& {Salviander}}{{Smith} et~al.}{2010}]{smith10}
{Smith} K.~L.,  {Shields} G.~A.,  {Bonning} E.~W.,  {McMullen} C.~C.,
  {Rosario} D.~J.,   {Salviander} S.,  2010, \mn@doi [\apj]
  {10.1088/0004-637X/716/1/866}, \href
  {http://adsabs.harvard.edu/abs/2010ApJ...716..866S} {716, 866}

\bibitem[\protect\citeauthoryear{{Thomas}, {Ma}, {McConnell}, {Greene},
  {Blakeslee}  \& {Janish}}{{Thomas} et~al.}{2016}]{thomas16}
{Thomas} J.,  {Ma} C.-P.,  {McConnell} N.~J.,  {Greene} J.~E.,  {Blakeslee}
  J.~P.,   {Janish} R.,  2016, \mn@doi [\nat] {10.1038/nature17197}, \href
  {http://adsabs.harvard.edu/abs/2016Natur.532..340T} {532, 340}

\bibitem[\protect\citeauthoryear{{Tsalmantza}, {Decarli}, {Dotti}  \&
  {Hogg}}{{Tsalmantza} et~al.}{2011}]{tsalmantza11}
{Tsalmantza} P.,  {Decarli} R.,  {Dotti} M.,   {Hogg} D.~W.,  2011, \mn@doi
  [\apj] {10.1088/0004-637X/738/1/20}, \href
  {http://adsabs.harvard.edu/abs/2011ApJ...738...20T} {738, 20}

\bibitem[\protect\citeauthoryear{{Vasiliev} \& {Merritt}}{{Vasiliev} \&
  {Merritt}}{2013}]{vasiliev13}
{Vasiliev} E.,  {Merritt} D.,  2013, \mn@doi [\apj]
  {10.1088/0004-637X/774/1/87}, \href
  {http://adsabs.harvard.edu/abs/2013ApJ...774...87V} {774, 87}

\bibitem[\protect\citeauthoryear{{Vasiliev}, {Antonini}  \&
  {Merritt}}{{Vasiliev} et~al.}{2014}]{vasiliev14}
{Vasiliev} E.,  {Antonini} F.,   {Merritt} D.,  2014, \mn@doi [\apj]
  {10.1088/0004-637X/785/2/163}, \href
  {http://adsabs.harvard.edu/abs/2014ApJ...785..163V} {785, 163}

\bibitem[\protect\citeauthoryear{{Vasiliev}, {Antonini}  \&
  {Merritt}}{{Vasiliev} et~al.}{2015}]{vasiliev15}
{Vasiliev} E.,  {Antonini} F.,   {Merritt} D.,  2015, \mn@doi [\apj]
  {10.1088/0004-637X/810/1/49}, \href
  {http://adsabs.harvard.edu/abs/2015ApJ...810...49V} {810, 49}

\bibitem[\protect\citeauthoryear{{Vaughan}, {Uttley}, {Markowitz},
  {Huppenkothen}, {Middleton}, {Alston}, {Scargle}  \& {Farr}}{{Vaughan}
  et~al.}{2016}]{vaughan16}
{Vaughan} S.,  {Uttley} P.,  {Markowitz} A.~G.,  {Huppenkothen} D.,
  {Middleton} M.~J.,  {Alston} W.~N.,  {Scargle} J.~D.,   {Farr} W.~M.,  2016,
  \mn@doi [\mnras] {10.1093/mnras/stw1412}, \href
  {http://adsabs.harvard.edu/abs/2016MNRAS.461.3145V} {461, 3145}

\bibitem[\protect\citeauthoryear{{Vecchio}, {Colpi}  \& {Polnarev}}{{Vecchio}
  et~al.}{1994}]{vecchio94}
{Vecchio} A.,  {Colpi} M.,   {Polnarev} A.~G.,  1994, \mn@doi [\apj]
  {10.1086/174683}, \href {http://adsabs.harvard.edu/abs/1994ApJ...433..733V}
  {433, 733}

\bibitem[\protect\citeauthoryear{{Volonteri}, {Haardt}  \& {Madau}}{{Volonteri}
  et~al.}{2003}]{volonteri03}
{Volonteri} M.,  {Haardt} F.,   {Madau} P.,  2003, \mn@doi [\apj]
  {10.1086/344675}, \href {http://adsabs.harvard.edu/abs/2003ApJ...582..559V}
  {582, 559}

\bibitem[\protect\citeauthoryear{{Wang} \& {Li}}{{Wang} \&
  {Li}}{2011}]{wangli11}
{Wang} J.,  {Li} Y.,  2011, \mn@doi [\apjl] {10.1088/2041-8205/742/1/L12},
  \href {http://adsabs.harvard.edu/abs/2011ApJ...742L..12W} {742, L12}

\bibitem[\protect\citeauthoryear{{Wang}, {Chen}, {Hu}, {Mao}, {Zhang}  \&
  {Bian}}{{Wang} et~al.}{2009}]{wang09}
{Wang} J.-M.,  {Chen} Y.-M.,  {Hu} C.,  {Mao} W.-M.,  {Zhang} S.,   {Bian}
  W.-H.,  2009, \mn@doi [\apjl] {10.1088/0004-637X/705/1/L76}, \href
  {http://adsabs.harvard.edu/abs/2009ApJ...705L..76W} {705, L76}

\bibitem[\protect\citeauthoryear{{Wang}, {Berczik}, {Spurzem}  \&
  {Kouwenhoven}}{{Wang} et~al.}{2014}]{wang14}
{Wang} L.,  {Berczik} P.,  {Spurzem} R.,   {Kouwenhoven} M.~B.~N.,  2014,
  \mn@doi [\apj] {10.1088/0004-637X/780/2/164}, \href
  {http://adsabs.harvard.edu/abs/2014ApJ...780..164W} {780, 164}

\bibitem[\protect\citeauthoryear{{Wang}, {Greene}, {Ju}, {Rafikov}, {Ruan}  \&
  {Schneider}}{{Wang} et~al.}{2016}]{wang16}
{Wang} L.,  {Greene} J.~E.,  {Ju} W.,  {Rafikov} R.~R.,  {Ruan} J.~J.,
  {Schneider} D.,  2016, \apj, p. submitted

\bibitem[\protect\citeauthoryear{{Yu}}{{Yu}}{2002}]{yu02a}
{Yu} Q.,  2002, \mn@doi [\mnras] {10.1046/j.1365-8711.2002.05242.x}, \href
  {http://adsabs.harvard.edu/abs/2002MNRAS.331..935Y} {331, 935}

\bibitem[\protect\citeauthoryear{{Yu}, {Lu}  \& {Kauffmann}}{{Yu}
  et~al.}{2005}]{yu05}
{Yu} Q.,  {Lu} Y.,   {Kauffmann} G.,  2005, \mn@doi [\apj] {10.1086/433166},
  \href {http://adsabs.harvard.edu/abs/2005ApJ...634..901Y} {634, 901}

\bibitem[\protect\citeauthoryear{{de Diego}, {Dultzin-Hacyan}, {Benitez}  \&
  {Thompson}}{{de Diego} et~al.}{1998}]{dediego98}
{de Diego} J.~A.,  {Dultzin-Hacyan} D.,  {Benitez} E.,   {Thompson} K.~L.,
  1998, \aap, \href {http://adsabs.harvard.edu/abs/1998A%26A...330..419D} {330,
  419}

\bibitem[\protect\citeauthoryear{{del Valle}, {Escala}, {Maureira-Fredes},
  {Molina}, {Cuadra}  \& {Amaro-Seoane}}{{del Valle} et~al.}{2015}]{delvalle15}
{del Valle} L.,  {Escala} A.,  {Maureira-Fredes} C.,  {Molina} J.,  {Cuadra}
  J.,   {Amaro-Seoane} P.,  2015, \mn@doi [\apj] {10.1088/0004-637X/811/1/59},
  \href {http://adsabs.harvard.edu/abs/2015ApJ...811...59D} {811, 59}

\makeatother
\end{thebibliography}


\label{lastpage}
\end{document}